\documentclass[preprint,3p]{elsarticle}

\makeatletter
\def\ps@pprintTitle{%
 \let\@oddhead\@empty
 \let\@evenhead\@empty
 \def\@oddfoot{\centerline{\thepage}}%
 \let\@evenfoot\@oddfoot}
\makeatother

\usepackage{graphics}
\usepackage{graphicx}
\usepackage{subfig}
\usepackage{epsfig}
\usepackage{amsmath,amssymb,esint}
\usepackage{amsthm}
\usepackage{float}   
\usepackage{caption}
\usepackage[dvipsnames]{xcolor}
\usepackage{multirow}
\usepackage[colorlinks]{hyperref}

\usepackage{tikz}
\usetikzlibrary{calc}
\usetikzlibrary{arrows.meta}
\tikzset{>={Latex[width=1mm,length=1mm]}}
\usetikzlibrary{shapes.geometric, arrows}

\tikzstyle{process} = [rectangle, minimum width=2.5cm, minimum height=1cm, text
centered, text width = 3cm, draw=black]
\tikzstyle{decision} = [diamond, minimum width=2.5cm, minimum height=1cm,
aspect=2,inner sep=-0.5ex,text centered, text width = 2.5cm, draw=black]
\tikzstyle{arrow} = [thick,->,>=stealth]
\tikzstyle{line} = [draw, -latex']

\newcommand{\revE}[1]{{\color{black} #1}}

\biboptions{sort&compress}

\journal{Journal of Computational Physics}

\begin{document}

\begin{frontmatter}

\title{Conservative finite-volume framework and pressure-based algorithm for flows of incompressible, ideal-gas and real-gas fluids at all speeds}

\author{Fabian Denner\corref{cor1}}
\ead{fabian.denner@ovgu.de}
\author{Fabien Evrard\corref{cor2}}
\author{Berend G.M.~van Wachem\corref{cor2}}

\address{Chair of Mechanical Process Engineering,
Otto-von-Guericke-Universit\"{a}t Magdeburg,\\ Universit\"atsplatz 2, 39106
Magdeburg, Germany}

\cortext[cor1]{Corresponding author: }

\begin{abstract}
A conservative finite-volume framework, based on a collocated variable arrangement, for the simulation of flows at all speeds, applicable to incompressible, ideal-gas and real-gas fluids is proposed in conjunction with a fully-coupled pressure-based algorithm. The applied conservative discretisation and implementation of the governing conservation laws as well as the definition of the fluxes using a momentum-weighted interpolation are identical for incompressible and compressible fluids, and are suitable for complex geometries represented by unstructured meshes. Incompressible fluids are described by predefined constant fluid properties, while the properties of compressible fluids are described by the Noble-Abel-stiffened-gas model, with the definitions of density and specific static enthalpy of both incompressible and compressible fluids combined in a unified thermodynamic closure model. The discretised governing conservation laws are solved in a single linear system of equations for pressure, velocity and \revE{temperature}. Together, the conservative finite-volume discretisation, the unified thermodynamic closure model and the pressure-based algorithm yield a conceptually simple, but versatile, numerical framework. 
The proposed numerical framework is validated thoroughly using a broad variety of test-cases, with Mach numbers ranging from 0 to 239, including viscous flows of incompressible fluids as well as the propagation of acoustic waves and transiently evolving supersonic flows with shock waves in ideal-gas and real-gas fluids.
These results demonstrate the accuracy, robustness and the convergence, as well as the conservation of mass and energy, of the numerical framework for flows of incompressible and compressible fluids at all speeds, on structured and unstructured meshes. 
In particular, the precise recovery of a divergence-free velocity field in the incompressible limit, the accurate prediction of acoustic waves, and the convergence to the correct weak solution for strong shock waves with the same finite-volume discretisation and pressure-based algorithm are important features of the proposed numerical framework. 
\end{abstract}
\begin{keyword}
Finite-volume methods \sep Pressure-based algorithms \sep Flows at all speeds \sep Compressible fluids \sep Incompressible fluids \sep Unstructured meshes \\~\\
\textcopyright~2020. This manuscript version is made available under the CC-BY-NC-ND 4.0 license. \href{http://creativecommons.org/licenses/by-nc-nd/4.0/}{http://creativecommons.org/licenses/by-nc-nd/4.0/}
\end{keyword}
\end{frontmatter}

\section{Introduction}

Since the seminal work of \citet{Harlow1968,Harlow1971a}, the formulation of numerical algorithms that can be applied for fluid flows at any speed is a central quest in {\em computational fluid dynamics} (CFD). Yet, despite extensive research efforts over the past 50 years, the development of numerical methods and algorithms that are able to provide an accurate and robust prediction of the behaviour of fluids with different compressibility and of fluid flows at all speeds has proven difficult. Although the flow of any fluid and at any speed is described by the governing equations describing the conservation of mass, momentum and energy, different modelling assumptions with respect to the compressibility of the fluid and the different physical mechanisms dominating at different flow speeds yield dissimilar mathematical characteristics of the governing equations. This in turn leads to distinct and often contrasting numerical requirements.

When developing numerical methods for flows at all speeds, it is important to recognise the numerical implications associated with the flow speed $U$, represented by the Mach number $M=U/a$, where $a = \sqrt{1/(\rho \beta_\textup{s})}$ is the speed of sound, and with the isentropic compressibility of the fluid, $\beta_\textup{s} = \{\textup{d}\rho / (\rho \, \textup{d}p)\}_\textup{s}$, that relates changes in pressure $p$ and density $\rho$ of a fluid at constant entropy.  
While pressure and density are strongly coupled for large flow speeds ($M > 0.1$), in particular for supersonic flows ($M > 1$), the pressure-density coupling diminishes at low Mach numbers and vanishes for $M \rightarrow 0$, where $\textup{d}\rho \rightarrow 0$. Founded on the observation that density changes are small at small speeds, a common assumption when modelling fluid flows is that the fluid is {\em incompressible}, with a constant density ($\textup{d}\rho = 0$) along the fluid particle trajectories and, consequently, $\beta_\textup{s} = 0$. Hence, pressure waves propagate with infinite speed ($a \rightarrow \infty$) in incompressible fluids, contrary to compressible fluids where $\beta_\textup{s} > 0$ and $0<a<\infty$. In fact, the convergence of solutions of the governing equations of the flow of compressible fluids to the governing equations of the flow of incompressible fluids for $M \rightarrow 0$ has been proven rigorously by \citet{Klainerman1981} and \citet{Hoff1998}.
In addition to the governing conservation laws, compressible fluids require a thermodynamic closure model that describes the relationship between density, pressure and energy. The ideal-gas model represents the most simple and most widely used thermodynamic closure model, with $p \propto (\rho,T)$, where $T$ is the temperature. More complex formulations, so-called {\em real-gas} models, further include the effects of intermolecular repulsion \citep{Toro2009}, intermolecular attraction \citep{Harlow1971, Saurel2007} or both \citep{Hill1986, LeMetayer2016}, or other material properties, {\em e.g.}~the acentric factor describing the shape of the molecules \citep{Soave1972, Peng1976}.
For an incompressible fluid, however, no closure model is required, since the density is not coupled to pressure, and an isothermal flow of an incompressible fluid is fully described only by the momentum and continuity equations, {\em i.e.}~the energy equation becomes redundant.

The challenge in developing numerical frameworks that are applicable to incompressible fluids and compressible fluids at all flow speeds is, therefore, to construct a numerical method that combines a unified thermodynamic closure model, a uniform set of interpolation functions, a consistent handling of the incompressible limit, shock capturing capabilities, a method to advect the solution that is applicable in all speed regimes, as well as a set of solution variables that are physically meaningful for incompressible and compressible fluids \citep{Hauke1998}. 

The choice of solution variables is of particular importance in constructing a numerical method that is applicable to flows at all speeds, since a unified algorithm is predicated on a single set of solution variables \citep{Hauke1998}. 
Choosing the conserved variables, {\em i.e.}~density, momentum and total energy, as solution variables for the continuity, momentum and energy equations, respectively, is desirable for compressible fluids at sufficiently large speeds ($M> 0.1$). However, the continuity equation is not effective as a transport equation for density in the incompressible limit, because $\textup{d} \rho \rightarrow 0$, and, instead, becomes a constraint on the velocity field with $\boldsymbol{\nabla} \cdot \boldsymbol{u} \rightarrow 0$ \citep{Chorin1993}.
An attractive choice of the solution variables for numerical algorithms applicable to predict flows at all speeds is, therefore, the primitive variables including pressure \citep{Harlow1971, Chen1991, Hauke1998, Chen2010}, {\em i.e.}~pressure, velocity and temperature.
Using pressure as a solution variable, the vanishing density differences in the incompressible limit do not pose a problem and the pressure acts as a Lagrange multiplier that enforces $\boldsymbol{\nabla} \cdot \boldsymbol{u} \rightarrow 0$ \citep{Ferziger2020, Ozanski2017, Toutant2017}. Conveniently, choosing primitive variables as solution variables still allows to discretise the governing equations in conservative form \citep{VanDoormaal1987}.
In practice, however, achieving accurate conservation of mass and energy, constructing robust shock capturing schemes and ensuring a stable numerical solution in the transonic regime has proven difficult in the context of primitive variables \citep{Bijl1998, Wesseling2001}.
It is, therefore, convenient to develop numerical algorithms {\em either} for incompressible fluids {\em or} for compressible fluids, which has led to two primary classes of algorithms: pressure-based algorithms and density-based algorithms.

{\em Pressure-based} algorithms, in which the continuity equation serves as an equation for pressure, while density is constant (incompressible fluid) or evaluated explicitly using an equation of state (compressible fluid), may be used to predict flows at all speeds, see {\em e.g.}~\citep{Karki1989, Rhie1989, Chen1991, Karimian1995, Demirdzic1995, Moukalled2001, Acharya2007, Javadi2008, Tsui2008, Darwish2009a, Chen2010, Darwish2014, Denner2014a, Denner2018c}.
For both incompressible and compressible fluids, the majority of pressure-based algorithms are founded on pressure-correction methods, such as projection methods \citep{Chorin1968, Bell1989}, the SIMPLE method \citep{Patankar1972,Patankar1980} and its subsequent derivatives, or the PISO method \citep{Issa1985, Issa1986}.
However, the weak coupling between density, pressure, velocity and energy of the discretised governing equations as a result of the iterative predictor-corrector solution procedure, which usually necessitates underrelaxation of the discretised equations to reach a converged solution, is a key shortcoming of segregated methods \citep{Kunz1999, Darwish2014}. This has motivated the development of coupled algorithms, where the discretised governing equations are solved in a single linear system of equations, for both incompressible fluids \citep{vanWachem2006a, vanWachem2007, Darwish2009a, Denner2014a} and compressible fluids \citep{Chen1991, Karimian1995, Chen2010, Darwish2014, Xiao2017, Denner2018c}, showing great potential in terms of versatility,
robustness and performance in all speed regimes. 
For instance, \citet{Darwish2009a} demonstrated substantial performance benefits for incompressible flows compared to pressure-correction methods and \citet{Denner2018b} reported robust results for flows with Mach numbers ranging from $0.001$ to $100$ with a fully-coupled pressure-based algorithm.

Contemporary numerical methods for the simulation of compressible flows are typically predicated on {\em density-based} algorithms, {\em e.g.}~\citep{Beam1978, MacCormack1982, Turkel1997}, where the conserved quantities are chosen as solution variables for the governing conservation equations and, in particular, the continuity equation serves as an equation for density. 
While density-based algorithms are naturally suited for compressible flows, they are poorly suited for low-Mach number flows \citep{Karimian1994, Wesseling2001}, where the coupling of pressure and density vanishes.
Although density-based algorithms have been applied to low-Mach number flows with some success, this requires pre-conditioning techniques \citep{Turkel1993, Turkel1993a, Turkel1997, Kadioglu2005, Turkel2006} that are computationally very expensive, especially for transient problems, and the success of which is typically determined, at least in parts, by predefined constants \citep{Turkel1987,Turkel1993}.
In order to improve the performance for low Mach number flows, recent work has been focusing on combining density-based methods with segregated pressure-correction algorithms \citep{Xiao2004,Nerinckx2005, Xiao2006, Fuster2018, Kraposhin2018} and/or reformulating the energy equation as an equation for pressure \citep{vanderHeul2003, Munz2003, Nerinckx2005, Park2005, Cordier2012}. 
These density-based algorithms have been applied successfully to a wide range of flows, including flows ranging from an incompressible flow to the propagation of strong shock waves, stationary and high-speed discontinuous waves as well as the propagation of linear acoustic waves \citep{vanderHeul2003, Xiao2004, Xiao2006, Moguen2019}.

An important aspect for the design of numerical frameworks for fluid flows at all speeds is that pressure plays an important role in all Mach number regimes \citep{Karki1989, Moukalled2016}; pressure changes are, contrary to density changes, always finite.
Exploiting this versatile role of pressure by including pressure as a primary solution variable in the numerical framework, thus, provides a seemingly distinct advantage for applications in all Mach number regimes: \revE{it provides a solution variable, {\em i.e.}~pressure, which is meaningful in all Mach number regimes and does not require particular pre-conditioning techniques.} This is further supported by the analysis of \citet{Hauke1998}, \revE{who identified the primitive variables (pressure, velocity and temperature) as particularly suitable solution variables to predict flows at all speeds.}
Remarkably, all of the numerical methods that stand out with respect to modelling fluid flows at all speeds, due to their versatility and robustness, incorporate the unique role of pressure, albeit in different ways.
In pressure-based algorithms, the special role of pressure can be taken into account through an appropriate linearisation of the discretised continuity equation \citep{VanDoormaal1987, Karki1989, Karimian1994, Denner2018c}: for compressible flows, the continuity equations serves as a transport equation for density, with density formulated as a function of pressure by an equation of state, whereas for incompressible flows, the continuity equation serves as a constraint on the divergence of the velocity field \citep{Xiao2017}, with pressure acting as a Lagrange multiplier.
The extension of density-based algorithms to low Mach numbers, either by introducing a pressure-Poisson equation \citep{Xiao2004, Xiao2006} or by reformulating the energy equation as an equation for pressure \citep{vanderHeul2003, Munz2003, Nerinckx2005}, provides a pressure-velocity coupling at low speeds and enforces a divergence-free velocity field in the incompressible limit. 
However, despite the broad variety of numerical methods able to simulate flows at all speeds, a numerical framework based on a unified conservative discretisation that is able to incorporate incompressible fluids as well as ideal-gas and real-gas compressible fluids, and which can predict flows at all speeds accurately and robustly, including low-Mach acoustics, Riemann problems and multidimensional flows ranging from the incompressible limit to supersonic flow, has not been presented in the literature yet.

In this article, a conservative, collocated, finite-volume framework in combination with a fully-coupled pressure-based algorithm for flows of incompressible, ideal-gas and real-gas fluids at all Mach numbers is proposed. The governing equations describing the conservation of continuity, momentum and energy are discretised using standard finite-volume methods and are solved for pressure, velocity and \revE{temperature} in a single linear system of equations. Incompressible fluids are described by predefined constant fluid properties, while compressible fluids are described by the Noble-Abel-stiffened-gas model \citep{LeMetayer2016}, with the definitions of density and specific static enthalpy of both incompressible and compressible fluids combined in a unified thermodynamic closure model.
This enables the design of a conceptually simple, but versatile, numerical algorithm that is able to predict flows of incompressible fluids as well as flows of compressible fluids at all speeds. 
The conservative discretisation and implementation of the governing equations are identical for incompressible and compressible fluids, employing a single definition of the fluxes based on a momentum-weighted interpolation \citep{Bartholomew2018}.
A broad variety of representative test-cases featuring flows of incompressible and compressible fluids in all Mach number regimes are considered to scrutinise and validate the proposed numerical framework: the propagation of acoustic waves, contact discontinuities and shock waves, shock tubes in different Mach number regimes, Taylor vortices in an inviscid fluid, diffusion-dominated problems, a lid-driven cavity, supersonic flow over a forward-facing step, and Stokes flow around a rotating sphere.
The presented results demonstrate the accuracy and robustness, as well as the conservation and convergence properties, of the numerical framework for all flow speeds on structured and unstructured meshes. In particular, the precise recovery of a divergence-free velocity field for $M \rightarrow 0$, the accurate prediction of acoustic waves and the convergence to the correct weak solution for $M \gg 1$ are important features of the proposed numerical framework. As such, the proposed numerical framework stands out for the simplicity of its discretisation in conjunction with the broad range of flows that can be predicted accurately and robustly.

The governing equations are introduced in Section \ref{sec:governingEqs}. Subsequently, the three main building blocks of the proposed finite-volume method are presented: a unified thermodynamic closure model in Section \ref{sec:thermodynamicClosure}, the finite-volume discretisation in Section \ref{sec:fvm}, and the pressure-based algorithm used to solve the discretised governing equations in Section \ref{sec:algorithm}. The results of representative test-cases are presented and discussed in Section \ref{sec:results}. The article is summarised and concluded in Section \ref{sec:conclusions}.

\section{Governing equations}
\label{sec:governingEqs}

The conservation laws governing fluid flows at all speeds, applicable to both incompressible and compressible flows, formulated in a Cartesian coordinate system, are the conservation of mass
\begin{equation}
\frac{\partial \rho}{\partial t} + \frac{\partial \rho u_i}{\partial x_i} = 0 
, \label{eq:continuity}
\end{equation}
the conservation of momentum 
\begin{equation}
\frac{\partial \rho u_j}{\partial t} + \frac{\partial \rho u_i u_j}{\partial
x_i} = - \frac{\partial p}{\partial x_j} + 
\frac{\partial \tau_{ji}}{\partial x_i}  , \label{eq:momentum}  
\end{equation}
and the conservation of energy 
\begin{equation}
\frac{\partial \rho h}{\partial t} + \frac{\partial \rho u_i h}{\partial x_i} =
\frac{\partial p}{\partial t} - \frac{\partial q_i}{\partial x_i} +
\frac{\partial}{\partial x_j} \left(\tau_{ji} \, u_i \right)  , \label{eq:energy}
\end{equation}
where $t$ is time, $\boldsymbol{u}$ is the velocity vector, $p$ is pressure, $\rho$ is the density of the fluid and  $h = h_\textup{s} + \boldsymbol{u}^2/2$ is the specific total enthalpy, with $h_\textup{s}$ the specific static enthalpy. 
The stress tensor $\boldsymbol{\tau}$ for the considered Newtonian fluids is given as
\begin{equation}
\tau_{ji} = \mu  \left(\frac{\partial
u_j}{\partial x_i} + \frac{\partial u_i}{\partial x_j}\right)
- \frac{2}{3} \mu \frac{\partial u_k}{\partial x_k} \delta_{ij},
\end{equation}
where $\mu$ is the dynamic viscosity of the fluid. Heat conduction is modelled by
Fourier's law, 
\begin{equation}
q_i = - k \frac{\partial T}{\partial x_i},
\end{equation}
where $k$ is the thermal conductivity of the fluid and $T$ is the temperature. 

The enthalpy formulation is chosen for the energy equation, rather than the more common internal energy formulation, because it leads to a straightforward application in the numerical algorithm, since the transient pressure term on the right-hand side of Eq.~(\ref{eq:energy}) does not require linearisation \citep{Denner2018c, Kraposhin2018}.
The governing conservation laws require closure through an appropriate model that defines the thermodynamic properties (see Section \ref{sec:thermodynamicClosure}).

\section{Thermodynamic closures}
\label{sec:thermodynamicClosure}

In order to close the governing conservation laws presented in Section \ref{sec:governingEqs}, the thermodynamic properties of the fluid have to be defined. In the proposed formulation, this is achieved by defining  the density $\rho$ and the specific static enthalpy
\begin{equation}
h_\textup{s} = c_p \, T + e^\ast, \label{eq:hs}
\end{equation}
where $c_p$ is the specific isobaric heat capacity and $e^\ast$ is the specific residual energy, through a set of input quantities ($\rho_0$, $c_{v}$, $c_{p}$, $\Pi$, $b$).
This approach enables the formulation of a unified thermodynamic closure for incompressible, ideal-gas and real-gas fluids, which facilitates a straightforward finite-volume discretisation that is applicable for incompressible flows as well as compressible flows in all Mach number regimes.

An incompressible fluid is characterised by a constant density, with
$\textup{d}\rho = 0$, defined as
\begin{equation}
\rho = \rho_0  . \label{eq:rhoIncomp}
\end{equation}
The specific isobaric heat capacity $c_p$ is assumed to be constant for incompressible fluids and the specific residual energy is $e^\ast=0$. The speed of sound for an incompressible fluid is given as
\begin{equation}
a=\sqrt{\left(\frac{\textup{d}p}{\textup{d}\rho}\right)_s} \rightarrow \infty,
\label{eq:soundSpeedIncomp}
\end{equation}
with subscript $s$ denoting constant entropy.

The Noble-Abel-stiffened-gas (NASG) model, originally proposed by \citet{LeMetayer2016}, is chosen to represent ideal and real gases. The NASG model is a combination of the stiffened-gas model \citep{Harlow1971,LeMetayer2004} and the Noble-Abel-gas model (also called co-volume gas model) \citep{Toro2009}, with the motivation of defining a simple gas model that accounts for molecular attraction and repulsion. The thermal and caloric equations of state of the NASG model are given as \citep{LeMetayer2016}
\begin{align}
p (v,T) & = (\gamma-1) \,  \frac{c_v \, T}{v-b} - \Pi  \label{eq:NASGthermal} \\
p (v,e) & = (\gamma-1) \, \frac{e-e_0}{v-b} - \gamma \Pi \label{eq:NASGcaloric} ,
\end{align}
respectively, where $\gamma = c_{p}/c_{v}$ is the heat capacity ratio, $c_v$ is the specific isochoric heat capacity, $v = 1/\rho$ is the specific volume, $e$ is the specific internal energy and $e_0$ is the specific reference energy. The pressure constant $\Pi$ represents attraction between molecules and is typically relevant for condensed phases, {\em e.g.}~to model liquids, while the co-volume $b$ accounts for the volume occupied by the individual molecules of the fluid. The density is given by rearranging Eq.~(\ref{eq:NASGthermal}) as
\begin{equation}
\rho = \frac{p + \Pi}{(\gamma - 1) \, c_{v} \, T + b \, (p + \Pi) } ,
\label{eq:rhoComp}
\end{equation} 
and the specific total enthalpy, $h = h_\textup{s} + \boldsymbol{u}^2/2$, follows from Eqs.~(\ref{eq:hs}), (\ref{eq:NASGthermal}) and (\ref{eq:NASGcaloric}) as
\begin{equation}
h = c_p \, T + e^\ast + \frac{\boldsymbol{u}^2}{2},
\label{eq:enthalpy}
\end{equation}
with specific residual energy
\begin{equation}
e^\ast = b \, p + e_0. \label{eq:qComp}
\end{equation}
In the following, the specific reference energy is assumed to be $e_0=0$, because only single-phase flows without phase transition and reactions are considered.
The specific heat capacities $c_v$ and $c_p$ are constant and the
speed of sound is \citep{LeMetayer2016}
\begin{equation}
a = \sqrt{\gamma \, \frac{p+\Pi}{\rho \, (1 - b \, \rho)}}.
 \label{eq:soundSpeed}
\end{equation}
Contrary to the van der Waals gas model, which also accounts for molecular attraction and repulsion, the coefficients $\Pi$ and $b$ are constant; the NASG model thus represents these molecular interactions in the simplest possible form. Furthermore, the NASG model is, unlike, for instance, the van der Waals gas model, unconditionally convex. With respect to liquids, such as water, the NASG model resolves the inaccuracy of specific heat capacities resulting from applying the classical stiffened-gas model \citep{LeMetayer2016}. The NASG model reduces to the ideal-gas (IG) model for $\Pi=0$ and $b=0$, to the Noble-Abel (NA) gas model for $\Pi = 0$ and $b > 0$, and to the stiffened-gas (SG) model  for $\Pi > 0$ and $b=0$.

In order to incorporate incompressible and compressible fluids in the same numerical framework, the definitions for the density $\rho$ and the specific residual energy $e^\ast$ are unified by the binary operators $\mathcal{C}$ and $\mathcal{I}= 1 - \mathcal{C}$.
The binary operator $\mathcal{C}$, given as
\begin{equation}
\mathcal{C} = 
\begin{cases}
0 \, , \ \text{for incompressible fluids} \\
1 \, ,  \ \text{for compressible fluids}
\end{cases}
\end{equation}
is used as a coefficient for the compressible part and, analogously, the binary operator $\mathcal{I}$ is used as a coefficient for the incompressible part of the unified closure model. The density of the fluid is then defined, based on Eqs.~(\ref{eq:rhoIncomp}) and (\ref{eq:rhoComp}), as
\begin{equation}
\rho = \mathcal{C} \, \left[\frac{p + \Pi}{(\gamma - 1) \, c_v \, T 
+ b \, (p + \Pi)} \right] + \mathcal{I} \, \rho_0,
\label{eq:rhoFull}
\end{equation}
and the specific residual energy is given based on Eq.~(\ref{eq:qComp}), and assuming $e_0=0$, as
\begin{equation}
e^\ast = \mathcal{C} \, b \, p . \label{eq:qFull}
\end{equation}
The type of fluid considered in a simulation can be simply specified through the binary operator $\mathcal{C}$, without changes to the thermodynamic closure model or the discretisation of the governing equations.
An incompressible fluid ($\mathcal{C}=0$) is, thereby, fully defined by setting $\rho_0$ and $c_{p}$, while a compressible fluid ($\mathcal{C}=1$) is defined by setting $c_v$, $c_p$, $\Pi$ and $b$.

\section{Finite-volume discretisation}
\label{sec:fvm}
The proposed numerical framework is founded on a collocated finite-volume discretisation, which is based on the integral formulation of the governing conservation laws, for unstructured meshes. Taking the convection-diffusion equation for the transport of a general flow variable, $\phi$, as an example, given as
\begin{equation}
\frac{\partial \rho \phi}{\partial t} + \frac{\partial \rho  u_i \phi}{\partial
x_i} = \frac{\partial}{\partial x_i} \left(\Gamma_\phi \frac{\partial
\phi}{\partial x_i}\right) , \label{eq:transportEq}
\end{equation}
where $\Gamma_\phi$ is the diffusion coefficient of $\phi$, its
integral form with respect to control volume $V$ is given as
\begin{equation}
\iiint_V \frac{\partial \rho \phi}{\partial t} \, \textup{d}V + \iiint_V
\frac{\partial \rho u_i \phi}{\partial x_i} \, \textup{d}V = \iiint_V 
\frac{\partial}{\partial x_i} \left(\Gamma_\phi \frac{\partial \phi}{\partial x_i}\right) \, \textup{d}V.
\label{eq:transportEqInt}
\end{equation}
The discretisation of each individual term is discussed in the following.

\subsection{Gradient evaluation}
\label{eq:spatialInterpolation}
The spatial gradient at cell centre $P$ is evaluated using the divergence theorem, given as
\begin{equation}
\left. \frac{\partial \phi}{\partial x_i} \right|_P \approx \frac{1}{V_P} \sum_f
\overline{\phi}_f \, n_{i,f} \, A_f,
\end{equation}
where $f$ denotes the faces bounding cell $P$, $V_P$ is the volume of cell $P$, $\boldsymbol{n}_f$ is the normal vector of face $f$ pointing outwards with respect to cell $P$, and $A_f$ is the area of face $f$. The face value $\overline{\phi}_f$ is interpolated from the adjacent cell centres $P$ and $Q$, schematically illustrated in Fig.~\ref{fig:discGeoGeneral}, as
\begin{equation}
\overline{\phi}_f = (1- l_{Pf}) \, \phi_P + l_{Pf} \, \phi_Q + r_{i,f}
\left.
\overline{\frac{\partial \phi}{\partial x_i}} \right|_f ,
\label{eq:cellFaceInterp}
\end{equation} 
where $l_{Pf}$ is the inverse-distance weighting coefficient, 
\begin{equation}
l_{Pf} = \frac{|\boldsymbol{r}_{Pf}|}{\Delta s_{f}} \label{eq:inverseDistWeight} ,
\end{equation}
with  $\Delta s_f$ the distance between cell centres $P$ and $Q$, and $\boldsymbol{r}_{Pf}$ is the vector connecting cell centre $P$ with face interpolation point $f'$. A formally second-order accurate gradient-based correction of mesh-skewness \citep{Demirdzic1995, Karimian2006} is included in Eq.~(\ref{eq:cellFaceInterp}), with $\boldsymbol{r}_f$ the vector connecting the interpolation point $f'$ of the face with face centre $f$ on meshes with skewness, see Fig.~\ref{fig:discGeoGeneral}.

\begin{figure}
\begin{center}
\subfloat[General discretisation]
{
\begin{tikzpicture}[scale=0.6]
\draw [-{>[scale=1.8]}, thick] (3.6,2.6) -- (5.5,2.25);
\node at (4.8,2) {${\boldsymbol{s}}_f$};
\draw [-{>[scale=1.8]}, thick] (3.6,2.6) -- (5.5,2.6);
\node at (4.8,3) {${\boldsymbol{n}}_f$};
\draw [semithick] (-1,-0.6) -- (2,-1.1);
\draw [semithick] (2,-1.1) -- (3.6,0.9);
\draw [semithick] (3.6,0.9) -- (3.6,4.3);
\draw [semithick] (3.6,4.3) -- (0.4,4.3);
\draw [semithick] (0.4,4.3) -- (-2,2.4);
\draw [semithick] (-2,2.4) -- (-1,-0.6);
\draw [semithick] (3.6,0.9) -- (5,-2);
\draw [semithick] (3.6,4.3) -- (7,3);
\draw [semithick] (7,3) -- (8,-1);
\draw [semithick] (5,-2) -- (7,-2.5);
\draw [semithick] (7,-2.5) -- (8,-1);
\draw [fill] (1.0,1.7) circle [radius=0.1];
\node [above] at (1.0,1.75) {$P$};
\draw [fill] (5.9,0.5) circle [radius=0.1];
\node [above] at (5.9,0.55) {$Q$};
\draw [dashed, semithick] (1.0,1.7) -- (5.9,0.5);
\draw [fill] (3.6,2.6) circle [radius=0.07];
\node at (3.35,2.8) {$f$};
\draw [thick] (3.6,1.07) circle [radius=0.07];
\node at (3.95,1.5) {$f'$};
\draw [-{>[scale=1.8]}, thick] (3.6,1.05) -- (3.6,2.6);
\node at (3.2,2) {$\boldsymbol{r}_f$};
\end{tikzpicture}  \label{fig:discGeoGeneral}
}
\qquad
\subfloat[TVD differencing]
{
\begin{tikzpicture}[scale=0.6]
\draw [->, thick] (3,4) -- (4,4);
\node [above] at (3.3,4) {$\boldsymbol{u}$};
\draw [thick] (-1,-1) -- (2,-1.5);
\draw [thick] (2,-1.5) -- (3.6,0.5);
\draw [thick] (3.6,0.5) -- (3.8,3.3);
\draw [thick] (3.8,3.3) -- (0.4,4);
\draw [thick] (0.4,4) -- (-2,2);
\draw [thick] (-2,2) -- (-1,-1);
\draw [thick] (3.6,0.5) -- (6.5,0.0);
\draw [thick] (3.8,3.3) -- (5.7,5);
\draw [thick] (5.7,5) -- (8.8,4.4);
\draw [thick] (6.5,0) -- (8.6,1.3);
\draw [thick] (8.6,1.3) -- (8.8,4.4);
\draw [fill] (1.2,1.2) circle [radius=0.1];
\node [above] at (1.2,1.2) {$U$};
\draw [fill] (6.25,2.6) circle [radius=0.1];
\node [above] at (6.25,2.6) {$D$};
\draw [fill] (3.7,1.9) circle [radius=0.1];
\node at (3.35,2.2) {$f$};
\end{tikzpicture} \label{fig:discGeoTVD}
} 
\caption{Schematic illustration of (a) cell $P$ with its neighbour cell $Q$ and the shared face $f$, where $\boldsymbol{n}_f$ is the unit normal vector of face $f$ and $\boldsymbol{s}_f$ is the unit vector connecting cell centres $P$ and $Q$ (both outward pointing with respect to cell $P$), with $f'$ the interpolation point associated with face $f$ and $\boldsymbol{r}_f$ the vector from interpolation point $f'$ to face centre $f$, and (b) upwind cell $U$ and downwind cell $D$ of face $f$, where $\boldsymbol{u}$ represents the velocity vector.}
\label{fig:discGeo}
\end{center}
\end{figure}

\subsection{Transient terms}
\label{sec:transientTerm}
The First-Order Backward Euler scheme, also widely known as BDF1 scheme, and the Second-Order Backward Euler scheme, also widely known as BDF2 scheme, are used to discretise the transient terms of the governing flow equations. The transient term of the transport equation (\ref{eq:transportEqInt}), with $\Phi = \rho \phi$, is given for cell $P$ discretised with the First-Order Backward Euler scheme as
\begin{equation}
\iiint_{V} \frac{\partial \Phi}{\partial t} \, \textup{d}V \approx \frac{\Phi_P
- \Phi_P^{(t-\Delta t_1)}}{\Delta t_1} \, V_P + \mathcal{O}(\Delta t_1),
\end{equation}
and discretised with the Second-Order Backward Euler scheme as \citep{Denner2019a}
\begin{equation}
\iiint_{V} \dfrac{\partial \Phi}{\partial t} \ \textup{d}V \approx \left[
\left(\dfrac{1}{\Delta t_1} + \dfrac{1}{\Delta \tau} \right)
\Phi_P - \left(\dfrac{1}{\Delta t_1} + \frac{1}{\Delta t_2} \right) \,
\Phi_P^{(t-\Delta t_1)} + \dfrac{\Delta t_1}{\Delta t_2 \Delta \tau} \,
\Phi_P^{(t-\Delta \tau)} \right] V_P + \mathcal{O}(\Delta t_1 \Delta \tau) ,
\label{eq:sobeRaw2}
\end{equation}
with $\Delta \tau = \Delta t_1 + \Delta t_2$, where $\Delta t_1$ is the current time-step, $\Delta t_2$ is the previous time-step, superscript $(t-\Delta t_1)$ denotes values of the previous time-level and superscript $(t-\Delta \tau)$ denotes values of the previous-previous time-level. If the time-step is constant, with $\Delta t_1 = \Delta t_2$, the transient term of Eq.~(\ref{eq:transportEqInt}) discretised with the Second-Order Backward Euler scheme simplifies to the more familiar form
\begin{equation}
\iiint_{V} \frac{\partial \Phi}{\partial t} \, \textup{d}V \approx \frac{3
\Phi_P - 4 \Phi_P^{(t-\Delta t_1)} + \Phi_P^{(t-2\Delta t_1)}}{2 \Delta t_1} \,
V_P + \mathcal{O}(\Delta t_1^2).
\end{equation} 
For consistency, all transient terms of the governing equations (\ref{eq:continuity})-(\ref{eq:energy}) are discretised with the same scheme \citep{Denner2018c}.

\subsection{Advection terms}
\label{sec:advectionTerm}
Applying the divergence theorem, the advection term of Eq.~(\ref{eq:transportEqInt}) is given as
\begin{align}
\iiint_V \frac{\partial \rho u_i \phi}{\partial x_i} \, \textup{d}V & =
\oiint_{\partial V} \rho u_i \phi \, \textup{d}S_i , 
\end{align}
where $\boldsymbol{S}$ is the outward-pointing surface vector on the surface $\partial V$ of control volume $V$. Assuming the surface of the control volume has a finite number of flat faces $f$ with area $A_f$, and applying the midpoint rule \citep{Ferziger2020, Moukalled2016}, the advection term follows in semi-discretised form as
\begin{align}
\oiint_{\partial V} \rho u_i \phi \, \textup{d}S_i & \approx \sum_f \tilde{\rho}_f
\vartheta_f \tilde{\phi}_f A_f ,
\end{align}
where $\vartheta_f = \boldsymbol{u}_f \cdot \boldsymbol{n}_f$ is the advecting velocity at face $f$, which will be discussed in detail in Section \ref{sec:advectingVel}. The advected variable $\tilde{\phi}_f$ and the density $\tilde{\rho}_f$ are interpolated using a TVD interpolation for three-dimensional unstructured meshes with an implicit correction of mesh skewness \citep{Denner2015a}, given as
\begin{equation}
\tilde{\phi}_f = \phi_U + \xi_f \frac{|\boldsymbol{r}_{Uf}|}{\Delta s_f} 
(\phi_D-\phi_U) \ , \label{eq:tvd}
\end{equation}
where subscripts $U$ and $D$ denote the upwind and downwind cells, as illustrated in Fig.~\ref{fig:discGeoTVD}, $\xi_f$ is the flux limiter and $\boldsymbol{r}_{Uf}$ is the vector connecting the cell centre of the upwind cell $U$ with face interpolation point $f'$. 
A detailed description of the implementation of this TVD interpolation using common TVD schemes on skewed and non-equidistant meshes can be found in \citep{Denner2015a}. 
In this study, the first-order upwind scheme, $\xi_f = 0$, the central differencing scheme, $\xi_f = 1$, and the Minmod scheme \citep{Roe1986}, $\xi_f(g_f) = \text{max} (0, \text{min}(1,g_f))$, where $g_f$ is the ratio of the upwind and downwind gradients of $\phi$ \citep{Denner2015a}, are considered. 

\subsection{Diffusion terms}
\label{sec:diffusionTerm}
Applying the divergence theorem and the midpoint rule, the diffusion term of the transport equation (\ref{eq:transportEqInt}) is given as
\begin{align}
\iiint_V  \frac{\partial}{\partial x_i}
\left(\Gamma_\phi \frac{\partial \phi}{\partial x_i}\right) \, \textup{d}V &
\approx \sum_f \Gamma_{\phi,f} \left. \frac{\partial \phi}{\partial x_i}
\right|_f n_{i,f} \, A_f.
\end{align}
Following \citet{Ferziger2003}, the diffusion coefficient $\Gamma_\phi$ at face $f$ is defined as
\begin{equation}
\frac{1}{\Gamma_{\phi,f}} = \frac{1 - l_{Pf}}{\Gamma_{\phi,P}} +
\frac{l_{Pf}}{\Gamma_{\phi,Q}} \ . \label{eq:diffCoeffFace}
\end{equation}
Considering an orthogonal mesh, where the unit normal vector $\boldsymbol{n}_f$ of face $f$ and the unit vector $\boldsymbol{s}_f$ connecting the adjacent cell centres $P$ and $Q$ are parallel, with $\boldsymbol{n}_f = \boldsymbol{s}_f$, the face-centred gradient is approximated with second-order accuracy as
\begin{equation}
\left. \frac{\partial \phi}{\partial x_i} \right|_f n_{i,f} \approx
\frac{\phi_Q-\phi_P}{\Delta s_f}.
\label{eq:gradFace}
\end{equation}
The decomposition and deferred correction approach of \citet{Demirdzic1982} is applied to correct for non-orthogonality of the mesh, as illustrated in Fig.~\ref{fig:discGeoGeneral}, with the face-centred gradient defined as \citep{Demirdzic1995}
\begin{equation}
\left. \frac{\partial \phi}{\partial x_i} \right|_f n_{i,f} \approx
\alpha_f \frac{\phi_Q-\phi_P}{\Delta s_f} +
\left. \overline{\frac{\partial \phi}{\partial x_i}} \right|_f  (n_{i,f} -
\alpha_f s_{i,f} ) .
\label{eq:gradFaceDeferred}
\end{equation}
The scaling factor $\alpha_f = ({\boldsymbol{n}}_f \cdot {\boldsymbol{s}}_f)^{-1}$ ensures a robust convergence even for large non-orthogonality of the mesh \citep{Mathur1997, Tsui2006}. Equation (\ref{eq:gradFaceDeferred}) reduces to Eq.~(\ref{eq:gradFace}) for an orthogonal mesh with $\boldsymbol{n}_f = \boldsymbol{s}_f$.

\section{Pressure-based algorithm}
\label{sec:algorithm}

A finite-volume framework with a pressure-based algorithm for the prediction of flows of incompressible fluids and compressible fluids at all speeds is proposed. To this end, the governing equations (\ref{eq:continuity})-(\ref{eq:energy}) are closed by the thermodynamic closure model and discretised using the finite-volume discretisation presented in Sections \ref{sec:thermodynamicClosure} and \ref{sec:fvm}, respectively. Once discretised and linearised as detailed below, the governing equations are solved simultaneously in a single linear system of equations, $\boldsymbol{\mathcal{A}} \boldsymbol{\psi} = \boldsymbol{\sigma}$, for the pressure $p$, the velocity vector $\boldsymbol{u} \equiv (u,v,w)^T$ and the \revE{temperature $T$}. For a three-dimensional computational mesh with $N$ cells, the linear system of governing equations is given as
\begin{equation}
\begin{pmatrix}
{\boldsymbol{\mathcal{A}}}^{\rho,p} &
\boldsymbol{\mathcal{A}}^{\rho,u} & 
\boldsymbol{\mathcal{A}}^{\rho,v} & 
\boldsymbol{\mathcal{A}}^{\rho,w} & 
{\boldsymbol{0}} \\
{\boldsymbol{\mathcal{A}}}^{\rho u,p} &
{\boldsymbol{\mathcal{A}}}^{\rho u,u} & 
{\boldsymbol{\mathcal{A}}}^{\rho u,v} & 
{\boldsymbol{\mathcal{A}}}^{\rho u,w} & 
{\boldsymbol{0}} \\
{\boldsymbol{\mathcal{A}}}^{\rho v,p} &
{\boldsymbol{\mathcal{A}}}^{\rho v,u} & 
{\boldsymbol{\mathcal{A}}}^{\rho v,v} & 
{\boldsymbol{\mathcal{A}}}^{\rho v,w} & 
{\boldsymbol{0}} \\
{\boldsymbol{\mathcal{A}}}^{\rho w,p} &
{\boldsymbol{\mathcal{A}}}^{\rho w,u} & 
{\boldsymbol{\mathcal{A}}}^{\rho w,v} & 
{\boldsymbol{\mathcal{A}}}^{\rho w,w} & 
{\boldsymbol{0}} \\
{\boldsymbol{\mathcal{A}}}^{\rho h,p} &
{\boldsymbol{\mathcal{A}}}^{\rho h,u} & 
{\boldsymbol{\mathcal{A}}}^{\rho h,v} & 
{\boldsymbol{\mathcal{A}}}^{\rho h,w} & 
{\boldsymbol{\mathcal{A}}}^{\rho h,\revE{T}}
\end{pmatrix} \cdot
\begin{pmatrix}
\boldsymbol{\psi}^p \\
\boldsymbol{\psi}^u \\
\boldsymbol{\psi}^v \\
\boldsymbol{\psi}^w \\
\boldsymbol{\psi}^\revE{T}
\end{pmatrix} =
\begin{pmatrix}
\boldsymbol{\sigma}^{\rho} \\
\boldsymbol{\sigma}^{\rho u} \\
\boldsymbol{\sigma}^{\rho v} \\
\boldsymbol{\sigma}^{\rho w} \\
\boldsymbol{\sigma}^{\rho h}
\end{pmatrix},
\label{eq:eqsys}
\end{equation}
where $\boldsymbol{\mathcal{A}}^{\zeta,\chi}$, with $\zeta$ the conserved quantity and $\chi$ the primary solution variable of a given governing equation, are the coefficient submatrices of size $N \times N$ of  the continuity equation (\ref{eq:continuityDiscFull}) for $\zeta = \rho$, the momentum equations (\ref{eq:momentumDiscFull}) for $\zeta \in \{ \rho u, \rho v, \rho w\}$, and the energy equation (\ref{eq:energyDiscFull}) for $\zeta = \rho h$. The subvectors $\boldsymbol{\psi}^\chi$ of length $N$ hold the solution for primary solution variable $\chi$ and the subvectors $\boldsymbol{\sigma}^\zeta$ of length $N$ hold all contributions from previous nonlinear iterations and previous time-levels.

The solution procedure performs nonlinear iterations in which the linear system of governing equations (\ref{eq:eqsys}) is solved using the {\em Block-Jacobi} preconditioner and the {\em BiCGSTAB} solver of the software library PETSc \citep{Balay1997, petsc-web-page, petsc-user-ref} until the residual of (\ref{eq:eqsys}) satisfies $\| \boldsymbol{\mathcal{A}} \boldsymbol{\psi} - \boldsymbol{\sigma} \| < \eta \, \| \boldsymbol{\sigma} \|$, where $\eta$ is the predefined solution tolerance and $\| \cdot \|$ denotes the $L_2$-norm, as presented and tested in detail by \citet{Denner2018c}.

\subsection{Advecting velocity}
\label{sec:advectingVel}
In the proposed numerical framework, the advecting velocity $\vartheta_f = \boldsymbol{u}_f \cdot \boldsymbol{n}_f$ is based on a momentum-weighted interpolation (MWI), originally introduced by \citet{Rhie1983}, and serves to advect the conserved quantities $\zeta = \{\rho, \rho \boldsymbol{u}, \rho h\}$. Furthermore, for flows of incompressible fluids and low Mach number flows of compressible fluids, the advecting velocity allows to solve the continuity equation for pressure (see Section \ref{sec:incompLimit}) and prevents pressure-velocity decoupling associated with the collocated variable arrangement \cite{Ferziger2020, Bartholomew2018}.

Following the work of \citet{Bartholomew2018}, the advecting velocity $\vartheta_f$ at face $f$ is given as
\begin{equation}
\begin{split}
\vartheta_f  = \overline{u}_{i,f} \, {n}_{i,f} - \hat{{d}}_f
\left[\frac{p_Q-p_P}{\Delta s_f} - \rho_f^\ast
\left(\left.
\frac{1-l_{Pf}}{\rho_P} \frac{\partial p}{\partial x_i} \right|_P +
\left. \frac{l_{Pf}}{\rho_Q} \frac{\partial p}{\partial x_i}
\right|_Q \right) {s}_{i,f}  -  \frac{\rho^{\ast
(t-\Delta t_1)}_f}{\Delta t_1} \left(\vartheta^{(t-\Delta t_1)}_f -
\overline{u}_{i,f}^{(t-\Delta t_1)} {n}_{i,f} \right) \right] ,
\label{eq:advVel}
\end{split}
\end{equation} 
where the interpolated face velocities $\overline{\boldsymbol{u}}_f$ and $\overline{\boldsymbol{u}}_f^{(t-\Delta t_1)}$ are obtained by linear interpolation, and $l_{Pf}$ is given by Eq.~(\ref{eq:inverseDistWeight}). As derived and discussed in detail by \citet{Bartholomew2018}, the coefficient $\hat{{d}}_f$ is defined as
\begin{equation}
\hat{{d}}_f = \dfrac{\left(\dfrac{V_P}{\mathcal{S}_P} +
\dfrac{V_Q}{\mathcal{S}_Q} \right)}{2+\dfrac{\rho^\ast_f}{\Delta t_1} \,
\left(\dfrac{V_P}{\mathcal{S}_P} +
\dfrac{V_Q}{\mathcal{S}_Q} \right)} ,
\end{equation}
where $\mathcal{S}_P = \sum_{j=1}^3 \mathcal{D}^{\rho u_j,u_j}_{P}$ and $\mathcal{S}_Q = \sum_{j=1}^3 \mathcal{D}^{\rho u_j,u_j}_{Q}$ are the sum of the diagonal matrix coefficients of the velocity arising from the advection and shear stress terms of the discretised momentum equations, see Eq.~(\ref{eq:coeffDiagonalVelP}) in \ref{sec:LGScoeff}. The face density is defined as
\begin{equation}
\frac{1}{\rho_f^\ast} = \frac{1-l_{Pf}}{\rho_P} + \frac{l_{Pf}}{\rho_Q} . 
\end{equation}

The MWI provides a robust pressure-velocity coupling for incompressible flows by introducing a cell-to-cell pressure coupling and applying a low-pass filter acting on the third derivative of pressure \cite{Demirdzic1995, Wesseling2001, Ferziger2020, Bartholomew2018}, thus avoiding pressure-velocity decoupling due to the collocated variable arrangement. The transient term of Eq.~(\ref{eq:advVel}) ensures a time-step independent contribution of the MWI in conjunction with the coefficient $\hat{{d}}_f$ \citep{Bartholomew2018} and is important for a correct temporal evolution of pressure waves \citep{Xiao2017, Bartholomew2018}. However, the MWI is known to introduce numerical dissipation that manifests in an unphysical dissipation of kinetic energy \citep{Ham2004, Bartholomew2018}, a conservation error that converges with $\Delta x^3$ and that is, assuming the consistent formulation given by Eq.~(\ref{eq:advVel}), independent of the applied time-step \citep{Bartholomew2018}. 

\subsection{Discretised governing equations}
\label{sec:discGovEq}
Applying the finite-volume methods described in Section \ref{sec:fvm} and, in particular, using the BDF1 scheme for the transient term in the interest of clarity, the discretised continuity equation (\ref{eq:continuity}) for cell $P$ is given as
\begin{equation}
\frac{\rho_P - \rho_P^{(t-\Delta t_1)}}{\Delta t_1} V_P + \sum_f
\tilde{\rho}_f \, \vartheta_f \, A_f = 0 .
\label{eq:continuityDisc}
\end{equation}

Similar to the discretisation of the continuity equation, applying the finite-volume scheme presented in Section \ref{sec:fvm}, the discretised momentum equations (\ref{eq:momentum}) in cell $P$ are given as
\begin{equation}
\begin{split}
\frac{\rho_P \, u_{j,P} - \rho^{(t-\Delta t_1)}_P \, u^{(t-\Delta
t_1)}_{j,P}}{\Delta t_1} V_P + \sum_f \tilde{\rho}_f \, \vartheta_f \,
\tilde{u}_{j,f} \, A_f = - \sum_f \overline{p}_f \, {n}_{j,f} \, A_f
\\ + \sum_f \mu_f \left(\left.
{\frac{\partial u_j}{\partial x_i}} \right|_f +
\left. {\frac{\partial u_i}{\partial x_j}} \right|_f \right) \,
{n}_{i,f} \, A_f 
 - \sum_f \frac{2}{3} \mu_f  \left. \overline{\frac{\partial u_k}{\partial
 x_k}} \right|_f {n}_{i,f} A_f, 
\end{split} \label{eq:momentumDisc}
\end{equation}
where  the viscosity $\mu_f$ at face $f$ is defined by Eq.~(\ref{eq:diffCoeffFace}).
In order to account for mesh non-orthogonality, the deferred correction approach given in Eq.~(\ref{eq:gradFaceDeferred}) is applied to decompose the shear-stress term as
\begin{equation}
\left(\left. \frac{\partial u_j}{\partial x_i} \right|_f + \left.\frac{\partial u_i}{\partial x_j} \right|_f \right) {n}_{i,f} \approx
\alpha_f \frac{u_{j,Q}-u_{j,P}}{\Delta s_f} + \left. \overline{\frac{\partial
u_j}{\partial x_i}} \right|_f ({n}_{i,f} - \alpha_f {s}_{i,f}) + \left. \overline{\frac{\partial u_i}{\partial x_j}} \right|_f {n}_{i,f} .
\label{eq:decompShearStress}
\end{equation}

The discretised energy equation (\ref{eq:energy}) in cell $P$, using the applied finite-volume discretisation, is given as
\begin{equation}
\begin{split}
\frac{\rho_P h_{P} - \rho^{(t-\Delta t_1)}_P h^{(t-\Delta
t_1)}_{P}}{\Delta t_1} + \sum_f \tilde{\rho}_f \, \vartheta_f \,
\tilde{h}_{f} \, A_f = \frac{p_{P} - p^{(t-\Delta t_1)}_P }{\Delta t_1}
V_P + \sum_f k_f  \left.
\frac{\partial T}{\partial x_i} \right|_f  {n}_{i,f} \,
 A_f \\ + 
  \sum_f \overline{u}_{i,f} \, \mu_f \left(\left.
\overline{\frac{\partial u_j}{\partial x_i}} \right|_f +
\left. \overline{\frac{\partial u_i}{\partial x_j}} \right|_f - \left. \frac{2}{3} \, \overline{\frac{\partial u_k}{\partial x_k}} \right|_f \right) \,
{n}_{j,f} \, A_f ,
\end{split} \label{eq:energyDisc}
\end{equation}
where the heat conduction term is decomposed as described by Eq.~(\ref{eq:gradFaceDeferred}) and the thermal conductivity $k_f$ at face $f$ is defined by Eq.~(\ref{eq:diffCoeffFace}). 

\subsection{Linearisation and implementation}
\label{sec:linearisation}
The details of the linearisation of the governing equations have been shown to be a critical aspect for all-Mach formulations and algorithms \citep{VanDoormaal1987, Karimian1994, Kunz1999, Denner2018c} and provides additional potential with respect to the performance of fully-coupled algorithms \citep{Denner2018c}. To this end, a Newton linearisation is applied to facilitate an implicit treatment of all dominant pressure, velocity and \revE{temperature} terms in the linear system resulting from the linearisation and discretisation of the governing equations (\ref{eq:continuity})-(\ref{eq:energy}), given for two generic fluid variables as
\begin{equation}
\phi_1 \, \phi_2 \Rightarrow \phi_1^{(n+1)}  \phi_2^{(n+1)} \approx \phi_1^{(n)}  \phi_2^{(n+1)} + \phi_1^{(n+1)} 
\phi_2^{(n)} - \phi_1^{(n)} \phi_2^{(n)}
\label{eq:newtonDouble}
\end{equation}
or for three generic fluid variables as
\begin{equation}
\phi_1 \, \phi_2 \, \phi_3 \Rightarrow \phi_1^{(n+1)}  \phi_2^{(n+1)} \phi_3^{(n+1)}  \approx \phi_1^{(n)} \phi_2^{(n)} \phi_3^{(n+1)} +
\phi_1^{(n)} \phi_2^{(n+1)} \phi_3^{(n)} + \phi_1^{(n+1)} \phi_2^{(n)}
\phi_3^{(n)} - 2 \phi_1^{(n)} \phi_2^{(n)} \phi_3^{(n)},
\label{eq:newtonTriple}
\end{equation}
where $n$ is the iteration counter associated with the nonlinear iterations performed to solve the system of discretised governing equations, Eq.~(\ref{eq:eqsys}), at each time-step. Superscript $(n)$ denotes the most recent available solution, which is the solution of the previous time-step during the first nonlinear iteration of a given time-step or, otherwise, the solution of the previous nonlinear iteration, and superscript $(n+1)$ denotes the solution that is sought implicitly. 

Applying the Newton linearisation given in Eq.~(\ref{eq:newtonDouble}) to the advection term and formulating the cell-centered density $\rho_P$ of the transient term as a semi-implicit function of pressure $p_P$, given as
\begin{equation}
\rho_P^{(n+1)} \approx \mathcal{C} \, \left[\frac{p_P^{(n+1)} + \Pi}{(\gamma - 1) \,
c_v \, T_P^{(n)} + b \, (p_P^{(n)} +  \Pi)}
\right] + \mathcal{I} \, \rho_0,
\label{eq:rhoFullImp}
\end{equation}
the discretised continuity equation (\ref{eq:continuityDisc}) follows as
\begin{equation}
\frac{\rho_P^{(n+1)} - \rho_P^{(t-\Delta t_1)}}{\Delta t_1} V_P + \sum_f \left(
\tilde{\rho}_f^{(n)} \, \vartheta_f^{(n+1)} + \tilde{\rho}_f^{(n+1)} \,
\vartheta_f^{(n)} - \tilde{\rho}_f^{(n)} \, \vartheta_f^{(n)} \right) \, A_f = 0 .
\label{eq:continuityDiscFull}
\end{equation} 
Following previous studies \citep{Denner2018b, Denner2018c}, the advecting velocity $\vartheta_f^{(n+1)}$ is defined by a semi-implicit formulation as
\begin{equation}
\begin{split}
\vartheta_f^{(n+1)} \approx \overline{u}_{i,f}^{(n+1)} \, {n}_{i,f} & - \hat{{d}}_f
\left[\frac{p_Q^{(n+1)} - p_P^{(n+1)}}{\Delta s_f} - \rho_f^{\ast (n)}
\left(\left. \frac{1-l_{Pf}}{\rho_P^{(n)}} \frac{\partial p}{\partial x_i}
\right|_P^{(n)} + \left. \frac{l_{Pf}}{\rho_Q^{(n)}} \frac{\partial p}{\partial
x_i} \right|_Q^{(n)} \right)  {s}_{i,f}  \right]\\ & + 
\hat{{d}}_f \, \frac{\rho^{\ast (t-\Delta t_1)}_f}{\Delta t_1}
\left(\vartheta^{(t-\Delta t_1)}_f - \overline{u}_{i,f}^{(t-\Delta t_1)} {n}_{i,f} \right)  .
\label{eq:advVelSemi}
\end{split}
\end{equation} 

Linearising the transient terms and the advection terms with the Newton linearisation given in Eqs.~(\ref{eq:newtonDouble}) and (\ref{eq:newtonTriple}), respectively, following the work of \citet{Denner2018c}, and treating cell-centered pressure and velocity contributions implicitly, the discretised momentum equations (\ref{eq:momentumDisc}) follow as
\begin{equation}
\begin{split}
& \frac{\rho_P^{(n)} u_{j,P}^{(n+1)} + \rho_P^{(n+1)} u_{j,P}^{(n)}
- \rho_P^{(n)} u_{j,P}^{(n)} - \rho^{(t-\Delta t_1)}_P u^{(t-\Delta
t_1)}_{j,P}}{\Delta t_1} V_P \\  & + \sum_f \left(\tilde{\rho}_f^{(n)}
\vartheta_f^{(n)} \tilde{u}_{j,f}^{(n+1)} + \tilde{\rho}_f^{(n)}
\vartheta_f^{(n+1)} \tilde{u}_{j,f}^{(n)}
+ \tilde{\rho}_f^{(n+1)}
\vartheta_f^{(n)} \tilde{u}_{j,f}^{(n)} - 2 \tilde{\rho}_f^{(n)}
\vartheta_f^{(n)} \tilde{u}_{j,f}^{(n)} \right) A_f 
 = - \sum_f \overline{p}_f^{(n+1)} \, {n}_{j,f} A_f 
\\ & +
\sum_f \mu_f \left( \alpha_f \frac{u_{j,Q}^{(n+1)} - u_{j,P}^{(n+1)}}{\Delta
s_f} + \left.
\overline{\frac{\partial u_j}{\partial x_i}} \right|_f^{(n)} ({n}_{i,f} - \alpha_f {s}_{i,f}) + \left.
\overline{\frac{\partial u_i}{\partial x_j}} \right|_f^{(n)} {n}_{i,f} -
\frac{2}{3} \left. \overline{\frac{\partial u_k}{\partial x_k}} \right|_f^{(n)}
{n}_{i,f} \right)  A_f  ,
\end{split} \label{eq:momentumDiscFull}
\end{equation}
and the discretised energy equation (\ref{eq:energyDisc}) becomes
\begin{equation}
\begin{split}
& \frac{\rho_P^{(n)}  h_P^{(n+1)} + \rho_P^{(n+1)}  h_P^{(n)}
- \rho_P^{(n)}  h_P^{(n)} - \rho^{(t-\Delta t_1)}_P  h^{(t-\Delta
t_1)}_{P}}{\Delta t_1} V_P \\ & + \sum_f \left(\tilde{\rho}_f^{(n)}
\vartheta_f^{(n)} \tilde{h}_{f}^{(n+1)} + \tilde{\rho}_f^{(n)}
\vartheta_f^{(n+1)} \tilde{h}_{f}^{(n)}
+ \tilde{\rho}_f^{(n+1)}
\vartheta_f^{(n)} \tilde{h}_{f}^{(n)} - 2 \tilde{\rho}_f^{(n)}
\vartheta_f^{(n)} \tilde{h}_{f}^{(n)} \right)  A_f  \\
& = \frac{p_{P}^{(n+1)} - p^{(t-\Delta t_1)}_P }{\Delta t_1} V_P
 + \sum_f k_f \left(\alpha_f \frac{T_{Q}^{(\revE{n+1})}-T_{P}^{(\revE{n+1})}}{\Delta s_f} +
 \left. \overline{\frac{\partial T}{\partial x_i}} \right|_f^{(n)} ({n}_{i,f} - \alpha_f {s}_{i,f}) \right)  A_f
 \\
& + \sum_f  \overline{u}_{i,f}^{(n+1)} \, \mu_f \left( 
\left. \overline{\frac{\partial u_j}{\partial x_i}} \right|_f^{(n)}  + \left.  \overline{\frac{\partial u_i}{\partial x_j}} \right|_f^{(n)} - \frac{2}{3} \left. \overline{\frac{\partial u_k}{\partial x_k}} \right|_f^{(n)}  \right) {n}_{j,f} \, A_f  ,
\end{split} \label{eq:energyDiscFull}
\end{equation}
with $\rho^{(n+1)}_P$ given by Eq.~(\ref{eq:rhoFullImp}) and $\vartheta^{(n+1)}_f$ given by Eq.~(\ref{eq:advVelSemi}). \revE{The implicitly computed specific total enthalpy $h_P^{(n+1)}$ at cell-centre $P$ is formulated, following Eq.~(\ref{eq:enthalpy}) and assuming $e_0 =0$, as an implicit function of temperature $T$ and pressure $p$, given as
\begin{equation}
h_P^{(n+1)} = c_p \, T_P^{(n+1)} + \mathcal{C} \, b \, p_P^{(n+1)} + \frac{\boldsymbol{u}_P^{(n),2}}{2}.
\end{equation}
This treatment enables the implicit solution of the energy equation for temperature, pressure and velocity, which allows to solve the cell-centred values of temperature of the heat conduction term implicitly, see Eq.~(\ref{eq:energyDiscFull}), and, thus, time-step restrictions associated with an explicit treatment of the heat conduction term \citep{Patankar1980} do not apply for the presented algorithm.}

Following the work of  \citet{Khosla1974}, the TVD interpolation of advected variables, see Eq.~(\ref{eq:tvd}), is implemented using a deferred correction approach, given as
\begin{equation}
\tilde{\phi}_f^{(n+1)} \approx \phi_U^{(n+1)} + \xi_f \frac{|\boldsymbol{r}_{Uf}|}{\Delta s_f}  (\phi_D^{(n)} - \phi_U^{(n)}) \ , \label{eq:tvdImpl}
\end{equation}
where the upwind contribution is treated implicitly and the high-order correction is based on the values of the previous nonlinear iteration. This interpolation is unconditionally stable \citep{Khosla1974, Wesseling2001, Moukalled2016}, which is essential for the simulation of convection-dominated flows with Peclet numbers of $\textup{Pe} =  \rho |\boldsymbol{u}| \Delta x/\mu \gg 1$ and, in particular, inviscid flows ($\textup{Pe} \rightarrow \infty$). 

The coefficients of the linear equation system $\boldsymbol{\mathcal{A}} \boldsymbol{\psi} = \boldsymbol{\sigma}$, Eq.~(\ref{eq:eqsys}), for cell $P$ follow after rearranging the discretised and linearised governing equations (\ref{eq:continuityDiscFull}), (\ref{eq:momentumDiscFull}) and (\ref{eq:energyDiscFull}) as
\begin{align}
\mathcal{A}^{\rho,p}_{P} p_P^{(n+1)}  + \mathcal{A}^{\rho,p}_{Q} p_Q^{(n+1)} + \mathcal{A}_{P}^{\rho,u_i} u_{i,P}^{(n+1)} + \mathcal{A}_{Q}^{\rho,u_i} u_{i,Q}^{(n+1)} & =  \sigma^\rho_P \label{eq:continuityCoeff} \\
\mathcal{A}^{\rho u_j,p}_{P} p_P^{(n+1)}  + \mathcal{A}^{\rho u_j,p}_{Q} p_Q^{(n+1)}  + \mathcal{A}_{P}^{\rho u_j,u_j} u_{j,P}^{(n+1)} + \mathcal{A}_{Q}^{\rho u_j,u_j} u_{j,Q}^{(n+1)} + \mathcal{A}_{P}^{\rho u_j,u_i} u_{i,P}^{(n+1)} + \mathcal{A}_{Q}^{\rho u_j,u_i} u_{i,Q}^{(n+1)} & =  \sigma^{\rho u_j}_P  \label{eq:momentumCoeff} \\
\mathcal{A}^{\rho h,p}_{P} p_P^{(n+1)}  + \mathcal{A}^{\rho h,p}_{Q} p_Q^{(n+1)} + \mathcal{A}_{P}^{\rho h,u_i} u_{i,P}^{(n+1)} + \mathcal{A}_{Q}^{\rho h,u_i} u_{i,Q}^{(n+1)} + \mathcal{A}_{P}^{\rho h,\revE{T}} \revE{T}_{P}^{(n+1)} + \mathcal{A}_{Q}^{\rho h,\revE{T}} \revE{T}_{Q}^{(n+1)} & =  \sigma^{\rho h}_P, \label{eq:energyCoeff}
\end{align}
respectively, with $Q$ the neighbour cells of cell $P$. The individual coefficients $\mathcal{A}$ and right-hand side contributions $\sigma$ are given in \ref{sec:LGScoeff}.

The strong implicit coupling of pressure, density and velocity through a Newton linearisation has been shown to be beneficial for the performance and stability of the solution algorithm in all Mach number regimes \citep{Denner2018c}. For instance, the Newton linearisation of the advection term of the continuity equation (\ref{eq:continuityDiscFull}) facilitates a smooth transition from low to high Mach number regions \citep{VanDoormaal1987, Karimian1994, Denner2018c}, with the term $\sum_f \tilde{\rho}_f^{(n)} \vartheta_f^{(n+1)} A_f$ of Eq.~(\ref{eq:continuityDiscFull}) dominant at low Mach numbers and the term $\sum_f \tilde{\rho}_f^{(n+1)} \vartheta_f^{(n)} A_f$ dominant in regions of high Mach numbers \citep{Xiao2017}. As a result, the Newton linearisation of the advection term also yields performance and stability benefits for flows with sharp changes in Mach number and strong compressibility \citep{Denner2018c}, and provides the necessary implicit pressure-velocity coupling for incompressible flows \citep{Denner2014a, Xiao2017}.

\subsection{Incompressible limit}
\label{sec:incompLimit}
The incompressible limit deserves special attention, as this is the {\em Achilles' heel} of many previously proposed numerical frameworks for flows at all speeds. From a numerical viewpoint, the {\em incompressible limit} includes both the flow of compressible fluids with very small Mach numbers ($M \rightarrow 0$) and the flow of incompressible fluids ($\rho = \textup{const.}$). As density changes of the fluid particles vanish in the incompressible limit, with $\textup{d}\rho \rightarrow 0$, the density is constant along the fluid particle trajectories \citep{Chorin1993}, with
\begin{equation}
\frac{\textup{D}\rho}{\textup{D}t} \equiv \frac{\partial \rho}{\partial t} + u_i
\frac{\partial \rho}{\partial x_i} = 0.
\label{eq:materialDerivativeDensity}
\end{equation}
Inserting Eq.~(\ref{eq:materialDerivativeDensity}) into the governing equations
(\ref{eq:continuity})-(\ref{eq:energy}) yields
\begin{align}
\frac{\partial u_i}{\partial x_i} & = 0 , \label{eq:continuityImp} \\
\rho \left(\frac{\partial  u_j}{\partial t} + \frac{\partial u_i
u_j}{\partial x_i} \right) & = - \frac{\partial p}{\partial x_j} + 
\frac{\partial \tau_{ji}}{\partial x_i}  , \label{eq:momentumImp} \\
\rho \left(\frac{\partial h}{\partial t} + \frac{\partial u_i h}{\partial
x_i} \right) & = \frac{\partial p}{\partial t} - \frac{\partial q_i}{\partial
x_i} + \frac{\partial}{\partial x_j} \left(\tau_{ji} \, u_i \right)  ,
\label{eq:energyImp}
\end{align}
for the continuity, momentum and energy equations, respectively, in the
incompressible limit. 

Applying the discretisation and linearisation schemes presented in the previous sections to the governing equations in the incompressible limit, Eqs.~(\ref{eq:continuityImp})-(\ref{eq:energyImp}), the discretised continuity equation follows as
\begin{equation}
\sum_f \vartheta_f^{(n+1)} \,  A_f = 0 ,
\label{eq:continuityDiscImp}
\end{equation}
the discretised momentum equations are given as
\begin{equation}
\begin{split}
& \rho_P \left[\frac{u_{j,P}^{(n+1)} - u^{(t-\Delta t_1)}_{j,P}}{\Delta t_1} V_P  + \sum_f  \left( \vartheta_f^{(n)} \tilde{u}_{j,f}^{(n+1)} + \vartheta_f^{(n+1)} \tilde{u}_{j,f}^{(n)} - \vartheta_f^{(n)} \tilde{u}_{j,f}^{(n)} \right) A_f \right] \\
& = - \sum_f \overline{p}_f^{(n+1)} \, {n}_{j,f} \, A_f  +
\sum_f \mu_f \left( \alpha_f \frac{u_{j,Q}^{(n+1)} - u_{j,P}^{(n+1)}}{\Delta s_f} + \left.
\overline{\frac{\partial u_j}{\partial x_i}} \right|_f^{(n)} ({n}_{i,f} - \alpha_f {s}_{i,f}) + \left.
\overline{\frac{\partial u_i}{\partial x_j}} \right|_f^{(n)} {n}_{i,f}  \right)  A_f 
\end{split} \label{eq:momentumDiscImp}
\end{equation}
and the discretised energy equation follows as
\begin{equation}
\begin{split}
& \rho_P \left[\frac{h_P^{(n+1)} - h^{(t-\Delta t_1)}_{P}}{\Delta t_1} V_P  + \sum_f \left( \vartheta_f^{(n)} \tilde{h}_{f}^{(n+1)} + \vartheta_f^{(n+1)} \tilde{h}_{f}^{(n)} - \vartheta_f^{(n)} \tilde{h}_{f}^{(n)} \right) A_f \right] \\
& = \frac{p_{P}^{(n+1)} - p^{(t-\Delta t_1)}_P }{\Delta t_1} V_P + \sum_f k_f \left(\alpha_f \frac{T_{Q}^{(\revE{n+1})}-T_{P}^{(\revE{n+1})}}{\Delta s_f} +  \left. \overline{\frac{\partial T}{\partial x_i}} \right|_f^{(n)} ({n}_{i,f} - \alpha_f {s}_{i,f}) \right)  A_f \\
& + \sum_f \overline{u}_{i,f}^{(n+1)} \, \mu_f \left( \left. \overline{\frac{\partial u_j}{\partial x_i}} \right|_f^{(n)} + \left. \overline{\frac{\partial u_i}{\partial x_j}} \right|_f^{(n)} \right) {n}_{j,f} \, A_f .
\end{split} \label{eq:energyDiscImp}
\end{equation}

The definition of the semi-implicit advecting velocity $\vartheta_f^{(n+1)}$, with the implicit treatment of the cell-centred pressure values, as defined in Eq.~(\ref{eq:advVelSemi}), yields a consistent discretisation of Eq.~(\ref{eq:continuityImp}) as a function of pressure. This allows pressure to enforce a divergence-free velocity field in the incompressible limit, as well as a robust implicit pressure-velocity coupling for the collocated variable arrangement.
Furthermore, Eqs.~(\ref{eq:continuityDiscImp})-(\ref{eq:energyDiscImp}) treat all the terms implicitly which \citet{Nerinckx2005} identified to carry acoustic information, thereby eliminating the acoustic time-step restriction and enabling an efficient  solution for $M \rightarrow 0$ and, specifically, for $M=0$.
In fact, Eqs.~(\ref{eq:continuityDiscImp}) and (\ref{eq:momentumDiscImp}) are identical to the discretised continuity and momentum equations of the fully-coupled pressure-based algorithm for incompressible interfacial flows of \citet{Denner2014a}. 
Thus, the discretised governing equations presented in Section \ref{sec:discGovEq} represent the incompressible limit accurately and facilitate the simulation of incompressible flows. If {\em isothermal} incompressible fluids are considered, the energy equation may be disregarded, removing Eq.~(\ref{eq:energyDiscFull}) from Eq.~(\ref{eq:eqsys}), although this simplification is not taken into account in the results presented in Section \ref{sec:results}.

\section{Validation}
\label{sec:results}
The results for a broad variety of test-cases are presented here to scrutinise each aspect of the thermodynamic closure, the finite-volume discretisation and the fully-coupled pressure-based algorithm, including the convergence and conservation properties. In Section \ref{sec:acoustics}, the propagation of acoustic waves is considered to test the accurate prediction of acoustic effects for both ideal-gas and real-gas fluids, in particular the amplitude of pressure waves and the speed of sound. The propagation of a moving contact discontinuity is considered in Section \ref{sec:movingContact} to test the convergence under mesh refinement for linearly degenerate waves, a distinct challenge for finite-volume methods \citep{Banks2008}. In Section \ref{sec:shockWave}, the propagation of a strong shock wave with Mach number 100 is considered to check if the proposed finite-volume framework converges to the correct weak solution of the governing equations, for both ideal-gas and real-gas fluids. Shock tubes with flows in different Mach number regimes, ranging from $M=8.5\times10^{-3}$ to $M=239$, are compared against the exact Riemann solution in Section \ref{sec:shockTubes}. The evolution of Taylor vortices in an inviscid fluid is considered in Section \ref{sec:TaylorVortices} to test the conservation of kinetic energy of the proposed numerical framework. In Section \ref{sec:diffusionDominated}, the Poiseuille flow of an incompressible fluid and the Couette flow of a compressible fluid are simulated to probe the prediction of diffusion-dominated flows, both momentum diffusion and heat conduction, by the proposed numerical framework. The flow of an incompressible fluid in a lid-driven cavity at different Reynolds numbers is considered in Section \ref{sec:ldc} to test the accurate prediction of flows in which both advection and diffusion play an important role, \revE{and to demonstrate the correct enforcement of $\boldsymbol{\nabla} \cdot \boldsymbol{u} = 0$ for incompressible fluids}. In Section \ref{sec:ffs}, a supersonic flow of an ideal gas and a real gas over a forward-facing step are simulated, predominantly to scrutinise the mass conservation for a complex flow in which different Mach number regimes coexist. Finally, in Section \ref{sec:rotatingSphere}, the Stokes flow around a rotating sphere is simulated to demonstrate the reliable prediction of flows in complex geometries.

\subsection{Acoustic waves}
\label{sec:acoustics}
As a first test, the propagation of acoustic waves in a one-dimensional domain is simulated. The formation and propagation of acoustic waves is an important feature of compressible flows and predicting acoustic waves reliably is known to be challenging \citep{Moguen2012, Xiao2017, Denner2018b, Moguen2019}. In these simulations, the acoustic waves are generated at the domain inlet by a sinusoidal velocity perturbation with amplitude $\Delta u_0$. For small perturbations to the flow, $\Delta u_0 \ll a_0$, the resulting wave is a sound wave propagating with the speed of sound $a_0$. According to linear acoustic theory, the pressure wave has an amplitude of $\Delta p_0 = Z \, \Delta u_0$ \citep{Anderson2003}, where $Z=\rho a$ is the acoustic impedance. Four different fluids, with the fluid properties given in Table \ref{tab:acousticWaves_Properties}, are considered. In each case, the unperturbed flow velocity is $u_0= 1 \, \textup{m} \, \textup{s}^{-1}$, the ambient pressure is $p_0 = 10^5 \, \textup{Pa}$ and the ambient temperature is $T_0 = 300 \, \textup{K}$, leading to the density and speed of sound given in Table \ref{tab:acousticWaves_Waves}. The computational domain has a length of $1 \, \textup{m}$, which is represented by an equidistant mesh with mesh spacing $\Delta x = 2 \times 10^{-3} \, \textup{m}$, and the applied time-steps, see Table \ref{tab:acousticWaves_Waves}, correspond to a Courant number of $\textup{Co} = a_0 \Delta t / \Delta x \simeq 0.43$. The velocity at the domain inlet is $u_\textup{in} = u_0 + \Delta u_0 \, \sin(2 \pi f t)$, with frequency $f$ as given in Table \ref{tab:acousticWaves_Waves} and amplitude $\Delta u_0 = 0.01 \, u_0$.

\begin{table}[t]
\begin{center}
\caption{Fluid properties considered for the propagation of acoustic waves.}
\label{tab:acousticWaves_Properties}
\begin{tabular}{lrrrr}
Fluid & $\gamma$ & $c_p \, [\textup{J} \, \textup{kg}^{-1} \,
\textup{K}^{-1}]$ & $b \, [\textup{m}^3 \,
\textup{kg}^{-1}]$ & $\Pi \, [\textup{Pa}]$  \\
\hline 
Air & $1.400$ & $1008$ & $0$ & $0$  \\
JA2 propellant gas \citep{Johnston2005} & $1.225$ & $1484$ & $1.00
\times 10^{-3}$ & $0$  \\
Water 1 \citep{Coralic2014} & $6.120$ & $1367$ & $0$ & $3.430 \times 10^8$ \\
Water 2 \citep{LeMetayer2016} & $1.187$ & $4285$ & $6.61 \times 10^{-4}$ &
$7.028 \times 10^8$  \\
\hline
\end{tabular}
\end{center} 
\end{table}

\begin{table}[t]
\begin{center}
\caption{Density $\rho$ and speed of sound $a_0$ of the fluids defined in Table \ref{tab:acousticWaves_Properties} for ambient pressure $p_0 = 10^5 \, \textup{Pa}$ and ambient temperature $T_0 = 300 \, \textup{K}$, the applied time-step $\Delta t$, the frequency $f$ of the acoustic waves, the wavelength $\lambda_0$ and pressure amplitude $\Delta p_0$ of the acoustic waves based on linear acoustic theory, as well as the wavelength $\lambda$ and pressure amplitude $\Delta p$ of the acoustic waves computed with the proposed numerical framework.}
\label{tab:acousticWaves_Waves}
\begin{tabular}{lrrrrrrrr}
Fluid & $\rho \, [\textup{kg} \, \textup{m}^{-3}]$ & $a_0 \,
[\textup{m} \, \textup{s}^{-1}]$ & $\Delta t \, [\textup{s}]$ & $f \,
[\textup{s}^{-1}]$ & $\lambda_0 \, [\textup{m}]$ & $\lambda \, [\textup{m}]$ & $\Delta p_0 \, [\textup{Pa}]$ &
$\Delta p \, [\textup{Pa}]$
\\
\hline 
Air & $1.1574$ & $347.8$ & $2.5 \times 10^{-6}$ & $1750$ & $0.199$ & $0.199$ &
$4.025$ & $4.025$ \\
JA2􏰄 propellant gas \citep{Johnston2005} & $1.2214$ & $316.9$ & $2.7 \times
10^{-6}$ & $1750$ & $0.181$ & $0.181$ & $3.871$ & $3.869$ \\
Water 1 \citep{Coralic2014} & $1000.0$ & $1449$ & $6.0 \times 10^{-7}$ & $7000$
& $0.207$ & $0.207$ & $14490$ & $14487$ \\
Water 2 \citep{LeMetayer2016}  & $1053.6$ & $1615$ & $5.4 \times 10^{-6}$ &
$7000$ & $0.230$ & $0.231$ & $17016$ & $17012$\\
\hline
\end{tabular}
\end{center} 
\end{table}

\begin{figure}
\begin{center}
\subfloat[Air, $t=2.5 \times 10^{-3} \, \textup{s}$]
{\includegraphics{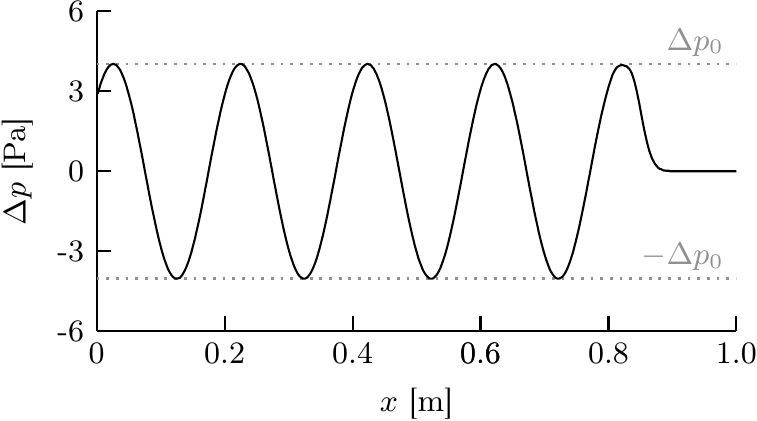}
\label{fig:acousticWaves_Air}
} 
\quad
\subfloat[JA􏰄2 propellant gas, $t=2.7 \times 10^{-3} \,
\textup{s}$] {\includegraphics{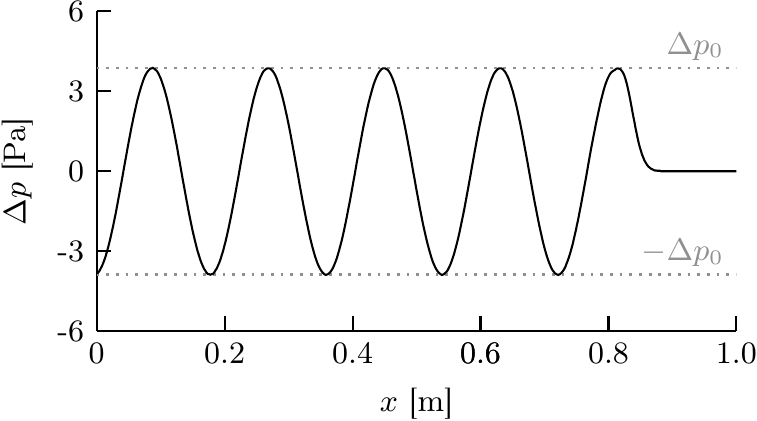}
\label{fig:acousticWaves_JA2}
}\\
\subfloat[Water 1, $t=6.0 \times 10^{-4} \, \textup{s}$]
{\includegraphics{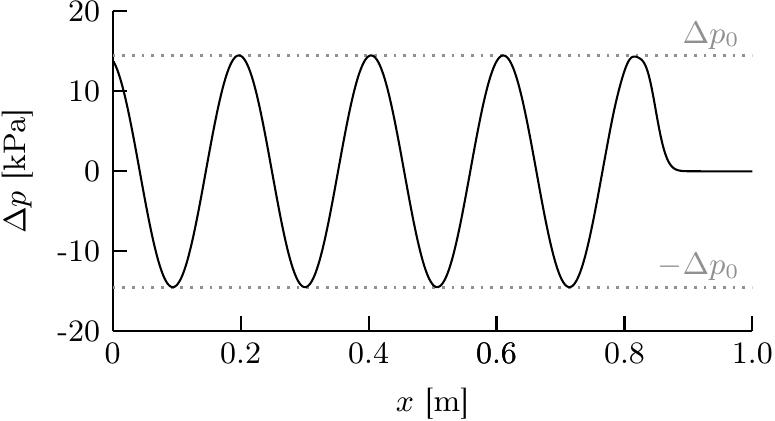}
\label{fig:acousticWaves_Water1}
}
\quad
\subfloat[Water 2, $t=5.4 \times 10^{-4} \, \textup{s}$]
{\includegraphics{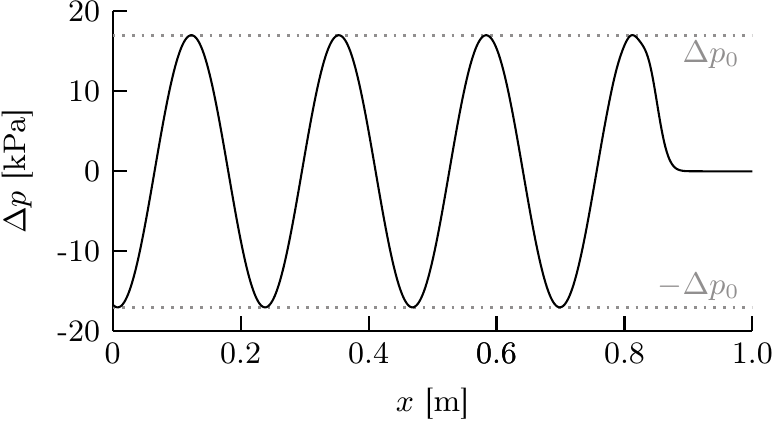}
\label{fig:acousticWaves_Water2}
}
\caption{Profiles of the pressure amplitude $\Delta p$ of acoustic waves in
different fluids, with the fluid properties given in Table
\ref{tab:acousticWaves_Properties} and the frequency given in Table
\ref{tab:acousticWaves_Waves}. The theoretical pressure amplitudes, $\pm \Delta
p_0$, based on linear acoustic theory are given as a reference.}
\label{fig:acousticWaves_dp}
\end{center}
\end{figure}

The computed pressure amplitude $\Delta p$ and the theoretical pressure amplitude $\Delta p_0$ based on linear acoustic theory, both given in Table \ref{tab:acousticWaves_Waves}, are in excellent agreement. Figure \ref{fig:acousticWaves_dp} shows the profiles of the pressure amplitude $\Delta p$ of the acoustic waves in the four considered fluids as a function of space, with good agreement of the minimum and maximum pressure amplitude with the theoretical pressure amplitude. In addition, the computed wavelength $\lambda$ is predicted accurately compared to the theoretical wavelength $\lambda_0$, given in Table \ref{tab:acousticWaves_Waves}, demonstrating a correct prediction of the speed of sound.

\subsection{Moving contact discontinuity}
\label{sec:movingContact}
A contact discontinuity is a linearly degenerate wave and represents the main source of error in terms of convergence of the applied finite-volume method under mesh refinement \citep{Harten1983, Banks2008}, \revE{with the contact discontinuity progressively smoothing over the course of the simulation \citep{Harten1977, Vorozhtsov1990}}. To test the accuracy of the proposed finite-volume framework in predicting contact discontinuities, a moving contact discontinuity in a one-dimensional domain with a length of $1 \, \textup{m}$ is simulated, as considered in previous studies \citep{vanderHeul2003, Moguen2019}. The contact discontinuity is initially located at $x_0 = 0.5 \, \textup{m}$, with the initial conditions of the left and right states given as
\begin{equation}
\begin{array}{ccc}
\rho_\textup{L} = 1.0 \, \textup{kg} \, \textup{m}^{-3}, & u_\textup{L} = 0.5
\, \textup{m} \, \textup{s}^{-1}, & p_\textup{L} = 0.5 \, \textup{Pa}, \\
\rho_\textup{R} = 0.5 \, \textup{kg} \, \textup{m}^{-3}, & u_\textup{R} = 0.5
\, \textup{m} \, \textup{s}^{-1}, & p_\textup{R} = 0.5 \, \textup{Pa}. \nonumber
\end{array}
\end{equation}
The contact discontinuity is simulated in an IG fluid with $\gamma = 1.4$ and $c_p = 1008 \, \textup{J} \, \textup{kg}^{-1} \, \textup{K}^{-1}$, as well as an NASG fluid with $\gamma = 2.0$, $c_p = 114.286  \, \textup{J} \, \textup{kg}^{-1} \, \textup{K}^{-1}$, $\Pi = 5.0 \, \textup{Pa}$ and $b = 10^{-3} \, \textup{m}^3 \, \textup{kg}^{-1}$. The transient terms are discretised using the BDF2 scheme and the applied time-step corresponds to $\textup{Co} = u_\textup{L} \Delta t / \Delta x = 0.5$.

\begin{figure}
\begin{center}
\subfloat[IG fluid]
{\includegraphics{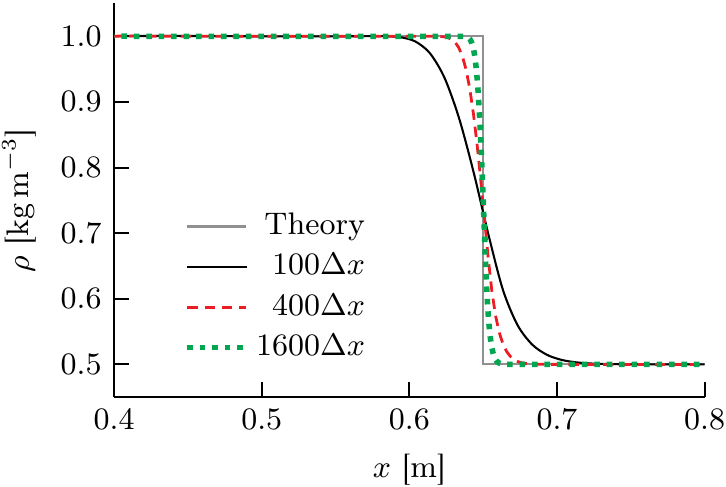}}
\qquad
\subfloat[NASG fluid]
{\includegraphics{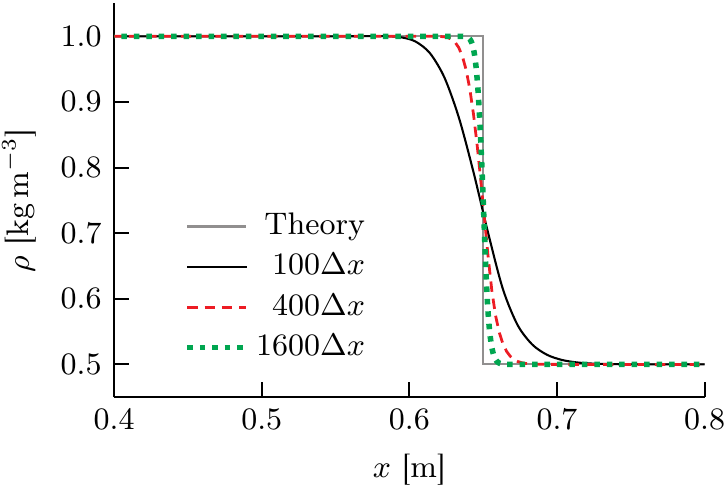}}
\caption{Profiles of the density $\rho$ of a moving contact discontinuity in (a) an IG fluid and (b) an NASG fluid, on equidistant meshes with different resolutions. The advection terms are discretised using the Minmod scheme, the transient terms are discretised using the BDF2 scheme and the time-step corresponds to $\textup{Co} = u_\textup{L} \Delta t / \Delta x = 0.5$.}
\label{fig:movingContact_RhoP}
\end{center}
\end{figure}

\begin{figure}
\begin{center}
\subfloat[IG fluid]
{\includegraphics{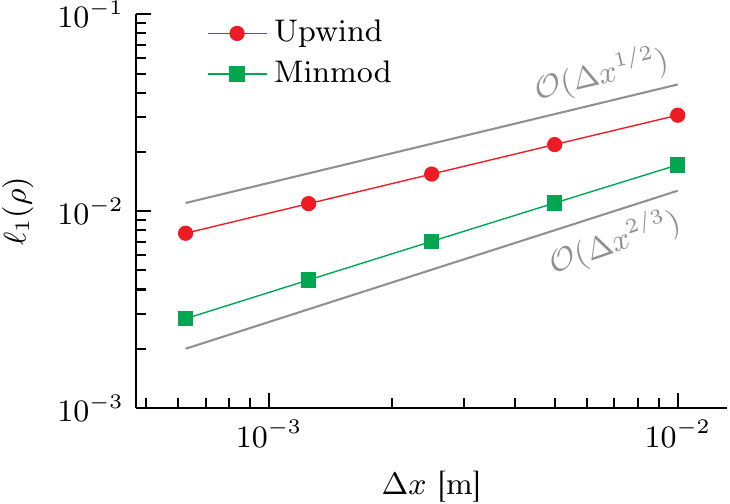}}
\qquad
\subfloat[NASG fluid]
{\includegraphics{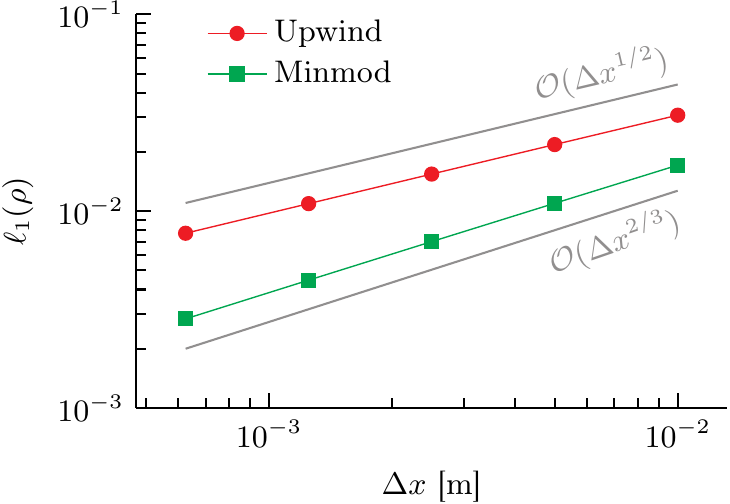}}
\caption{Spatial convergence of the $L_1$-norm of the density error, $\ell_1(\rho)$, as defined in Eq.~(\ref{eq:ell1Rho}), of a moving contact discontinuity in (a) an IG fluid and (b) an NASG fluid, using the first-order upwind scheme and the Minmod scheme. The transient terms are discretised using the BDF2 scheme and the time-step corresponds to $\textup{Co} = u_\textup{L} \Delta t / \Delta x = 0.5$.}
\label{fig:movingContact_SpatialConvergence}
\end{center}
\end{figure}

\begin{figure}
\begin{center}
\subfloat[Profiles of the density $\rho$]
{\includegraphics{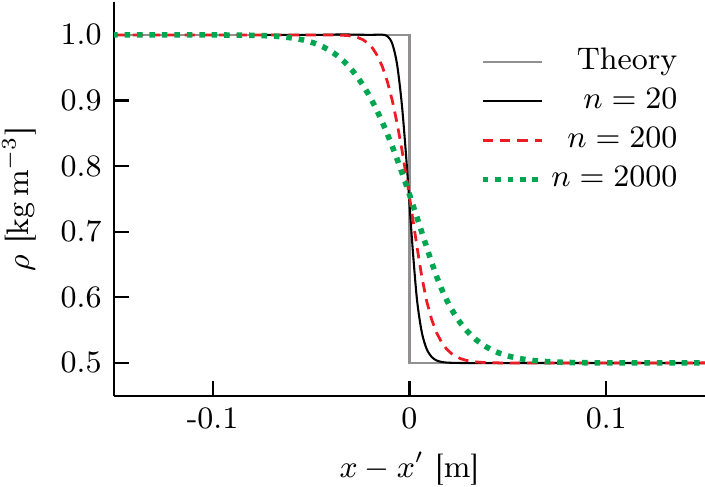}}
\qquad
\subfloat[Width $d$ of the contact discontinuity]
{\includegraphics{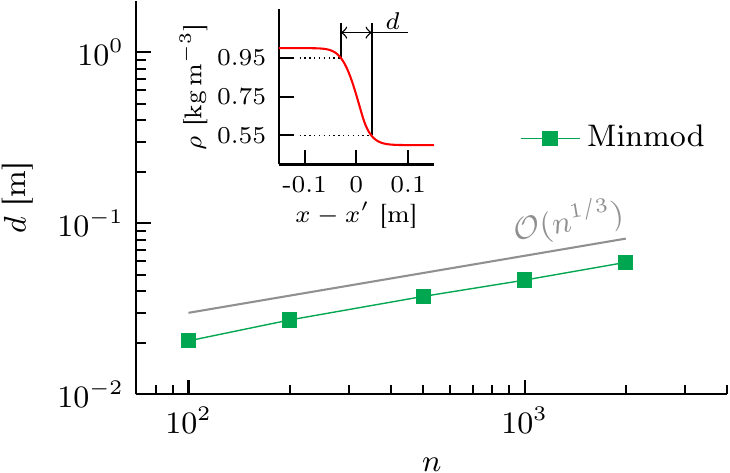}}
\caption{\revE{(a) Profiles of the density $\rho$ of a moving contact discontinuity in an IG fluid, after a different number of time-steps $n$, with $x'$ the position of the contact discontinuity, and (b) the width $d$ of the contact discontinuity, with its definition illustrated in the inset, as a function of time-steps $n$. The advection terms are discretised using the Minmod scheme, the transient terms are discretised using the BDF2 scheme and the time-step corresponds to $\textup{Co} = u_\textup{L} \Delta t / \Delta x = 0.5$.}}
\label{fig:movingContact_TemporalDivergence}
\end{center}
\end{figure}

Figure \ref{fig:movingContact_RhoP} shows the density profiles at $t=0.3 \, \textup{s}$ for both the IG fluid and the NASG fluid, using different mesh resolutions. The results of both fluids are in very good agreement and, irrespective of the mesh resolution, the contact discontinuity propagates with the correct velocity. The convergence of the $L_1$-norm of the solution error associated with a linearly degenerate wave for a $q$-th order advection scheme without compressive limiting is of order $q/(q+1)$ \citep{Banks2008}. The spatial convergence of the $L_1$-norm of the density error,
\begin{equation}
\ell_1(\rho) = \frac{1}{N} \sum_{P=1}^N \left| \frac{\rho_P^\textup{comp.} -
\rho_P^\textup{exact}}{\rho_\textup{L} - \rho_\textup{R}} \right|, \label{eq:ell1Rho}
\end{equation}
where $\rho_P^\textup{comp.}$ is the computed density at cell $P$ and $\rho_P^\textup{exact}$ is the corresponding exact density value, obtained with the upwind scheme ($q=1$) and the Minmod scheme ($q=2$) matches the theoretical order of convergence closely in both fluids, as observed in Fig.~\ref{fig:movingContact_SpatialConvergence}, with convergence order $1/2$ using the upwind scheme and order $2/3$ using the Minmod scheme.
Furthermore, the self-similarity of the transport of the contact discontinuity is not affected by the choice of fluid model, resulting in only minute differences in the L1-norm $\ell_1(\rho)$ between the IG fluid and the NASG fluid for a given mesh resolution, with $|\ell_1(\rho)_\textup{IG} - \ell_1(\rho)_\textup{NASG}| / \ell_1(\rho)_\textup{IG} < 10^{-2}$ using the Minmod scheme and $|\ell_1(\rho)_\textup{IG} - \ell_1(\rho)_\textup{NASG}| / \ell_1(\rho)_\textup{IG} < 10^{-7}$ using the upwind scheme.

\revE{In order to test the progressive smearing of the contact discontinuity during the course of the simulation, the contact discontinuity in the IG fluid is simulated in a domain with a length of $3 \, \mathrm{m}$ and with the contact discontinuity initially located at $x_0 = 0.1 \, \textup{m}$. The computational domain is resolved by $1200$ equidistant mesh cells and the advection terms are discretised using the Minmod scheme. All other settings remain the same as above. The computed density profiles after $n \in \{20, 200, 2000\}$ time-steps are plotted in Fig.~\ref{fig:movingContact_TemporalDivergence}a, clearly showing a progressive smearing of the contact discontinuity. The width of the contact discontinuity should be proportional to $n^{1/(q+1)}$ for a $q$-th order finite-difference or finite-volume method \citep{Harten1977, Vorozhtsov1990}, where $n$ is the number of time-steps. Figure \ref{fig:movingContact_TemporalDivergence}b shows the width $d$ of the contact discontinuity as a function of the time-step $n$, with the width $d$ given by the distance between the points at which the density takes the values $0.55 \, \text{kg} \, \text{m}^{-3}$ and $0.95 \, \text{kg} \, \text{m}^{-3}$, as illustrated in the inset of Fig.~\ref{fig:movingContact_TemporalDivergence}b, since the density changes abruptly between $0.5 \, \text{kg} \, \text{m}^{-3}$ and $1.0 \, \text{kg} \, \text{m}^{-3}$ at the considered contact discontinuity. As shown in Fig.~\ref{fig:movingContact_TemporalDivergence}b, the width of the contact discontinuity increases with the number of time-steps with a slope closely matching $n^{1/3}$, which is the increase expected for a consistently second-order finite-volume method \citep{Harten1977}.}

\subsection{Shock waves}
\label{sec:shockWave}
The propagation of a  shock wave poses particular challenges for finite-volume methods, because a shock wave is discontinuous and valid solutions of the governing conservation laws are not guaranteed to satisfy the second law of thermodynamics across shock waves \citep{Laney1998}. As such, simulating the propagation of a shock wave is well suited to test whether a numerical scheme reliably converges to the physically-correct weak solution of the governing conservation laws, which is a prerequisite for the accurate prediction of both the speed and strength of shock waves \citep{Hou1994, Laney1998}. To this end, the {\em Lax-Wendroff theorem} \citep{Lax1960} stipulates that if a conservative numerical scheme for hyperbolic conservation laws converges, the computed solution converges towards a weak solution of the conservation laws.

The propagation of a strong shock wave with Mach number $M_\textup{s} = 100$ in air and water in a one-dimensional domain with a length of $1 \, \textup{m}$ is simulated. Air is described by the IG model using the fluid properties given in Table \ref{tab:acousticWaves_Properties} and water is described by the NASG model using the fluid properties proposed by \citet{LeMetayer2016}, also given in Table \ref{tab:acousticWaves_Properties} (see properties of {\em Water 2}).
Viscous stresses and heat conduction are neglected, {\em i.e.}~$\mu=k=0$, so the governing equations (\ref{eq:continuity})-(\ref{eq:energy}) reduce to the Euler equations \citep{Swann1971}, which are hyperbolic. 
From the Rankine-Hugoniot relations, the pressure and density ratios across a shock wave propagating with velocity $u_\textup{s}$ in a quiescent NASG fluid are given as
\begin{align}
\frac{p_\textup{I}}{p_\textup{II}} & = 1 + \frac{2 \, \gamma}{\gamma+1} \left(M_\textup{s} - 1 \right) \left(1 + \frac{\Pi}{p_\textup{II}} \right) \label{eq:NASG_RH_p} \\
\frac{\rho_\textup{I}}{\rho_\textup{II}} & =
\dfrac{\dfrac{p_\textup{I} + \Pi}{p_\textup{II} + \Pi} + \dfrac{\gamma -
1}{\gamma + 1}}{\dfrac{\gamma - 1 + 2 \, b \, \rho_\textup{II}}{\gamma+1} \,
\dfrac{p_\textup{I} + \Pi}{p_\textup{II} + \Pi} +
\dfrac{\gamma+1-2 \, b \, \rho_\textup{II}}{\gamma+1}}, \label{eq:NASG_RH_rho}
\end{align}
where subscript $\textup{I}$ denotes the post-shock state, subscript $\textup{II}$ denotes the pre-shock state and $M_\textup{s} = u_\textup{s}/a_\textup{II}$ is the Mach number of the shock wave. 
With the initial conditions of the pre-shock state (II) for both cases given as
\begin{equation}
\begin{array}{ccc}
p_\textup{II} = 10^5 \, \textup{Pa}, & u_\textup{II} = 0
\, \textup{m} \, \textup{s}^{-1}, & T_\textup{II} = 300 \, \textup{K},
\nonumber
\end{array}
\end{equation}
the shock relations yield the initial conditions of the post-shock state (I) for air,
\begin{equation}
\begin{array}{ccc}
p_\textup{I} = 1.16665 \times 10^9 \, \textup{Pa}, & u_\textup{I} = 28979.9
\, \textup{m} \, \textup{s}^{-1}, & T_\textup{I} = 58616.7 \, \textup{K}, 
\nonumber
\end{array}
\end{equation}
and for water,
\begin{equation}
\begin{array}{ccc}
p_\textup{I} = 7.62925 \times 10^{12} \, \textup{Pa}, & u_\textup{I} = 44833.0
\, \textup{m} \, \textup{s}^{-1}, & T_\textup{I} = 278744 \, \textup{K}.
\nonumber
\end{array}
\end{equation}
The shock wave is initially located at $x_\textup{s,0} = 0.25 \, \textup{m}$ and the applied time-step corresponds to $\textup{Co} = u_\textup{s} \Delta t / \Delta x = 0.5$. 

\begin{figure}
\begin{center}
\subfloat[Air (IG fluid), $t=14.38 \, \mu\textup{s}$]
{\includegraphics{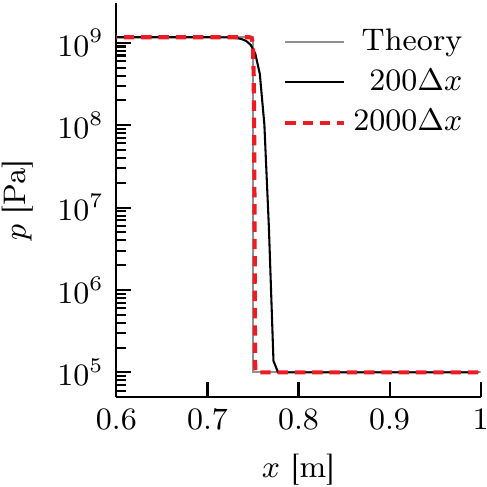}
\label{fig:shockWave_M100_IG}}
\quad
\subfloat[Water (NASG fluid), $t=3.096 \, \mu\textup{s}$]
{\includegraphics{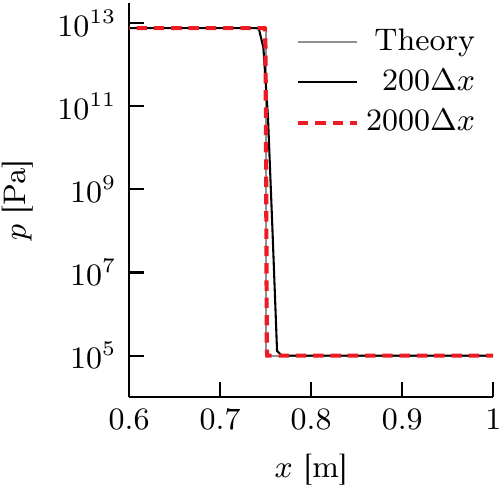}
\label{fig:shockWave_M100_NASG}}
\quad
\subfloat[$L_1$-norm of the density]
{\includegraphics{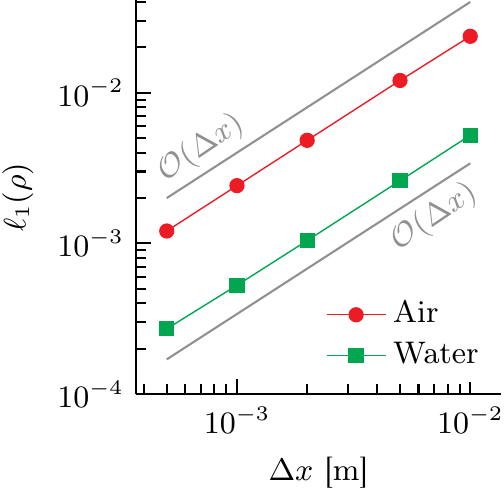}
\label{fig:shockWave_SpatialConvergence}}
\caption{Profiles of the pressure $p$ of a shock wave with Mach number $M_\textup{s} =100$ in (a) air, described as an IG fluid, and (b) water, described as an NASG fluid, and (c) spatial convergence of the $L_1$-norm of the density \revE{error}, $\ell_1(\rho)$, as defined in Eq.~(\ref{eq:ell1Rho}). The exact solution given by the Rankine-Hugoniot relations, Eq.~(\ref{eq:NASG_RH_p}), is shown as a reference in (a) and (b). The applied time-step corresponds to $\textup{Co} = u_\textup{L} \Delta t / \Delta x = 0.5$.}
\label{fig:shockWave_M100}
\end{center}
\end{figure}

The Rankine-Hugoniot relations are reproduced accurately in both air and water, as seen in Fig.~\ref{fig:shockWave_M100}, despite the very large pressure discontinuities with pressure ratios of more than four and seven orders of magnitude, respectively. In both fluids the $L_1$-norm of the density error, $\ell_1(\rho)$, converges with first order under mesh refinement, as seen in Fig.~\ref{fig:shockWave_SpatialConvergence}.
The first order convergence is imposed by the applied monotone discretisation schemes, in this case the Minmod scheme, and is expected for an oscillation-free numerical simulation of a shock wave \citep{Osher1984}. The robust convergence for strong shock waves further implies accurate conservation properties as well as convergence to the correct weak solution of the governing conservation laws using the proposed finite-volume framework and pressure-based algorithm.

\subsection{Shock tubes}
\label{sec:shockTubes}
Shock tubes are routinely and extensively used to test numerical frameworks and schemes for compressible flows, because they feature shock waves, rarefaction fans as well as contact discontinuities and because an exact reference solution based on the associated Riemann problem exists. Three different shock tubes, covering Mach numbers over five orders of magnitude, are considered. In all cases, the fluid has a heat capacity ratio of $\gamma = 1.4$ and a specific gas constant of $c_p = 1008 \, \textup{J} \, \textup{kg}^{-1} \, \textup{K}^{-1}$.

\begin{figure}[t]
\begin{center}
\subfloat[Pressure $p$]
{\includegraphics{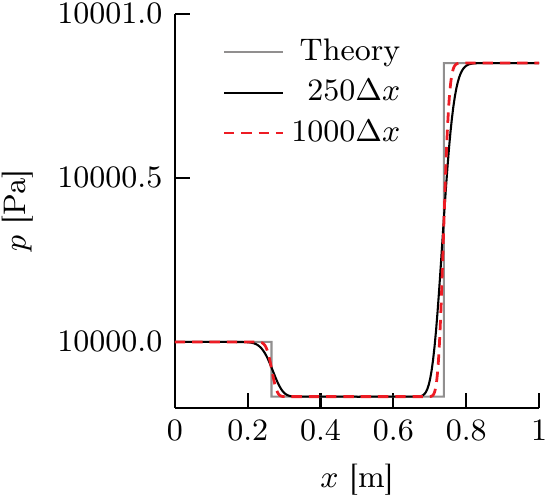}}
\
\subfloat[Density $\rho$]
{\includegraphics{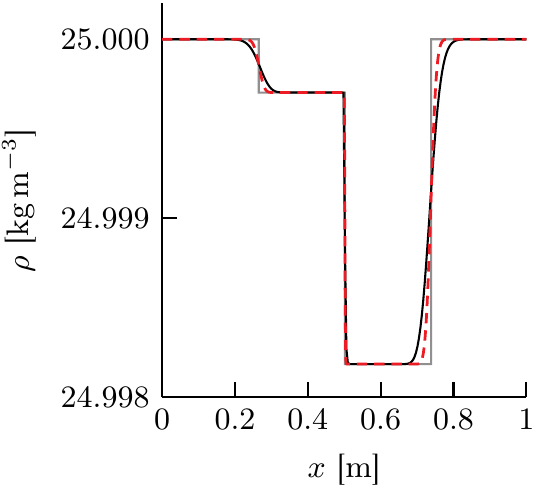}}
\
\subfloat[Mach number $M$]
{\includegraphics{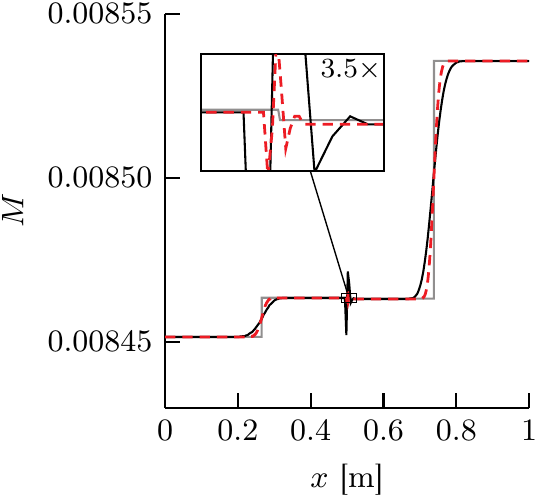}}
\caption{Profiles of pressure, density and Mach number of the low-Mach shock tube at $t=0.01 \, \textup{s}$, compared against the theoretical Riemann solution. \revE{In addition, a magnified view of the minute change in Mach number at the contact discontinuity is shown in (c)}.}
\label{fig:lowMachShockTube}
\end{center}
\end{figure}

A low-Mach shock tube, as proposed by \citet{Moguen2015a}, is considered. The discontinuity is initially located at $x_0 = 0.5 \, \textup{m}$, with the initial conditions of the left and right states given as
\begin{equation}
\begin{array}{ccc}
\rho_\textup{L} = 25.0 \, \textup{kg} \, \textup{m}^{-3}, & u_\textup{L} = 0.200
\, \textup{m} \, \textup{s}^{-1}, & p_\textup{L} = 10000.00 \, \textup{Pa}, \\
\rho_\textup{R} = 25.0 \, \textup{kg} \, \textup{m}^{-3}, & u_\textup{R} = 0.202
\, \textup{m} \, \textup{s}^{-1}, & p_\textup{R} = 10000.85 \, \textup{Pa}.
\nonumber
\end{array}
\end{equation}
The applied time-step corresponds to a Courant number of $\textup{Co}=
(u_\textup{L} + a_\textup{L}) \Delta t/\Delta x = 0.5$. Overall, the results obtained on both meshes are in very good agreement with the theoretical Riemann solution, as seen in Fig.~\ref{fig:lowMachShockTube}. Because the particle velocity is very small, $u_\textup{max} =0.202 \, \textup{m} \, \textup{s}^{-1}$, the contact discontinuity only moves by $0.002 \, \textup{m}$ in the studied time frame and, thus, remains very sharp, as evident by the density profile in Fig.~\ref{fig:lowMachShockTube}. A small wiggle is observed in the Mach number profile at the contact discontinuity, which however has no impact on the overall result.
 
\begin{figure}[t]
\begin{center}
\subfloat[Pressure]
{\includegraphics{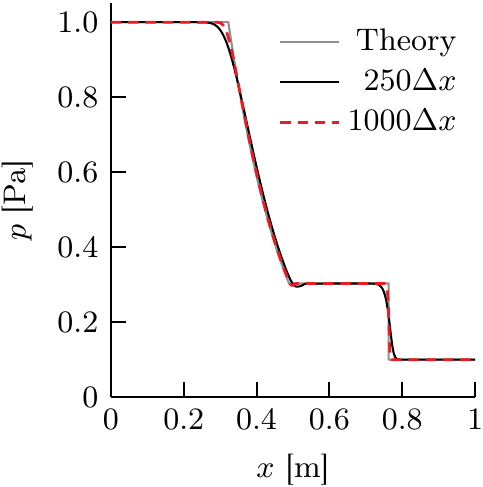}}
\
\subfloat[Density]
{\includegraphics{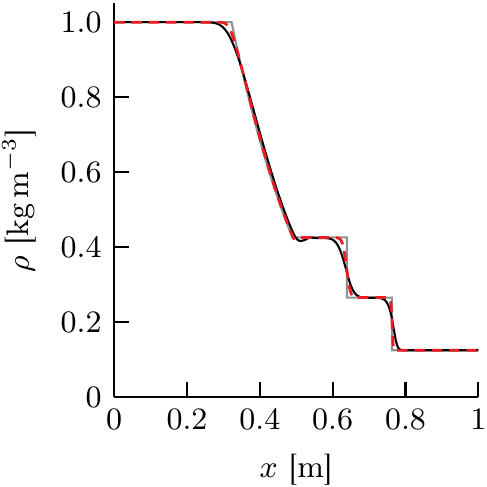}}
\
\subfloat[Mach number]
{\includegraphics{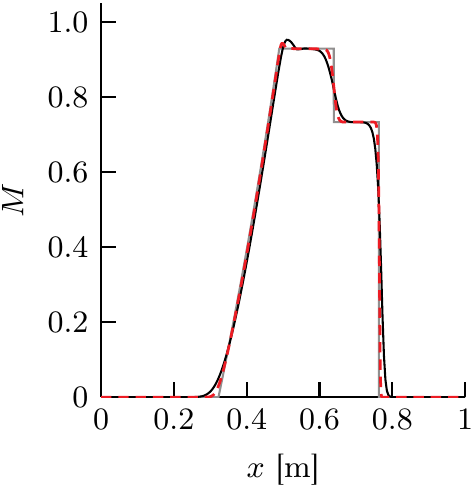}}
\caption{Profiles of pressure, density and Mach number of Sod's shock tube  at $t=0.15 \, \textup{s}$, compared against the theoretical Riemann solution.}
\label{fig:sodShockTube_b0}
\end{center}
\end{figure}

The shock tube initially introduced by \citet{Sod1978} is considered as a shock tube with intermediate Mach number, with initial conditions
\begin{equation}
\begin{array}{ccc}
\rho_\textup{L} = 1.0 \, \textup{kg} \, \textup{m}^{-3}, & u_\textup{L} = 0
\, \textup{m} \, \textup{s}^{-1}, & p_\textup{L} = 1.0 \, \textup{Pa}, \\
\rho_\textup{R} = 0.125 \, \textup{kg} \, \textup{m}^{-3}, & u_\textup{R} =
0 \, \textup{m} \, \textup{s}^{-1}, & p_\textup{R} = 0.1 \, \textup{Pa}.
\nonumber
\end{array}
\end{equation}
The discontinuity is initially located at $x_0 = 0.5 \, \textup{m}$ and the applied time-step corresponds to a Courant number of $\textup{Co}= a_\textup{L} \Delta t/\Delta x = 0.6$. The results obtained on both meshes, shown in Fig.~\ref{fig:sodShockTube_b0}, are in very good agreement with the theoretical Riemann solution.

\begin{figure}[t]
\begin{center}
\subfloat[Pressure]
{\includegraphics{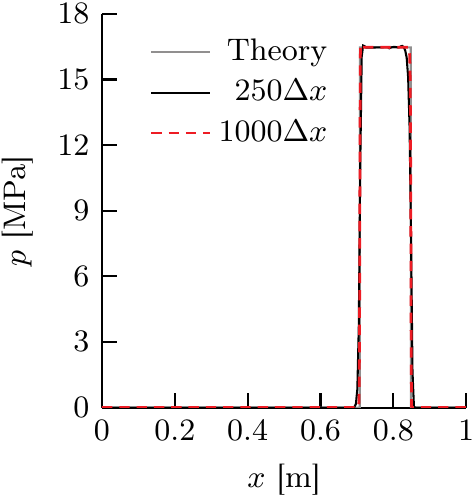}}
\
\subfloat[Density]
{\includegraphics{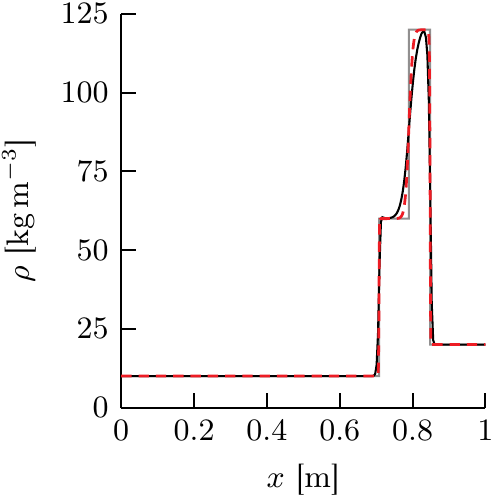}}
\
\subfloat[Mach number]
{\includegraphics{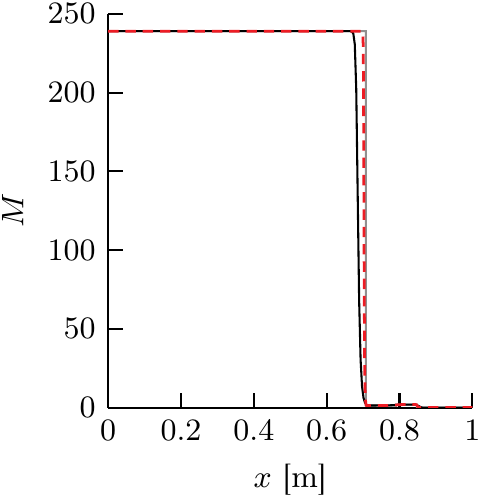}}
\caption{Profiles of pressure, density and Mach number of the high-Mach shock tube at $t=3.5 \times 10^{-4} \, \textup{s}$, compared against the theoretical Riemann solution.}
\label{fig:highMachShockTube}
\end{center}
\end{figure}

The high-Mach shock tube proposed by \citet{Xiao2004} is considered. The discontinuity is initially located at $x_0 = 0.5 \, \textup{m}$, with the initial conditions
\begin{equation}
\begin{array}{ccc}
\rho_\textup{L} = 10 \, \textup{kg} \, \textup{m}^{-3}, & u_\textup{L} = 2000
\, \textup{m} \, \textup{s}^{-1}, & p_\textup{L} = 500 \, \textup{Pa}, \\
\rho_\textup{R} = 20 \, \textup{kg} \, \textup{m}^{-3}, & u_\textup{R} =
0 \, \textup{m} \, \textup{s}^{-1}, & p_\textup{R} = 500 \, \textup{Pa}.
\nonumber
\end{array}
\end{equation}
Notably, the flow of the left state has a Mach number of $M_\textup{L}=239$. The applied time-step corresponds to a Courant number of $\textup{Co}= u_\textup{L} \Delta t/\Delta x = 0.5$. 
As observed in Fig.~\ref{fig:highMachShockTube}, although the profile of the Mach number is not predicted very accurately on the coarse mesh, the density and pressure profiles are in good agreement with the theoretical Riemann solution. On the fine mesh, the computed results are in very good agreement with the theoretical Riemann solution, demonstrating the accurate prediction of high-Mach Riemann problems with the proposed numerical framework.

\subsection{Taylor vortices}
\label{sec:TaylorVortices}
\begin{figure}[t]
\begin{center}
\subfloat[Velocity $u$]
{\includegraphics[width=0.35\textwidth]{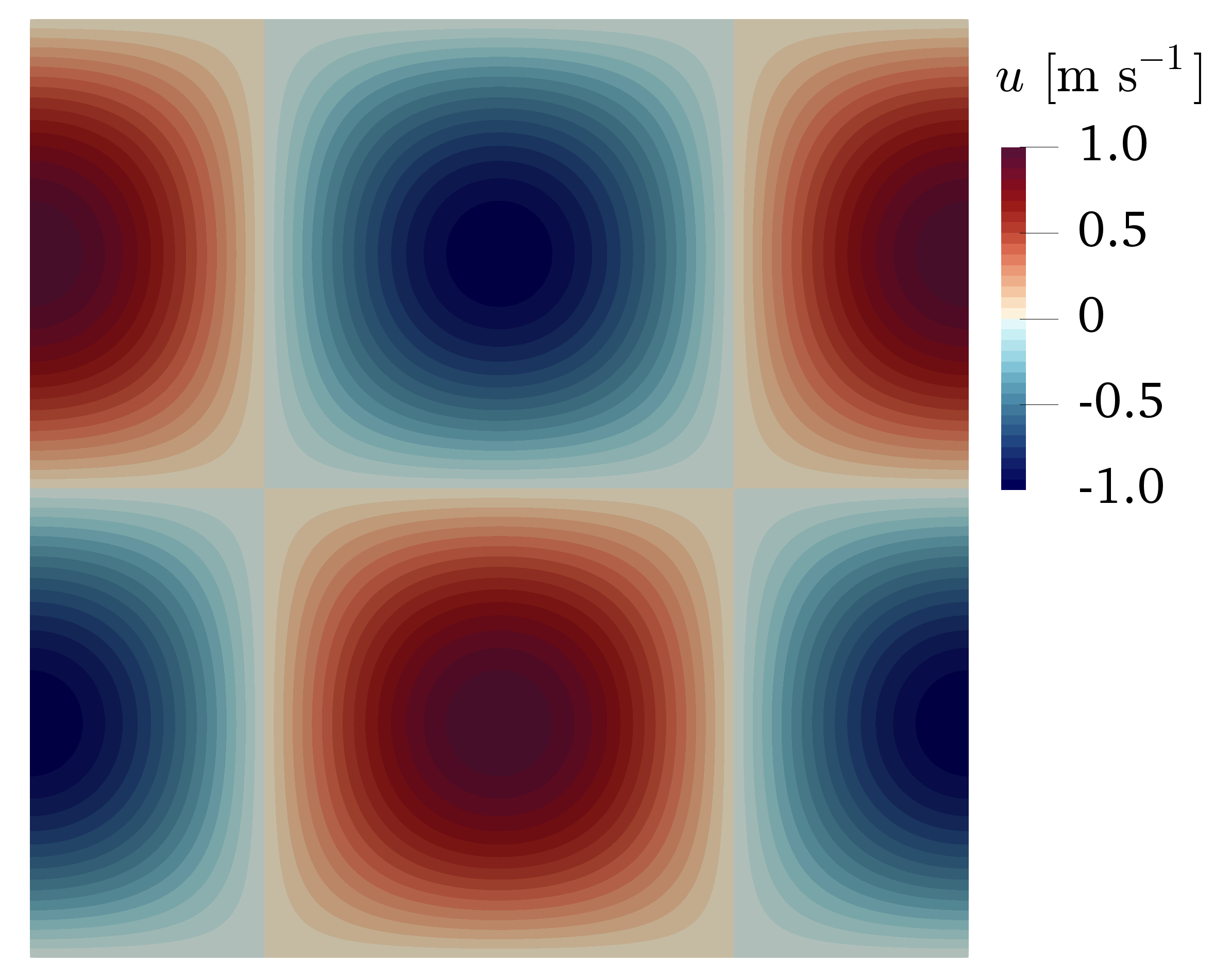}}
\qquad
\subfloat[Pressure $p$]
{\includegraphics[width=0.35\textwidth]{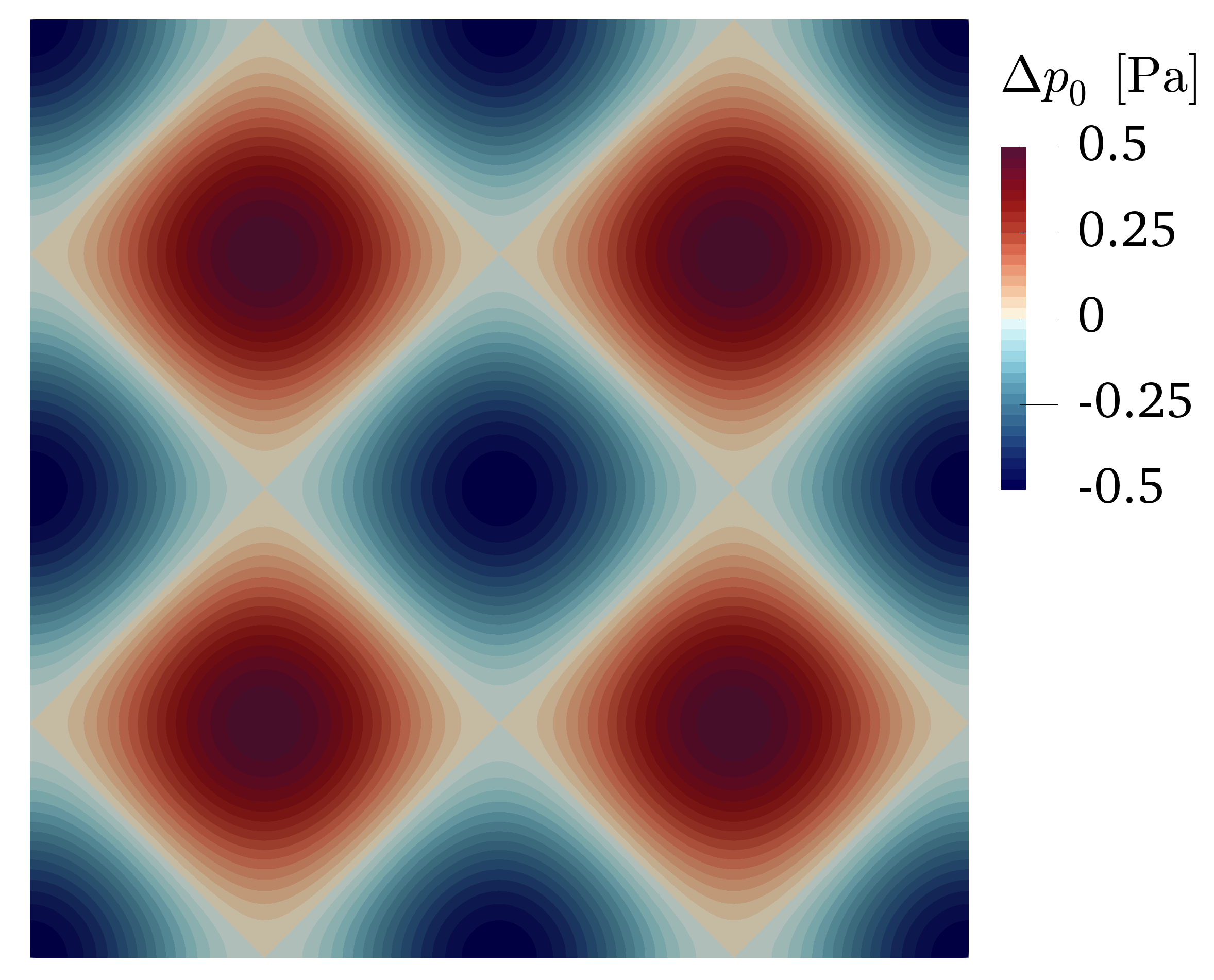}}
\caption{Contours of the initial velocity $u$ along the $x$-axis and the initial pressure $p$ of the Taylor vortices.}
\label{fig:taylorVortices_contours}
\end{center}
\end{figure}

The conservation of kinetic energy is a fundamental property arising from the conservation of mass and momentum. Two-dimensional Taylor vortices in an inviscid ($\mu=0$), non-conducting ($k=0$) fluid are simulated to analyse the conservation of kinetic energy by the proposed numerical framework. The domain has the dimensions $2 \, \textup{m} \times 2 \, \textup{m}$ and is periodic in all directions, so that energy transfer across the domain boundaries does not have to be considered. The initial conditions, shown in Fig.~\ref{fig:taylorVortices_contours}, are $u = -\cos(\pi x) \sin(\pi y)$, $v = \sin(\pi x) \cos(\pi y)$ and $p = - 0.25 \left[\cos(2\pi x) + \cos(2 \pi y) \right]$. Since $\mu=k=0$, the Taylor vortices are steady and no energy dissipation occurs naturally, with a constant kinetic energy of
\begin{equation}
E_\textup{kin} =  \frac{1}{2}  \int_{\Omega} \rho \, \boldsymbol{u}^2
\, \textup{d}\Omega \approx \frac{1}{2} \sum_{P=1}^N \rho_P \,
\boldsymbol{u}_P^2 \, V_P ,
\end{equation}
where $\Omega$ is the volume of the computational domain. Any dissipation of kinetic energy is, thus, the result of numerical dissipation induced by the applied discretisation.

\begin{figure}[t]
\begin{center}
\subfloat[Evolution of $\varepsilon_\textup{kin}$ with MWI.]
{\includegraphics{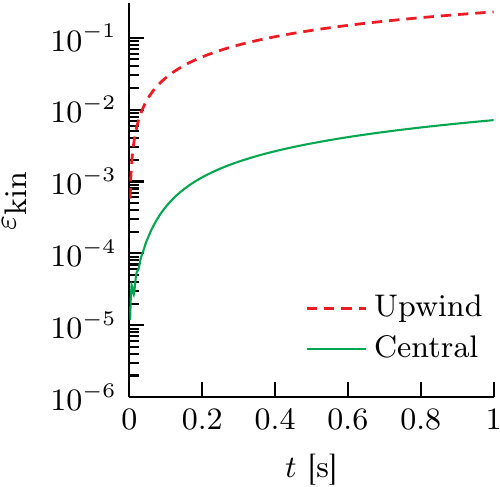}
\label{fig:taylorVortices_Evolution}}
\quad
\subfloat[Evolution of $\varepsilon_\textup{kin}$ without MWI.]
{\includegraphics{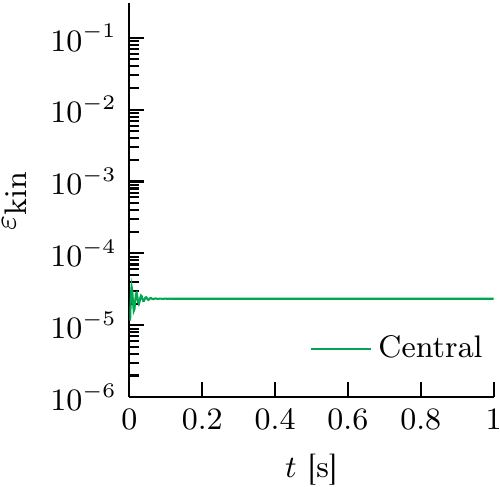}
\label{fig:taylorVortices_EvolutionNoMWI}}
\quad
\subfloat[Convergence of $\varepsilon_\textup{kin}$ at $t=1\, \textup{s}$.]
{\includegraphics{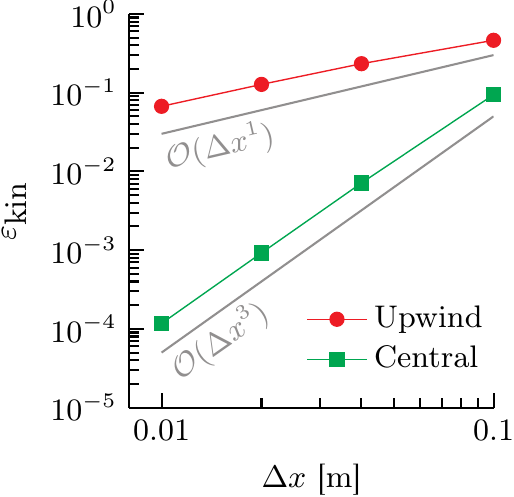}
\label{fig:taylorVortices_Convergence}}
\caption{Temporal evolution of $\varepsilon_\textup{kin}$, Eq.~(\ref{eq:kinEnergyError}), on an equidistant Cartesian mesh with $\Delta x = 0.04 \, \textup{m}$ defining the advecting velocity $\vartheta_f$ (a) with the MWI as described in Section \ref{sec:advectingVel} and (b) without the MWI as $\vartheta_f = \overline{\boldsymbol{u}}_f \cdot \boldsymbol{n}_f$, and (c) convergence of the error in kinetic energy $\varepsilon_\textup{kin}$ of the Taylor vortices with the MWI as described in Section \ref{sec:advectingVel}. The first-order upwind scheme or the central differencing scheme are applied for the discretisation of the advection term. The applied time-step in all cases is $\Delta t = 2 \times 10^{-3} \, \textup{s}$.}
\label{fig:taylorVortices_plots}
\end{center}
\end{figure}

Figure \ref{fig:taylorVortices_Evolution} shows the evolution of the error in kinetic energy of the Taylor vortices,
\begin{equation}
\varepsilon_\textup{kin} = \dfrac{E_{\textup{kin}}^{(0)} -
E_\textup{kin}}{E_{\textup{kin}}^{(0)}} \ , \label{eq:kinEnergyError}
\end{equation} 
with $E_{\textup{kin}}^{(0)}$ the kinetic energy of the initialised ($t=0$) flow field, in an IG fluid with $\gamma=1.4$ and $c_p = 1008 \, \textup{J} \, \textup{kg}^{-1} \, \textup{K}^{-1}$, and Mach number $M = 0.01$. As expected, the error in kinetic energy is substantially larger using the first-order upwind scheme compared to the error in kinetic energy obtained using the central differencing scheme. Interestingly, the applied transient discretisation scheme, {\em i.e.}~BDF1 or BDF2, does not affect the error in kinetic energy, which is consistent with the Taylor vortices being a steady flow in the absence of molecular viscosity and heat conduction.
However, even with central differencing, kinetic energy is dissipated as a result of the MWI formulation of the advecting velocity \citep{Bartholomew2018}, see Eq.~(\ref{eq:advVel}). \revE{No appreciable distortion of the vortices is observed for the considered simulations when central differencing is applied, which is consistent with the only small error in kinetic energy ($\varepsilon_\text{kin} < 1\%$) in theses cases.}

The flow is sufficiently compressible ($M=0.01$) and smooth, that pressure and velocity remain coupled even without MWI \citep{Bartholomew2018}. Exploiting this by omitting the correction introduced by the MWI, with the advecting velocity simply defined as $\vartheta_f = \overline{\boldsymbol{u}}_f \cdot \boldsymbol{n}_f$, the error in kinetic energy remains constant for $t \gtrsim 0.08 \, \textup{s}$, as seen in Fig.~\ref{fig:taylorVortices_EvolutionNoMWI}, which indicates that the numerical dissipation of kinetic energy is negligible. This is to be expected when simulating a sufficiently smooth flow with a second-order accurate finite-volume framework without any explicitly introduced physical or numerical dissipation. Only a small error in kinetic energy is observed at the beginning of the simulation, caused by the initial conditions \citep{Bartholomew2018}.

The error in kinetic energy converges with third order using central differencing under mesh refinement, as shown in Fig.~\ref{fig:taylorVortices_Convergence}, which is consistent with the third-order convergence of the error in kinetic energy introduced by the MWI \citep{Bartholomew2018}. On the other hand, when the first-order upwind scheme is applied, the kinetic energy dissipated artificially by the MWI is insignificant compared to the numerical diffusion introduced by the upwind scheme, as evident by the first-order convergence of the error in kinetic energy shown in Fig.~\ref{fig:taylorVortices_Convergence}. 

These results, therefore, suggest that the MWI is the only source of numerical dissipation in the proposed finite-volume discretisation, assuming a  consistent second-order (or higher-order) interpolation of spatial and transient terms is applied, {\em e.g.}~central differencing and BDF2.

\subsection{Diffusion-dominated flows}
\label{sec:diffusionDominated}
The test-cases discussed in the previous sections only test the discretisation of the transient and advection terms, not taking into account diffusion terms, {\em i.e.}~viscous stresses, heat conduction and viscous heating. Two well-defined diffusion-dominated flows, a planar Poiseuille flow of an incompressible fluid and a planar Couette flow of a compressible fluid, are considered to test the discretisation and implementation of the diffusion of momentum and heat.

\begin{figure}[t]
\begin{center}
\subfloat[Schematic]
{\includegraphics{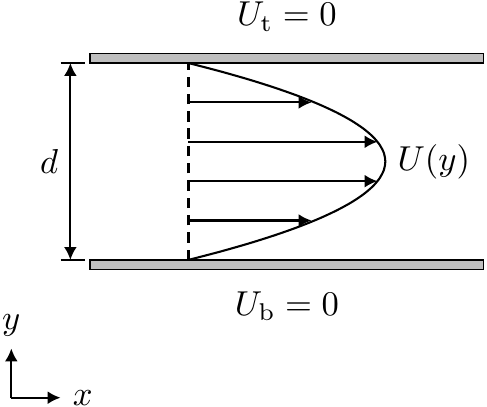}
\label{fig:diffusionDominated_schematicPoiseuille}}
\quad
\subfloat[Axial velocity profile]
{\includegraphics{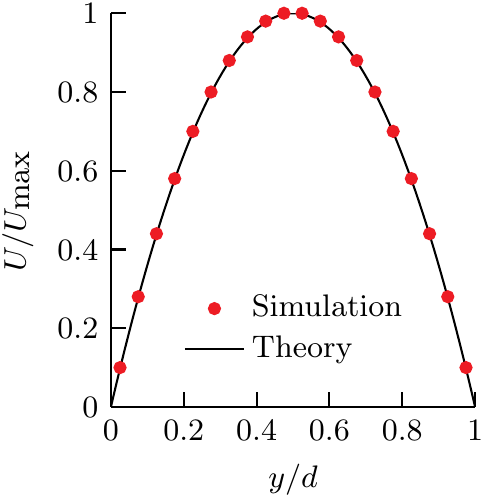}
\label{fig:diffusionDominated_PoiseuilleProfile}}
\quad
\subfloat[$L_\infty$-norm of the axial velocity]
{\includegraphics{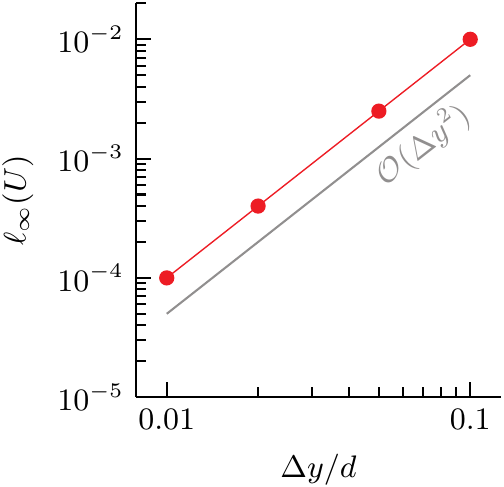}
\label{fig:diffusionDominated_PoiseuilleConvergence}}
\caption{Schematic of a planar Poiseuille flow, as well as the profile of the  axial velocity compared against the analytical solution, Eq.~(\ref{eq:poiseuilleU}), and spatial convergence of the $L_\infty$-norm of the error in axial velocity, Eq.~(\ref{eq:linfty_U}), of the planar Poiseuille flow of an incompressible fluid. The axial velocity profile in (b) is obtained on a mesh with $\Delta y = d/20$, with each dot representing a cell-centred value.}
\label{fig:diffusionDominated_Poiseuille}
\end{center}
\end{figure}

The planar Poiseuille flow of an incompressible fluid between two parallel plates of infinite length separated by a constant distance $d$, illustrated schematically in Fig.~\ref{fig:diffusionDominated_schematicPoiseuille}, is a flow that is entirely governed by viscous stresses. Assuming the viscosity $\mu$ is constant and the flow is laminar, the velocity profile is readily given as
\begin{equation}
U (y) = -\frac{\textup{d}p}{\textup{d}x} \, \frac{y (d-y)}{2 \, \mu},
\label{eq:poiseuilleU}
\end{equation}
where $- \textup{d}p/\textup{d}x$ is the driving pressure gradient. This type of flow, thus, allows a straightforward quantification of the solution error associated with the axial velocity. \revE{The computational domain is taken to be periodic in the streamwise direction, to circumvent any influence of inlet and outlet boundary conditions, and the flow is driven by a constant momentum source corresponding to the driving pressure gradient $- \textup{d}p/\textup{d}x$.}
The profile of the axial velocity $U$ of the planar Poiseuille flow obtained on a mesh with a resolution of $\Delta y = d/20$ is shown in Fig.~\ref{fig:diffusionDominated_PoiseuilleProfile}, alongside the spatial convergence of the $L_\infty$-norm of the error in axial velocity,
\begin{align}
\ell_\infty(U) = \textup{max} \left|
\frac{U_{P}^\textup{comp.} - U_{P}^\textup{exact}}{U_\textup{max}^\textup{exact}} \right|, \label{eq:linfty_U}
\end{align}
in Fig.~\ref{fig:diffusionDominated_PoiseuilleConvergence}. The axial velocity profile is in excellent agreement with the analytical solution, Eq.~(\ref{eq:poiseuilleU}), and the $L_\infty$-norm of the error in axial velocity converges with second order under mesh refinement, as expected given the second-order discretisation of the viscous stresses in Eq.~(\ref{eq:momentumDisc}).

\begin{figure}
\begin{center}
\subfloat[Schematic]
{\includegraphics{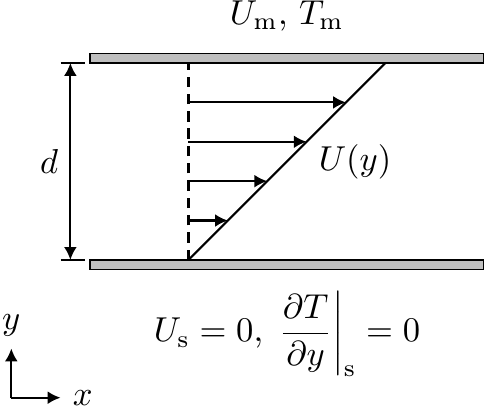}
\label{fig:diffusionDominated_schematicCouette}}
\quad
\subfloat[Temperature profile]
{\includegraphics{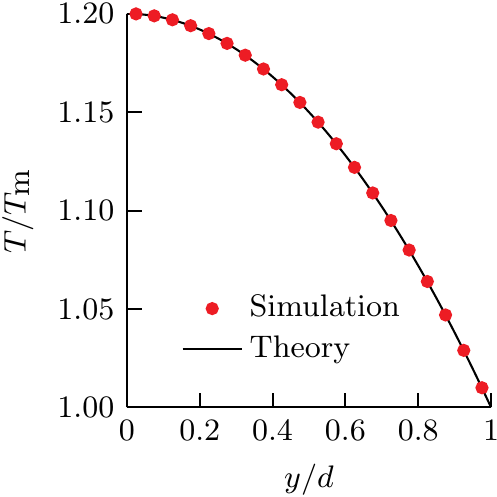}
\label{fig:diffusionDominated_CouetteProfile}}
\quad
\subfloat[$L_\infty$-norm of the temperature]
{\includegraphics{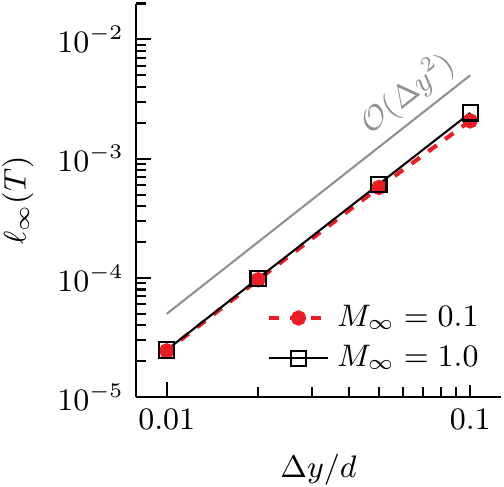}
\label{fig:diffusionDominated_CouetteConvergence}}
\caption{Schematic of a planar Couette flow, as well as the profile of the temperature compared against the analytical solution, Eq.~(\ref{eq:couetteT}), and spatial convergence of the $L_\infty$-norm of the temperature, Eq.~(\ref{eq:linfty_T}), of the planar Couette flow of a compressible fluid for both considered Mach numbers. The temperature profile in (b) is obtained on a mesh with $\Delta y = d/20$, with each dot representing a cell-centred value.}
\label{fig:diffusionDominated_Couette}
\end{center}
\end{figure}

The planar Couette flow of a compressible fluid between two parallel plates of infinite length separated by a constant distance $d$, illustrated schematically in Fig.~\ref{fig:diffusionDominated_schematicCouette}, is a compressible flow that is dominated by viscous stresses and heat conduction. Assuming the viscosity $\mu$ is constant and the stationary wall is adiabatic, the velocity and temperature profiles only depend on the Prandtl number $\textup{Pr} = \mu \, c_p / k$ and the Mach number $M_\textup{m} = U_\textup{m}/a_\textup{m}$ at the moving wall, with the velocity given as $U(y)/U_\textup{m}  = y/d$ and the temperature given as \citep{Malik2008}
\begin{equation}
\frac{T(y)}{T_\textup{m}} = 1 + \frac{\gamma-1}{2} \,
\textup{Pr} \, M_\textup{m}^2 \left[1- \left(\frac{y}{d} \right)^2 \right].
\label{eq:couetteT}
\end{equation}
This type of flow, thus, allows a straightforward quantification of the solution error associated with the temperature. The considered fluid is an ideal gas with a Prandtl number of $\textup{Pr}=1$ and a heat capacity ratio of $\gamma=1.4$. \revE{The computational domain is taken to be periodic in the streamwise direction, to circumvent any influence of inlet and outlet boundary conditions.}
The profile of the temperature $T$ of the planar compressible Couette flow with $M_\textup{m}=1.0$ obtained on a mesh with a resolution of $\Delta y = d/20$ is shown in Fig.~\ref{fig:diffusionDominated_CouetteProfile}, alongside the spatial convergence of the $L_\infty$-norm of the error in temperature,
\begin{align}
\ell_\infty(T) = \textup{max} \left|
\frac{T_{P}^\textup{comp.} - T_{P}^\textup{exact}}{T_\textup{s}^\textup{exact} - T_\textup{m}^\textup{exact}} \right|, \label{eq:linfty_T}
\end{align}
at Mach numbers $M_\textup{m} \in \{0.1,1.0\}$ in Fig.~\ref{fig:diffusionDominated_CouetteConvergence}.
The temperature profile is in excellent agreement with the analytical solution, Eq.~(\ref{eq:couetteT}), and the $L_\infty$-norm of the error in temperature converges with second order under mesh refinement for both Mach numbers, as expected given the second-order discretisation of the heat conduction term in Eq.~(\ref{eq:energyDisc}). Notably, $\ell_\infty(T)$ is independent of the Mach number $M_\textup{m}$ for a sufficiently high spatial resolution, as seen in Fig.~\ref{fig:diffusionDominated_CouetteConvergence}.

\subsection{Lid-driven cavity}
\label{sec:ldc}
\begin{figure}[t]
\begin{center}
\subfloat[Schematic]
{\includegraphics{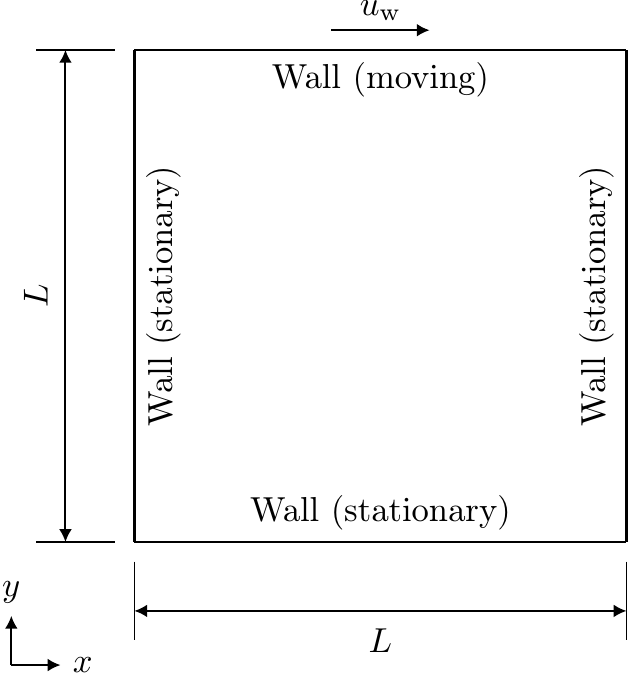}
\label{fig:ldc_schematic}}
\qquad
\subfloat[Polygonal mesh]
{\includegraphics[width=0.5\textwidth]{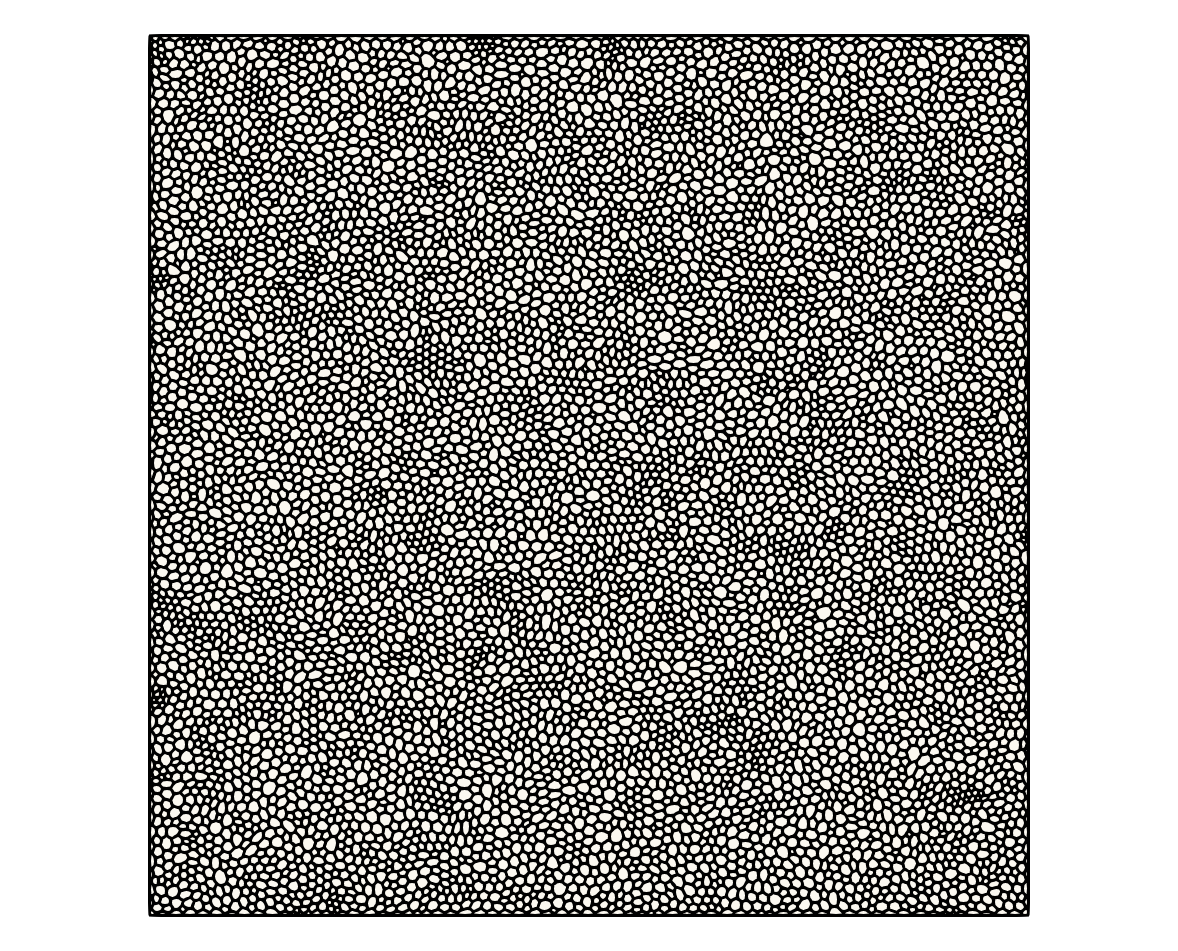}
\label{fig:ldc_poly}}
\caption{Schematic illustration and polygonal mesh of the lid-driven cavity.}
\label{fig:ldc_schematicPoly}
\end{center}
\end{figure}

The lid-driven cavity, schematically shown in Fig.~\ref{fig:ldc_schematic}, is a common test case to validate numerical methods for fluid flows, since it captures convective and diffusive momentum transport of the fluid. The considered two-dimensional domain is of size $L \times L$, with no-slip boundary conditions imposed on all four walls. The top wall moves with velocity $u_\textup{w}$ and the flow of the incompressible fluid has a Reynolds numbers of $\textup{Re} = \rho \, L \, u_\textup{w} / \mu \in \{100, 1000\}$.  \revE{A polygonal mesh with $8708$ cells, shown in Fig.~\ref{fig:ldc_poly}, represents the computational domain.}

\begin{figure}
\begin{center}
\subfloat[$u$-velocity]
{\includegraphics{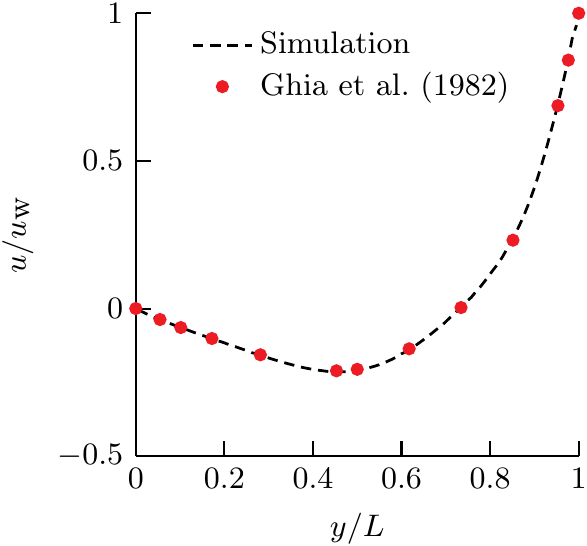}}
\qquad
\subfloat[$v$-velocity]
{\includegraphics{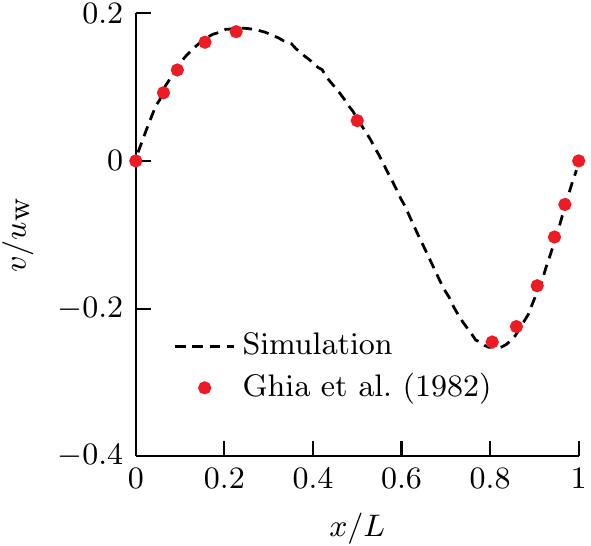}}
\caption{Profiles of (a) the $u$-velocity along the $y$-centreline of the domain and (b) the $v$-velocity along the $x$-centreline of the domain of the lid-driven cavity with $\textup{Re} = 100$. The results of \citet{Ghia1982} are shown as a reference.}
\label{fig:ldc_Re100}
\end{center}
\end{figure}

\begin{figure}
\begin{center}
\subfloat[$u$-velocity]
{\includegraphics{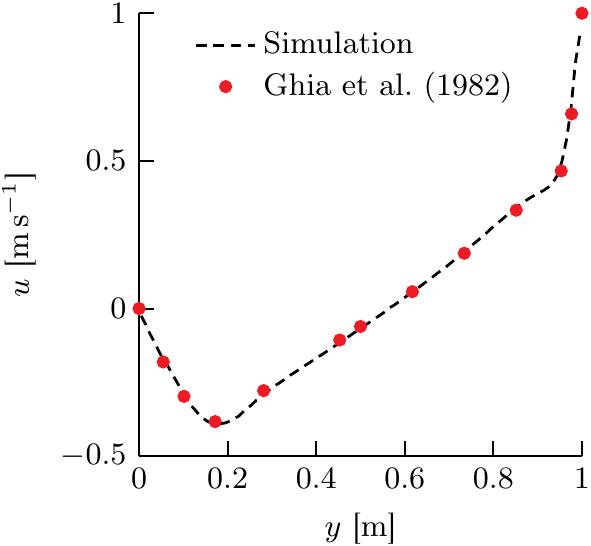}}
\qquad
\subfloat[$v$-velocity]
{\includegraphics{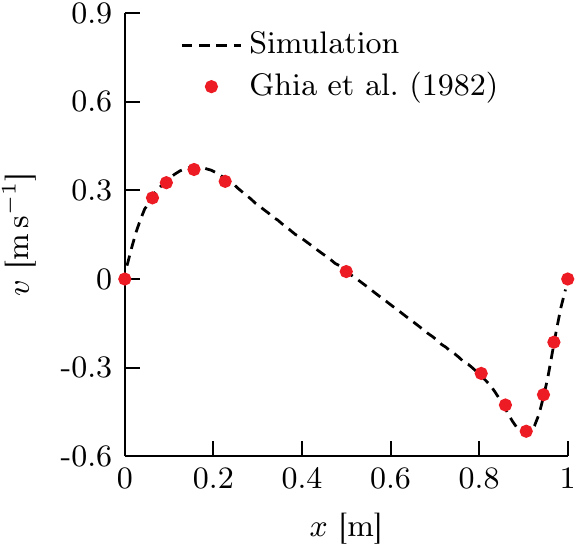}}
\caption{Profiles of (a) the $u$-velocity along the $y$-centreline of the domain and (b) the $v$-velocity along the $x$-centreline of the domain of the lid-driven cavity with $\textup{Re} = 1000$. The results of \citet{Ghia1982} are shown as a reference.}
\label{fig:ldc_Re1000}
\end{center}
\end{figure}

\begin{figure}
\begin{center}
\subfloat[Divergence of the velocity field at steady state]
{\includegraphics[width=0.41\textwidth]{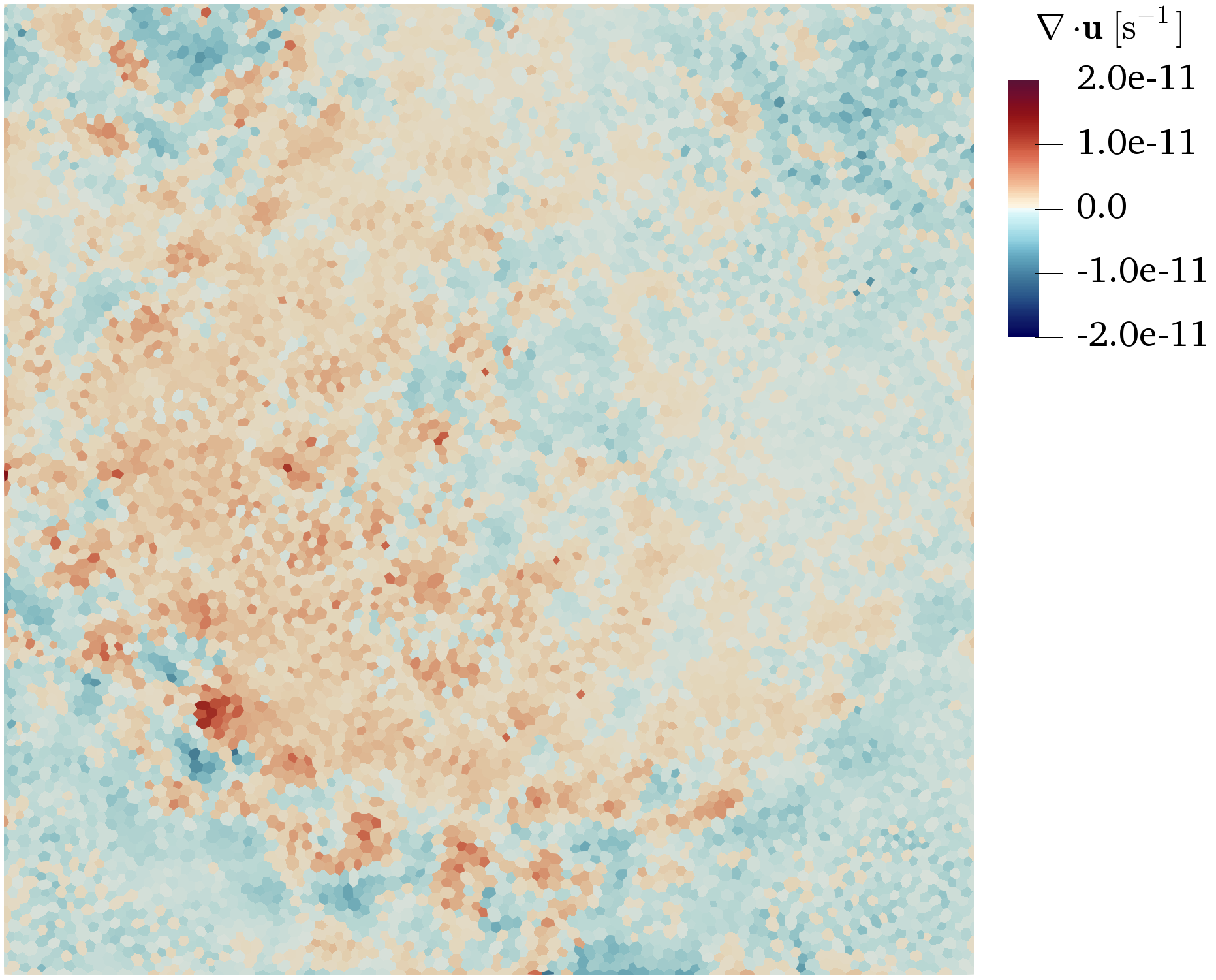}}
\qquad
\subfloat[$L_1$-norm of the error in the divergence of the velocity field]
{\includegraphics{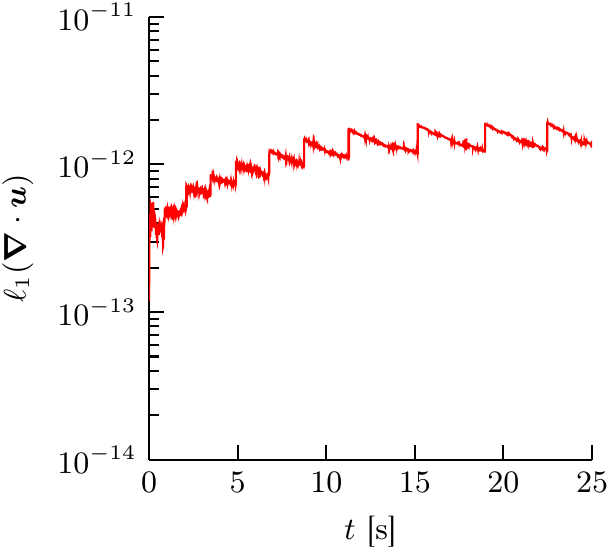}}
\caption{\revE{(a) Contours of the divergence of the velocity field, $\boldsymbol{\nabla} \cdot \boldsymbol{u}$, at steady state and b) $L_1$-norm of the error in the divergence of the velocity field, $\ell_1(\boldsymbol{\nabla}\cdot \boldsymbol{u})$, for the lid-driven cavity with $\text{Re} = 1000$.}}
\label{fig:ldc_divU}
\end{center}
\end{figure}

Figures \ref{fig:ldc_Re100} and \ref{fig:ldc_Re1000} show the $u$-velocity profile in the $y$-direction and the $v$-velocity profile in the $x$-direction along lines that pass through the centre of the domain for the two considered Reynolds numbers, compared against the reference results of \citet{Ghia1982}. 
The results are in \revE{very good} agreement with the reference results of \citet{Ghia1982}, \revE{as well as other studies that have previously considered this test-case \citep{Lilek1995,Munz2003, Karimian2006, Xiao2017}}, for both considered Reynolds numbers, demonstrating the accurate prediction of the convective-diffusive transport of momentum on unstructured meshes using the proposed algorithm. 
\revE{The contours of the divergence of velocity, $\boldsymbol{\nabla}\cdot \boldsymbol{u}$, at steady state are shown in Fig.~\ref{fig:ldc_divU} for the lid-driven cavity with $\text{Re} = 1000$, alongside the transient evolution (considering an initially quiescent fluid) of the $L_1$-norm of the error in the divergence of the velocity field, given as
\begin{equation}
\ell_1(\boldsymbol{\nabla}\cdot \boldsymbol{u}) = \frac{1}{N} \sum_{P=1}^N | \boldsymbol{\nabla}\cdot \boldsymbol{u}_P| = \frac{1}{N} \sum_{P=1}^N \left| \frac{1}{V_P} \sum_f \vartheta_f A_f \right|,
\end{equation}
where $f$ are the faces of cell $P$.
The divergence-free condition of the velocity field imposed by the conservation of mass in conjunction with the considered incompressible fluid, see Eq.~(\ref{eq:continuityImp}), is satisfied accurately, with only marginal errors subject to the applied tolerance of the iterative solver (see Section \ref{sec:algorithm}). This is to be expected from the proposed algorithm, as $\boldsymbol{\nabla}\cdot \boldsymbol{u} = 0$ is implicitly enforced by Eq.~(\ref{eq:continuityDiscImp}).}

\subsection{Forward-facing step}
\label{sec:ffs}
\begin{figure}[t]
\begin{center}
\subfloat[Mach number $M$]
{\includegraphics[width=0.47\textwidth]{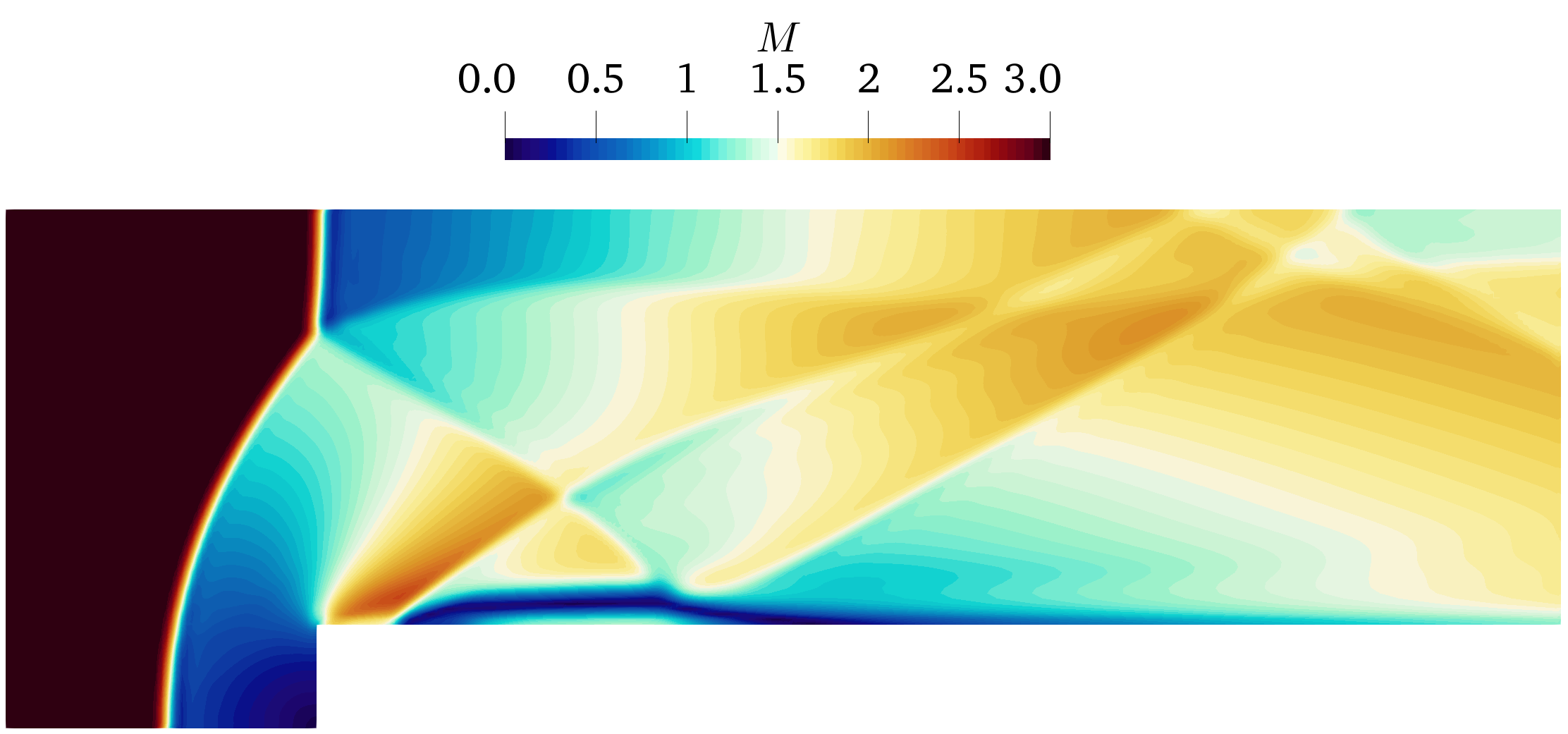}}
\quad
\subfloat[Pressure $p$]
{\includegraphics[width=0.47\textwidth]{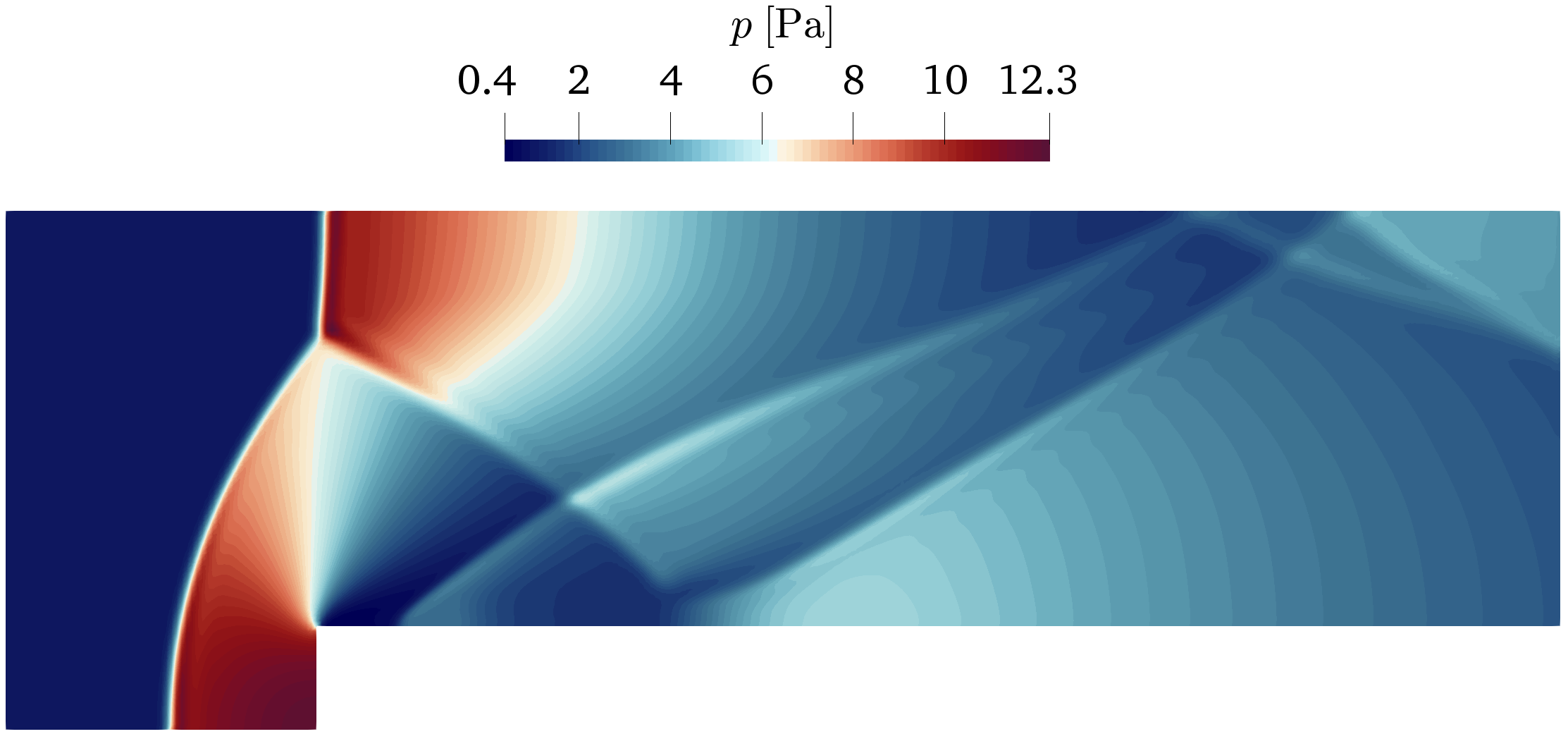}}
\caption{Contours of Mach number and pressure of the supersonic flow over a forward-facing step at $t=4 \, \textup{s}$  with co-volume $b=0$.}
\label{fig:ffsContours_b0}
\end{center}
\end{figure}

\begin{figure}
\begin{center}
\subfloat[Mach number $M$]
{\includegraphics[width=0.47\textwidth]{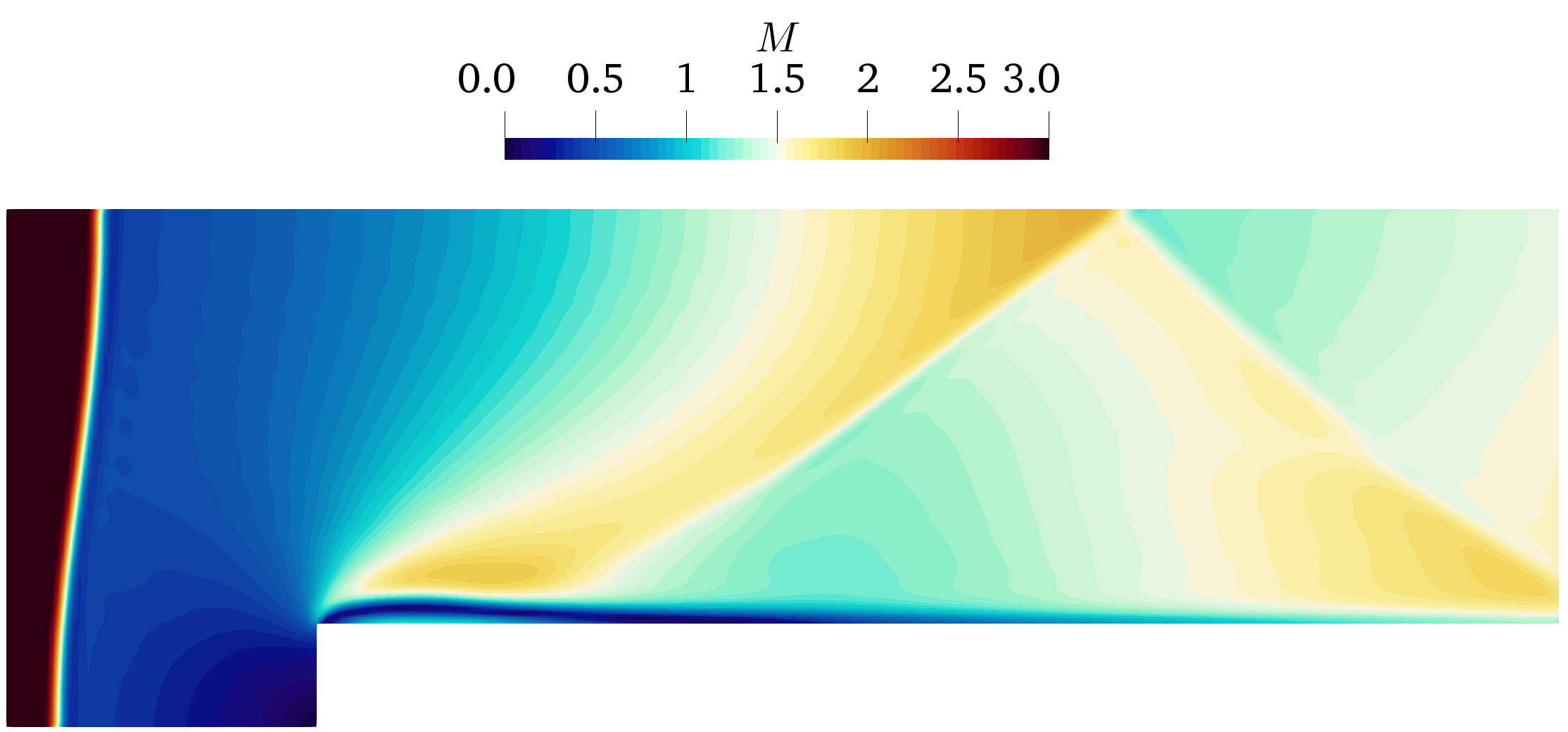}}
\quad
\subfloat[Pressure $p$]
{\includegraphics[width=0.47\textwidth]{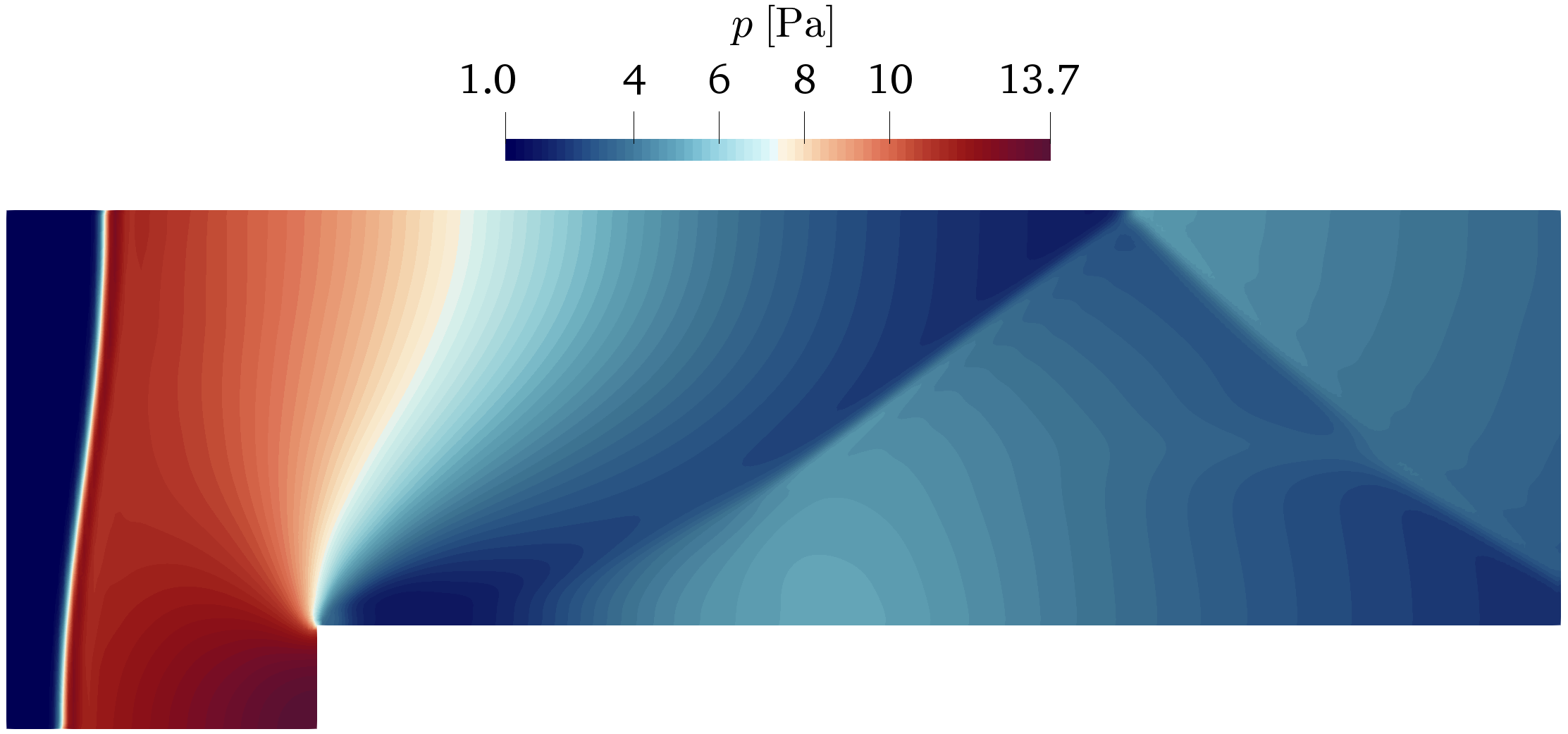}}
\caption{Contours of Mach number and pressure of the supersonic flow over a forward-facing step at $t=4 \, \textup{s}$  with co-volume $b=0.1 \, \textup{m}^3 \, \textup{kg}^{-1}$.}
\label{fig:ffsContours_b0p1}
\end{center}
\end{figure}

The two-dimensional supersonic flow over a forward-facing step of an initially uniform flow features the spatiotemporal evolution of shock waves, developing transonic flow and large pressure gradients.
This test-case is, thus, well suited to test the conservation properties of the finite-volume discretisation as well as the stability of the pressure-based algorithm during the transient development of large pressure gradients. Following \citet{Woodward1984}, the height of the computational domain is $1 \, \textup{m}$, and the step has of height $0.2 \, \textup{m}$ and is positioned at $0.6 \, \textup{m}$ from the inlet of the domain. The flow entering the domain has a Mach number of $M = u/a_0 = 3$ and a pressure of $p_0 = 1 \, \textup{Pa}$. The two-dimensional domain is represented by an equidistant Cartesian mesh with $\Delta x = 0.01 \, \textup{m}$ and the applied time-step corresponds to $\textup{Co} = u \, \Delta t/ \Delta x = 0.75$. The considered fluid has a heat capacity ratio of $\gamma=1.4$ and a specific isobaric heat capacity of $c_p = 1008 \, \textup{J} \, \textup{kg}^{-1} \, \textup{K}^{-1}$, with a co-volume of either $b=0$ or $b=0.1 \, \textup{m}^3 \, \textup{kg}^{-1}$. Figure \ref{fig:ffsContours_b0} shows the contours of the Mach number and the pressure at $t=4\, \textup{s}$ for $b=0$, which are in good agreement with previously reported results \cite{Woodward1984, Jasak1996, Denner2018c}. Changing the co-volume to $b=0.1 \, \textup{m}^3 \, \textup{kg}^{-1}$, the position of the primary shock wave in front of the forward-facing step moves further upstream and fewer reflected shock waves can be observed, as seen in Fig.~\ref{fig:ffsContours_b0p1}.  

\begin{figure}
\begin{center}
\subfloat[$b=0$]
{\includegraphics{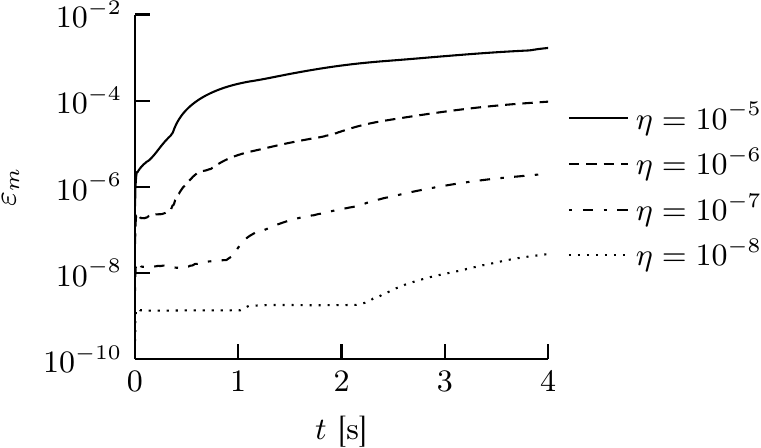}}
\qquad
\subfloat[$b=0.1 \, \textup{m}^3 \, \textup{kg}^{-1}$]
{\includegraphics{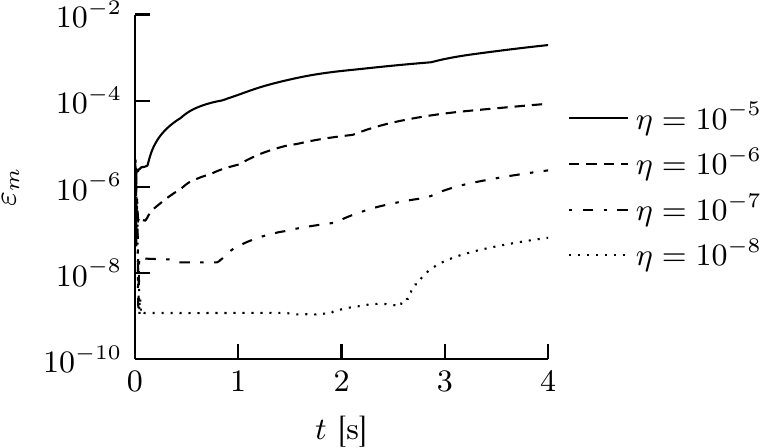}}
\caption{Temporal evolution of the mass conservation error, $\varepsilon_m$, as defined in Eq.~(\ref{eq:massError}), of the supersonic flow over a forward-facing step, obtained with different solution tolerances $\eta$.}
\label{fig:ffs_mass}
\end{center}
\end{figure}

Based on the initial mass $m^{(0)}$ at $t=0$, the mass in the domain $\Omega$ and the mass entering and leaving the domain over its boundaries $\partial \Omega$, the conservation error of mass at time $t$ is given as
\begin{equation}
\varepsilon_m (t) = \dfrac{1}{m^{(0)}} \left( m^{(0)} - \int_{\Omega} \rho(t) \, \textup{d}\Omega - \int_{0}^{t} \oiint_{\partial \Omega} \rho \, u_i \, \textup{d}\Sigma_{i} \, \textup{d}t \right), \label{eq:massError}
\end{equation}
where $\boldsymbol{\Sigma}$ is the outward-pointing surface vector of the surface $\partial \Omega$ of the computational domain $\Omega$.
The temporal evolution of the mass conservation error of the supersonic flow over the forward-facing step is shown in Fig.~\ref{fig:ffs_mass}, obtained with both considered co-volumes,  $b \in \{ 0, 0.1 \} \, \textup{m}^3 \, \textup{kg}^{-1}$, with different solution tolerances, $\eta \in \{10^{-5}, 10^{-6}, 10^{-7}, 10^{-8} \}$, applied for the solution of the system of governing equations (\ref{eq:eqsys}). Overall, the proposed finite-volume framework conserves mass accurately and the mass conservation error is predominantly a function of the solution tolerance, with a decreasing mass conservation error for a decreasing solution tolerance.

\subsection{Rotating sphere}
\label{sec:rotatingSphere}

\begin{figure}[t]
\begin{center}
\subfloat[Schematic (not to scale)]
{\includegraphics{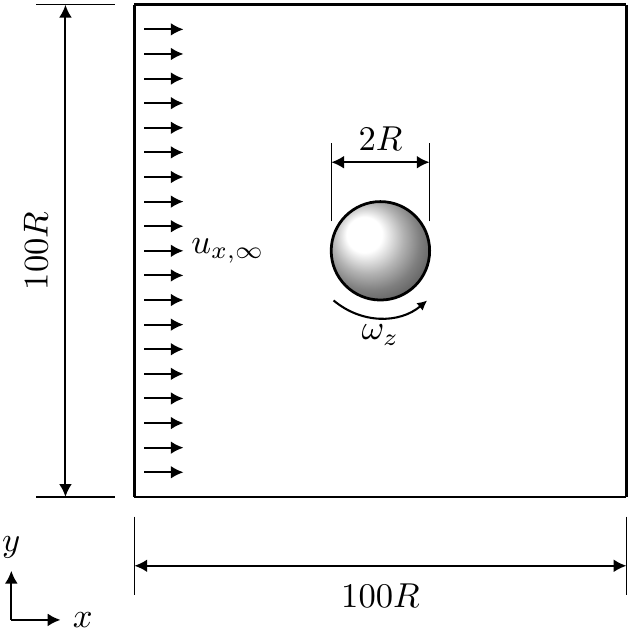}
\label{fig:rubinow_schematic}}
\qquad
\subfloat[Mesh with velocity contours]
{\includegraphics[width=0.445\textwidth]{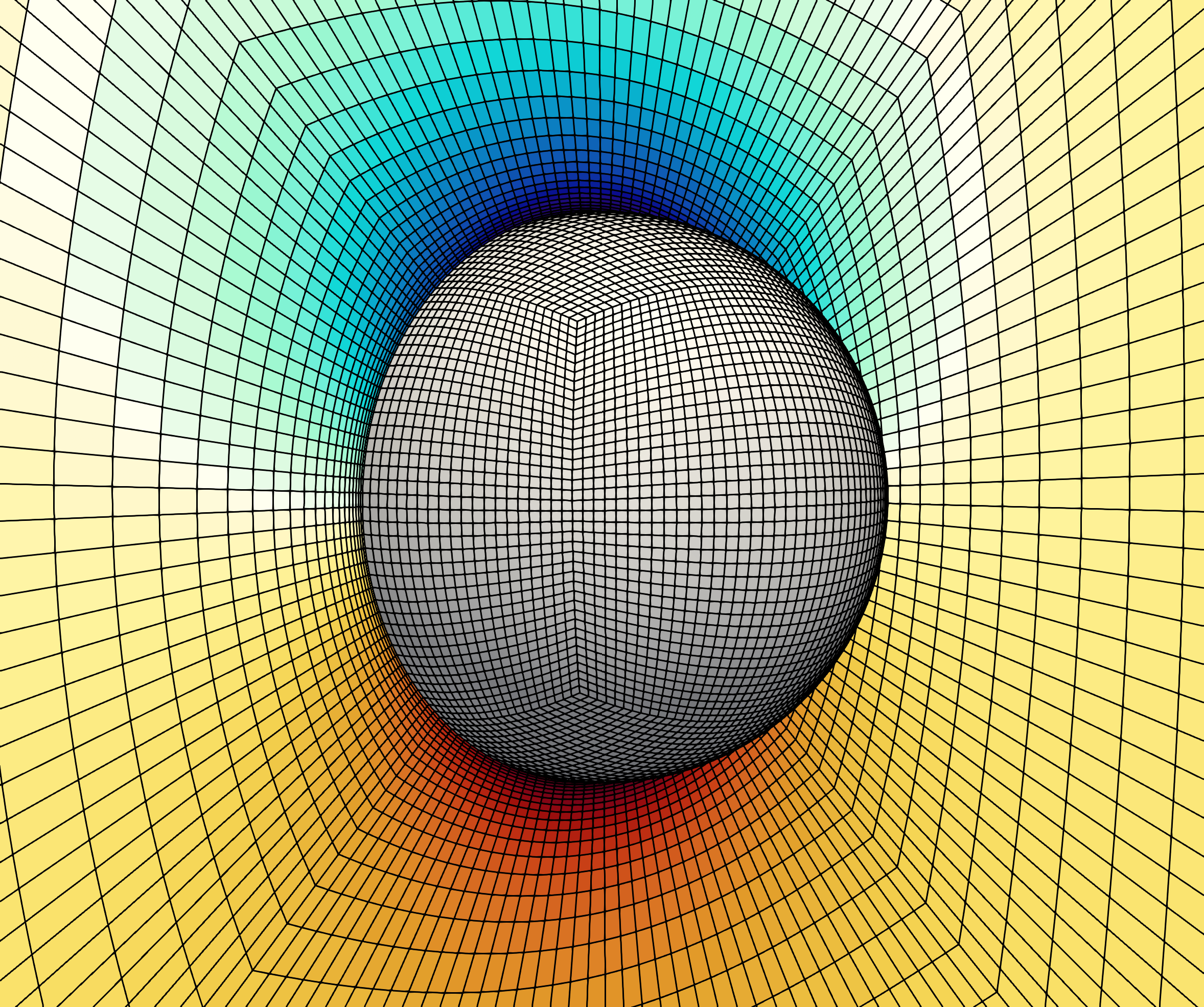}
\label{fig:rubinow_mesh}}
\caption{Schematic of the flow around a rotating sphere (in the $xy$-plane through the centre of the domain) and applied mesh in the vicinity of the sphere together with the contours of the axial velocity.}
\label{fig:rubinow_setup}
\end{center}
\end{figure}

The flow of an incompressible fluid around a sphere with radius $R$, rotating at angular velocity $\boldsymbol{\omega}$, in a Stokes flow with Reynolds number $\textup{Re} = \rho R |\boldsymbol{u}_\infty|/\mu \ll 1$, where $\boldsymbol{u}_\infty$ is the free-stream velocity, is considered. As a result of the rotation, a lift force is acting on the sphere, also known as Magnus effect, with the analytical solution for the force on the sphere given as \citep{Rubinow1961}
\begin{equation}
\boldsymbol{F} = 
- 6 \, \pi \, \mu \, R \, \boldsymbol{u}_\infty \left(1 + \frac{3}{8} \textup{Re} \right) +  \pi \, R^3 \rho \, \boldsymbol{\omega} \times \boldsymbol{u}_\infty \label{eq:sphereForceRubinow},
\end{equation}
where the first term on the right-hand side represents the drag force and the second term represents the lift force. 
The sphere is simulated in a cubical three-dimensional domain of size $100 R \times 100 R \times 100 R$, illustrated schematically in Fig.~\ref{fig:rubinow_schematic}, with the sphere placed at the centre of the domain. The considered flow has the free-stream velocity $\boldsymbol{u}_\infty = (u_{x,\infty},0,0)^T$, corresponding to $\textup{Re} = 0.05$, and the sphere rotates around its $z$-axis with $\boldsymbol{\omega} = (0, 0 , \omega_z)^T$. The computational domain is represented with a boundary-fitted hexahedral mesh with $384\,000$ cells, shown in Fig.~\ref{fig:rubinow_mesh}, which is strongly refined in the vicinity of the sphere and gradually coarsened (growth factor $1.2$) with increasing distance from the sphere. The applied time-step is $\Delta t = 100 \, t_\mu$, where $t_\mu = \rho R^2/\mu$ is the viscous timescale, which corresponds to a maximum Courant number of $\textup{Co} = 49-1559$, dependent on the angular velocity $\omega_z$, for the considered simulations. The transient term is discretised with the BDF2 scheme and the advection terms are discretised using the Minmod scheme.

\begin{figure}
\begin{center}
\subfloat[Drag coefficient]
{\includegraphics{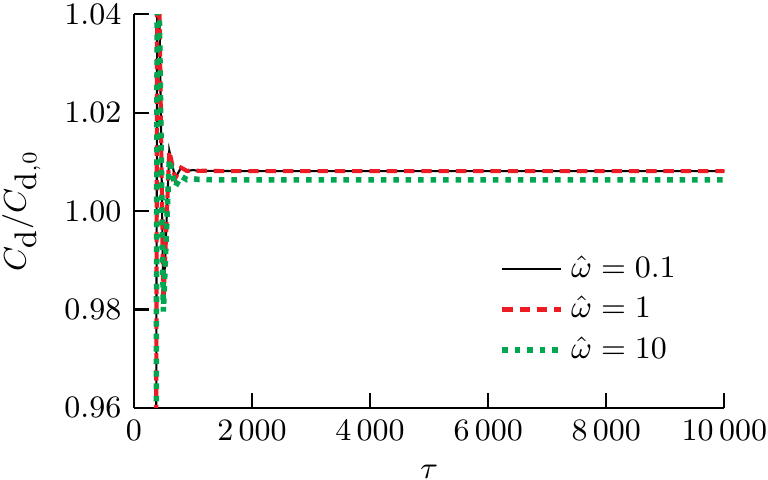}}
\
\subfloat[Lift coefficient]
{\includegraphics{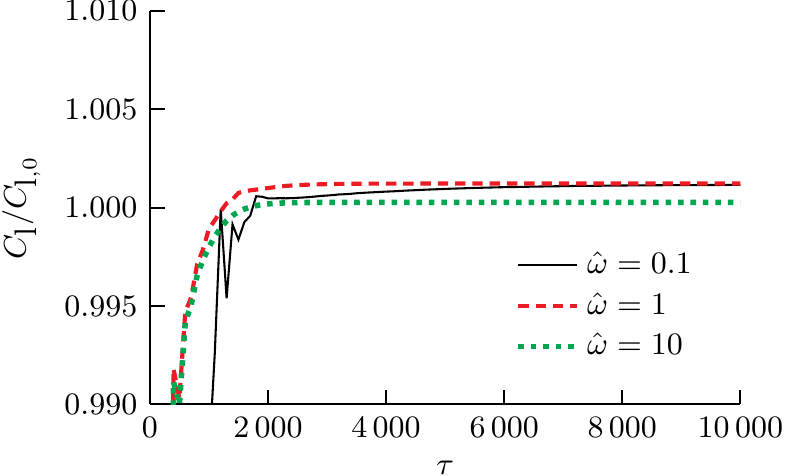}}
\caption{Drag coefficient $C_\textup{d}$ and lift coefficient $C_\textup{l}$ of the rotating sphere in Stokes flow for different dimensionless angular velocities $\hat{\omega}$ as a function of the dimensionless time $\tau = t/t_\mu$, normalised with the theoretical values, $C_{\textup{d},0}$ and $C_{\textup{l},0}$, based on Eq.~(\ref{eq:sphereForceRubinow}).}
\label{fig:rubinow_dragLift}
\end{center}
\end{figure}

Fig.~\ref{fig:rubinow_dragLift} shows the transient evolution of the drag coefficient, $C_\textup{d} = 2 F_\textup{d}/\rho A_\textup{p} u_{x,\infty}^2$, and the lift coefficient, $C_\textup{l} = 2 F_\textup{l}/\rho A_\textup{p} u_{x,\infty}^2$, with $A_\textup{p} = \pi R^2$ the projected area of the sphere, for three different dimensionless angular velocities, $\hat{\omega} = R \, \omega_z/u_{x,\infty}$, as a function of the dimensionless time $\tau = t/t_\mu$. For all three angular velocities the drag and lift coefficients are predicted accurately compared to the analytical solution, Eq.~(\ref{eq:sphereForceRubinow}), with errors $<1\%$ for both drag and lift coefficients.

\section{Conclusions}
\label{sec:conclusions}

A conservative numerical framework for the prediction of flows of incompressible, ideal-gas and real-gas fluids at all speeds has been presented. This numerical framework is founded on a unified thermodynamic closure model for incompressible and compressible fluids, a standard finite-volume discretisation applicable to structured and unstructured meshes, a single flux definition based on a momentum-weighted interpolation, as well as a fully-coupled pressure-based algorithm with collocated variable arrangement. 
The proposed unified thermodynamic closure model combines the definitions of incompressible fluids with the Noble-Abel-stiffened-gas model \citep{LeMetayer2016} for ideal-gas and real-gas fluids, which facilitates a straightforward finite-volume discretisation that is applicable to incompressible flows as well as compressible flows in all Mach number regimes. Since the thermodynamic closure model requires only the definition of the density and specific static enthalpy, it can be extended to more complex gas models, such as the Peng-Robinson model \citep{Peng1976}, without changes to the finite-volume discretisation or the pressure-based algorithm. 
The employed finite-volume framework combines well-established conservative discretisation schemes to yield a consistently second-order accurate discretisation that is applicable to structured and unstructured meshes. The discretised governing equations are solved in a single linear system of equations for pressure, velocity and \revE{temperature}, which enables a robust solution for flows at any speed.

The main feature of the proposed finite-volume discretisation and pressure-based algorithm is the accurate and robust simulation of flows of incompressible and compressible fluids at all speeds without changes to the discretisation or the solution procedure.
Using a Newton linearisation of the continuity equation in conjunction with the semi-implicit discretisation of the fluxes through the mesh faces by a momentum-weighted interpolation method, the discretised continuity equation acts as a transport equation for density in compressible flows and as a constraint on the  velocity field in incompressible flows. This allows this numerical framework to represent the incompressible limit correctly and enables the simulation of flows of both incompressible and compressible fluids with the same algorithm.

The proposed numerical framework has been validated using a broad variety of test-cases, demonstrating accurate and robust results, irrespective whether the considered flow was of an incompressible fluid, an ideal-gas fluid or a real-gas fluid, with an error convergence consistent of a second-order finite-volume discretisation. The propagation of acoustic waves demonstrated an accurate prediction of the speed of sound and acoustic effects in general, while the propagation of a moving contact discontinuity demonstrated convergence for linearly degenerate waves. The propagation of a strong shock wave as well as the shock tubes in different Mach number regimes scrutinised the resolution of strongly nonlinear and discontinuous flow features, which are predicted accurately in all Mach number regimes. In particular, the speed, position and strength of strong shock waves are predicted accurately, demonstrating that the finite-volume framework converges to the correct weak solution of the governing equations \citep{Hou1994}, further suggesting that the proposed algorithm implicitly satisfies the second law of thermodynamics. The evolution of Taylor vortices in an inviscid fluid offered the possibility to test the conservation of energy of the proposed numerical framework, showing that the momentum-weighted interpolation is the only source of numerical energy dissipation, an error which however converges with third order under mesh refinement. 
The Poiseuille flow of an incompressible fluid and the Couette flow of a compressible fluid demonstrated the accurate simulation of flows in which viscous stresses and heat conduction play a dominant role. 
\revE{The flow of an incompressible fluid in a lid-driven cavity at different Reynolds numbers further demonstrated the accurate simulation of flows in which both advection and diffusion play an important role, and demonstrated the correct enforcement of $\boldsymbol{\nabla} \cdot \boldsymbol{u} = 0$ for incompressible fluids to any chosen solver tolerance (within the limit of machine precision), on unstructured meshes}.
The results presented for the supersonic flow of an ideal gas and a real gas over a forward-facing step demonstrated accurate mass conservation, even for complex flows in which different Mach number regimes coexist. Lastly, the Stokes flow around a rotating sphere demonstrated that flows in complex three-dimensional geometries can be predicted accurately with the proposed numerical framework.

In this paper we have put forward a thermodynamic closure model, a finite-volume discretisation and a fully-coupled pressure-based algorithm for the prediction of the behaviour of the flow of incompressible fluids as well as compressible fluids described by ideal- or real-gas models on arbitrary meshes. We have combined these constituent parts into a fully-coupled pressure-based framework and have shown that this framework is able to predict realistic flows at any speed. However, these parts can also be used individually, for instance in existing frameworks.


\appendix

\section{Coefficients of the linear equation system} 
\label{sec:LGScoeff}

The coefficients of the discretised governing equations, Eqs.~(\ref{eq:continuityCoeff})-(\ref{eq:energyCoeff}), are given below. In order to simplify the presentation, the coefficients are given based on the assumption that cell $P$ is the upwind cell $U$ of face $f$ and using the BDF1 scheme for the discretisation of the transient terms.

For the discretised continuity equation (\ref{eq:continuityCoeff}), the pressure coefficients associated with cell $P$ and its neighbour cells $Q$ are
\begin{align}
\mathcal{A}^{\rho,p}_{P} & = \mathcal{C} \frac{V_P}{\left[(\gamma-1) \, c_v \, T_P^{(n)} + b \, (p_P^{(n)} + \Pi) \right] \Delta t_1} + \sum_f \left\{ \frac{\tilde{\rho}_f^{(n)} \hat{d}_f}{\Delta s_f} +  \mathcal{C}  \frac{ \vartheta_f^{(n)}}{(\gamma-1) \, c_v \, T_P^{(n)} + b \, (p_P^{(n)} + \Pi)} \right\} A_f \label{eq:A_p_rho_P} \\ 
\mathcal{A}^{\rho,p}_{Q} & = \sum_f - \frac{\tilde{\rho}_f^{(n)} \hat{d}_f}{\Delta s_f} A_f \label{eq:A_p_rho_Q} ,
\end{align}
respectively. 
The velocity coefficients, which arise from the implicit treatment of the advecting velocity of the advection term, associated with cell $P$ and its neighbour cells $Q$, are
\begin{align}
\mathcal{A}_{P}^{\rho,u_i} & = \sum_f \tilde{\rho}_f^{(n)} \, (1-l_{Pf}) \, n_{i,f} \, A_f \\ 
\mathcal{A}_{Q}^{\rho,u_i} & = \sum_f \tilde{\rho}_f^{(n)} \, l_{Pf} \, n_{i,f} \, A_f, 
\end{align}
respectively. 
The coefficient of the right-hand side vector, $\boldsymbol{\sigma}^{\rho}$, associated with cell $P$ is given as
\begin{equation}
\begin{split}
& \sigma^\rho_{P} = \left\{\rho_P^{(t-\Delta t_1)} - \mathcal{C} \frac{\Pi}{(\gamma-1) \, c_v \, T_P^{(n)} + b \, (p_P^{(n)} + \Pi)} - \mathcal{I} \rho_0 \right\} \frac{V_P}{\Delta t_1} + \sum_f \left\{ \vartheta_f^{(n)} - r_{j,f} \left. \overline{\frac{\partial u_i}{\partial x_j}} \right|_f^{(n)} n_{i,f} \right\}  \tilde{\rho}_f^{(n)} A_f\\
& - \sum_f \left\{ \hat{d}_f \left[ \rho_f^{\ast (n)} \left(\left. \frac{1-l_{Pf}}{\rho_P^{(n)}} \frac{\partial p}{\partial x_i} \right|_p^{(n)} + \left. \frac{l_{Pf}}{\rho_Q^{(n)}} \frac{\partial p}{\partial x_i} \right|_Q^{(n)} \right)  {s}_{i,f} + \frac{\rho^{\ast (t-\Delta t_1)}_f}{\Delta t_1} \left(\vartheta^{(t-\Delta t_1)}_f - \overline{u}_{i,f}^{(t-\Delta t_1)} {n}_{i,f} \right) \right] \right\} \tilde{\rho}_f^{(n)} A_f \\
& - \sum_f \left\{ \mathcal{C} \left[ \frac{\Pi - \delta_f (p_P^{(n)} + \Pi)}{(\gamma-1) c_v T_P^{(n)} + b (p_P^{(n)} + \Pi)} +  \frac{\delta_f (p_Q^{(n)} + \Pi) }{(\gamma-1) c_v T_Q^{(n)} + b  (p_Q^{(n)} + \Pi)}  \right]   + \mathcal{I} \, \rho_0  \right\} \vartheta_f^{(n)} A_f , \label{eq:S_rho}
\end{split} 
\end{equation}
where $\delta_f = \xi_f |\boldsymbol{r}_{Pf}|/\Delta s_f$ is the weighting coefficient that follows from the TVD discretisation of the advection term, see Section \ref{sec:advectionTerm}. 
 
For the discretised momentum equations (\ref{eq:momentumCoeff}), the pressure coefficients are given as
\begin{align}
\mathcal{A}^{\rho u_j, p}_{P} & = \mathcal{C} \frac{u_{j,P}^{(n)} \, V_P}{\left[(\gamma-1) \, c_v \, T_P^{(n)} + b \, (p_P^{(n)} + \Pi) \right] \Delta t_1} \nonumber
\\ & + \sum_f \left\{ \frac{\tilde{\rho}_f^{(n)} \tilde{u}_{j,f}^{(n)} \, \hat{d}_f}{\Delta s_f} +  \mathcal{C}  \frac{ \vartheta_f^{(n)}  \, \tilde{u}_{j,f}^{(n)}}{(\gamma-1) \, c_v \, T_P^{(n)} + b \, (p_P^{(n)} + \Pi)} + (1-l_{Pf}) \, n_{j,f} \right\} A_f \\
\mathcal{A}^{\rho u_j, p}_{Q} & = \sum_f \left\{-\frac{\tilde{\rho}_f^{(n)}  \tilde{u}_{j,f}^{(n)} \, \hat{d}_f}{\Delta s_f} + l_{Pf} \, n_{j,f} \right\} A_f.
\end{align}
The coefficients associated with velocity $u_j$ are given as
\begin{align}
\mathcal{A}^{\rho u_j, u_j}_{P} & = \frac{\rho_P^{(n)} \, V_P}{\Delta t_1} +  \mathcal{D}^{\rho u_j,u_j}_{P}
\\
\mathcal{A}^{\rho u_j, u_j}_{Q} & = -\sum_f \frac{\alpha_f \mu_f}{\Delta s_f} \, A_f,
\end{align}
where 
\begin{align}
\mathcal{D}^{\rho u_j,u_j}_{P} & = \sum_f \left\{ \tilde{\rho}_f^{(n)} \vartheta_f^{(n)}  + \frac{\alpha_f \mu_f}{\Delta s_f} \right\} A_f \label{eq:coeffDiagonalVelP} 
\end{align}
is the coefficient arising from the advection of velocity and the implicit velocity contribution of the decomposed shear stress term, which is used for the definition of the advection velocity $\vartheta_f$, see Section \ref{sec:advectingVel}.
The coefficients of the velocity components that arise from the implicit treatment of the advecting velocity of the advection term are
\begin{align}
\mathcal{A}^{\rho u_j, u_i}_{P} & = \sum_f \tilde{\rho}_f^{(n)} \, \tilde{u}_{j,f}^{(n)} \, (1-l_{Pf}) \, n_{i,f} \, A_f \\
\mathcal{A}^{\rho u_j, u_i}_{Q} & = \sum_f \tilde{\rho}_f^{(n)} \, \tilde{u}_{j,f}^{(n)} \, l_{Pf} \, n_{i,f} \, A_f.
\end{align}
The coefficient of the right-hand side subvector $\boldsymbol{\sigma}^{\rho u_j}$ follows as
\begin{equation}
\begin{split}
& \sigma^{\rho u_j}_{P} =  \left\{ \rho_P^{(t-\Delta t_1)} u_{j,P}^{(t-\Delta t_1)}  - \mathcal{C} \frac{u_{j,P}^{(n)} \, \Pi}{(\gamma-1) \, c_v \, T_P^{(n)} + b \, (p_P^{(n)} + \Pi)} - \mathcal{I} \rho_0 u_{j,P}^{(n)} + \rho_P^{(n)} u_{j,P}^{(n)} \right\} \frac{V_P}{\Delta t_1} \\
& - \sum_f \left\{ \vartheta_f^{(n)} \delta_f \left(u_{j,Q}^{(n)} - u_{j,P}^{(n)} \right)  +  \tilde{u}_{j,f}^{(n)} r_{k,f} \left. \overline{\frac{\partial u_i}{\partial x_k}} \right|_f^{(n)} n_{i,f} \right\}  \tilde{\rho}_f^{(n)} A_f \\
& - \sum_f \left\{ \hat{d}_f \left[ \rho_f^{\ast (n)} \left(\left. \frac{1-l_{Pf}}{\rho_P^{(n)}} \frac{\partial p}{\partial x_i} \right|_p^{(n)} + \left. \frac{l_{Pf}}{\rho_Q^{(n)}} \frac{\partial p}{\partial x_i} \right|_Q^{(n)} \right)  {s}_{i,f} + \frac{\rho^{\ast (t-\Delta t_1)}_f}{\Delta t_1} \left(\vartheta^{(t-\Delta t_1)}_f - \overline{u}_{i,f}^{(t-\Delta t_1)} {n}_{i,f} \right) \right] \right\} \tilde{\rho}_f^{(n)} \tilde{u}_{j,f}^{(n)} A_f \\
& - \sum_f \left\{ \mathcal{C} \left[ \frac{\Pi - \delta_f (p_P^{(n)} + \Pi)}{(\gamma-1) c_v T_P^{(n)} + b (p_P^{(n)} + \Pi)} +  \frac{\delta_f (p_Q^{(n)} + \Pi) }{(\gamma-1) c_v T_Q^{(n)} + b  (p_Q^{(n)} + \Pi)}  \right]  + \mathcal{I} \, \rho_0 \right\} \vartheta_f^{(n)} \tilde{u}_{j,f}^{(n)} A_f \\
& + \sum_f \left\{ 2 \tilde{\rho}_f^{(n)} \vartheta_f^{(n)} \tilde{u}_{j,f}^{(n)} -  r_{i,f} \left. \overline{\frac{\partial p}{\partial x_i}} \right|_f^{(n)} n_{j,f}   +  \mu_f \left.
\overline{\frac{\partial u_j}{\partial x_i}} \right|_f^{(n)} ({n}_{i,f} - \alpha_f {s}_{i,f}) + \mu_f \left.
\overline{\frac{\partial u_i}{\partial x_j}} \right|_f^{(n)} {n}_{i,f} -
\frac{2}{3} \, \mu_f \left. \overline{\frac{\partial u_k}{\partial x_k}} \right|_f^{(n)}
{n}_{i,f} \right\}  A_f .
\end{split}
\end{equation}

The coefficients of the discretised energy equation (\ref{eq:energyCoeff}) follow in a similar fashion, with the pressure coefficients given as
\begin{align}
\mathcal{A}^{\rho h, p}_{P} & = \left\{\mathcal{C} \left[\revE{\rho_P^{(n)} \, b} + \frac{h_P^{(n)}}{(\gamma-1)  c_v  T_P^{(n)} + b (p_P^{(n)} + \Pi)} \right] - 1 \right\} \frac{V_P}{\Delta t_1} \nonumber \\ & + \sum_f \left\{ \frac{\tilde{\rho}_f^{(n)} \tilde{h}_{f}^{(n)} \, \hat{d}_f}{\Delta s_f} +  \mathcal{C} \left[\revE{\tilde{\rho}_f^{(n)} \, \vartheta_f^{(n)} \, b} + \frac{ \vartheta_f^{(n)}  \, \tilde{h}_{f}^{(n)}}{(\gamma-1) c_v T_P^{(n)} + b (p_P^{(n)} + \Pi)} \right] \right\} A_f \\
\mathcal{A}^{\rho h, p}_{Q} & = \sum_f - \frac{\tilde{\rho}_f^{(n)}  \tilde{h}_{f}^{(n)} \, \hat{d}_f}{\Delta s_f} A_f,
\end{align}  
the velocity coefficients given as
\begin{align}
\mathcal{A}^{\rho h,u_i}_{P} & = \sum_f (1-l_{Pf}) \left\{\tilde{\rho}_f^{(n)} \, \tilde{h}_f^{(n)} \, {n}_{i,f} - \mu_f \left( \left. \overline{\frac{\partial u_j}{\partial x_i}} \right|_f^{(n)} + \left. \overline{\frac{\partial u_i}{\partial x_j}} \right|_f^{(n)}  - \frac{2}{3} \left. \overline{\frac{\partial u_k}{\partial x_k}} \right|_f^{(n)} {n}_{j,f} \right)\right\}  A_f 
\\
\mathcal{A}^{\rho h,u_i}_{Q} & = \sum_f l_{Pf} \left\{\tilde{\rho}_f^{(n)} \, \tilde{h}_f^{(n)} \, {n}_{i,f} - \mu_f \left( \left. \overline{\frac{\partial u_j}{\partial x_i}} \right|_f^{(n)} + \left. \overline{\frac{\partial u_i}{\partial x_j}} \right|_f^{(n)}  - \frac{2}{3} \left. \overline{\frac{\partial u_k}{\partial x_k}} \right|_f^{(n)} \right){n}_{j,f} \right\}  A_f 
\end{align}
and the coefficients of the \revE{temperature} given as
\begin{align}
\mathcal{A}^{\rho h, \revE{T}}_{P} & = \revE{c_p} \left( \frac{\rho_P^{(n)} \, V_P}{\Delta t_1} + \sum_f \tilde{\rho}_f^{(n)} \vartheta_f^{(n)} A_f \right) \revE{+ \sum_f \frac{\alpha_f \, k_f}{\Delta s_f} \, A_f} \\
\mathcal{A}^{\rho h, \revE{T}}_{Q} & = \revE{- \sum_f \frac{\alpha_f \, k_f}{\Delta s_f} \, A_f}.
\end{align}
The coefficient of the right-hand side subvector $\boldsymbol{\sigma}^{\rho h}$ follows as
\begin{equation}
\begin{split}
& \sigma^{\rho h}_P =  \left\{ \rho_P^{(t-\Delta t_1)} h_{P}^{(t-\Delta t_1)} \revE{- \rho_P^{(n)} \frac{\boldsymbol{u}_P^{(n),2}}{2}} - \mathcal{C} \frac{h_{P}^{(n)} \, \Pi}{(\gamma-1) \, c_v \, T_P^{(n)} + b \, (p_P^{(n)} + \Pi)} - \mathcal{I} \rho_0 h_{P}^{(n)} + \rho_P^{(n)} h_{P}^{(n)}  - p^{(t-\Delta t_1)}_P \right\} \frac{V_P}{\Delta t_1} \\
& - \sum_f \left\{\revE{ \vartheta_f^{(n)} \, \frac{\boldsymbol{u}_P^{(n),2}}{2}} + \vartheta_f^{(n)} \delta_f \left(h_{Q}^{(n)} - h_{P}^{(n)} \right) + \tilde{h}_{f}^{(n)} r_{j,f} \left. \overline{\frac{\partial u_i}{\partial x_j}} \right|_f^{(n)} n_{i,f} \right\} \tilde{\rho}_f^{(n)}  A_f \\
& - \sum_f \left\{ \hat{d}_f \left[ \rho_f^{\ast (n)} \left(\left. \frac{1-l_{Pf}}{\rho_P^{(n)}} \frac{\partial p}{\partial x_i} \right|_p^{(n)} + \left. \frac{l_{Pf}}{\rho_Q^{(n)}} \frac{\partial p}{\partial x_i} \right|_Q^{(n)} \right)  {s}_{i,f} + \frac{\rho^{\ast (t-\Delta t_1)}_f}{\Delta t_1} \left(\vartheta^{(t-\Delta t_1)}_f - \overline{u}_{i,f}^{(t-\Delta t_1)} {n}_{i,f} \right) \right] \right\} \tilde{\rho}_f^{(n)} \tilde{h}_{f}^{(n)} A_f \\
& - \sum_f \left\{ \mathcal{C} \left[ \frac{\Pi - \delta_f (p_P^{(n)} + \Pi)}{(\gamma-1) c_v T_P^{(n)} + b (p_P^{(n)} + \Pi)} +  \frac{\delta_f (p_Q^{(n)} + \Pi) }{(\gamma-1) c_v T_Q^{(n)} + b  (p_Q^{(n)} + \Pi)}  \right]  + \mathcal{I} \, \rho_0 \right\} \vartheta_f^{(n)} \tilde{h}_{f}^{(n)} A_f \\
& + \sum_f \left\{ 2 \tilde{\rho}_f^{(n)} \vartheta_f^{(n)} \tilde{h}_{f}^{(n)} 
+ k_f 
 \left. \overline{\frac{\partial T}{\partial x_i}} \right|_f^{(n)} ({n}_{i,f} - \alpha_f {s}_{i,f})
+ r_{l,f} \left. \overline{\frac{\partial u_i}{\partial x_l}} \right|_f^{(n)} \mu_f \left( \left. \overline{\frac{\partial u_j}{\partial x_i}} \right|_f^{(n)} + \left. \overline{\frac{\partial u_i}{\partial x_j}} \right|_f^{(n)}  - \frac{2}{3} \left. \overline{\frac{\partial u_k}{\partial x_k}} \right|_f^{(n)} \right) {n}_{j,f}  \right\}  A_f .
\end{split}
\end{equation}

\end{document}